\documentclass[12pt]{article}
\title{PhD Thesis\\Screening in Hot non-Abelian Plasma}
\author{P\'eter Petreczky}
\date{\today}
\usepackage{defs,amssymb,epsfig}
\begin{document}
\thispagestyle{empty}
\newpage
\null
{\Large
\vfill
\centerline{\cim Screening in Hot}
\vskip0.5truecm
\centerline{\cim Non-Abelian Plasma}
\vskip1truecm
\centerline{(PhD Thesis)}
\vskip1.5truecm
\centerline{P\'eter Petreczky}
\centerline{June 25, 1999}
\vskip4truecm
\hbox to \hsize{
\null\hfill
\parbox{6.5truecm}{
E\"otv\"os Lor\'and University
Budapest\\
Supervisor:\\
Andr\'as Patk\'os
}}
\vfill}
\newpage
\tableofcontents
\clearpage
\newpage
\section{Introduction}
According to the standard cosmological model after the Big-Bang the
density and the temperature of the 
early Universe  were very high and it consisted actually of
a plasma of different elementary particles.
This fact  is a great opportunity for particle physics since the early
Universe thus serves as a laboratory for particle physics. In order to 
describe the matter under such extreme conditions a new formalism has
been developed in the past 25 years 
which is nowadays called {\em finite temperature quantum field
theory} (see Refs. \cite{kapusta, lebellac} for detailed monographies
on the subject). 

As the expanding  Universe was cooling down it underwent several phase 
transitions. Two of these phase transitions,
namely the {\em electroweak phase transition} (EWPT), which occured when
the temperature was $T \sim 200 GeV$ and the {\em QCD deconfinement phase
transition}, which happened at temperature $T \sim 200 MeV$ are of
special interest. The underlying
theories are well established theoretically and tested experimentally 
\footnote{ The Higgs particle is not yet discovered and supersymmetric
extensions of the electroweak theory might be relevant}. The
neighborhood of EWPT is relevant for the understanding the evolution
of the
baryon asymmetry of the Universe (see \cite{dolgov92,rubakov96}
for reviews). 
The theoretical study of QCD deconfinement phase
transition is of great practical interest because there is a hope to create 
{\em quark-gluon plasma} in relativistic heavy ion collisions.
Actually, the possibility of producing the quark-gluon plasma state
experimentally was  one
of
the main motivations for detailed studies  of this phase
transition. One may hope to suggest signals of its existence
which can be tested experimentally and to measure some of the 
thermodynamical properties of the new phase which can be confronted
with the expectations based on theoretical models.

One of the most natural signatures of the deconfined phase is the screening of
static chromoelectric fields. Actually this was the first signature
based on which the existence of the new phase was proven theoretically
\cite{kuti81,mclerran81}. 
More precisely in the confined phase the potential between
heavy quark-antiquark pair rises linearly with the distance while
in the deconfined phase this potential is exponentially screened,
with an inverse screening length equal to the Debye screening mass.
The screening of static chromoelectric
fields is also of great phenomenological interest in detecting the
quark-gluon plasma \cite{wang,satz}.

In the ideal case when the temperature is sufficiently high the
perturbative approach seems to be adequate. The coupling constant is
small as the consequence of asymptotic freedom : at high temperature 
the only scale which is available is the temperature scale itself.
The coupling constant is small at large temperature $T$:
$g(T)\sim 1/\ln(T/\Lambda_{QCD})$. Thus naively one would think that the
quark-gluon plasma 
is
close to the ideal gas of quarks and gluons
(the situation is similar for the high temperature electroweak plasma). 
However, this simple 
picture is spoiled by infrared divergencies. It turns out that beyond
the natural length scale $1/T$ there are different screening legth
scales, namely the electric, or the Debye scale 
$1/gT$ and  the magnetic length scale
$1/g^2T$. While the electric scale is similar to some extent to the
Debye screening scale of QED plasma the magnetic scale is present 
only in the non-Abelian case. Over the past few years important progress has
been made in our understanding of the role of these scales
\cite{braaten90,jako94,kajantie96a,braaten95,braaten96}. 
Electric screening scale is very important in phenomenological
investigation of the thermodynamics of the quark-gluon plasma and
calculating the signals of the existance of this state
\cite{levai98,levai97}.
The existence of the 
electric screening scale sets the boundary of applicability of the of naive
perturbative approach and provides a possibility to cure
some infrared divergencies and to perform  a consistent resummation of
the perturbative series \cite{braaten90,braaten95}. In the non-Abelian 
gauge theory there are, however, also infrared divergencies due 
to the presence of the magnetic mass scale. This leads to the breakdown of
the perurbative approach beyond some order of perturbation theory
\cite{linde80}. It is believed that the existence of the magnetic mass
scale is responsible for the termination of the $1^{st}$ order phase
transition in the electroweak theory.  This occurs when the  Higgs mass 
is larger than
some critical value \cite{buchmuller97}.

In the calculation of static properties at high temperature
the most efficient way to deal with the presence of different legth
scales is the procedure of gradual reduction of the theory
\cite{ginsparg80,appelquist81,landsman89}
\footnote{The gradual reduction approach is based on integrating
out consecutively the heavy fields in the Euclidean
path integral. In the first step the non-zero Matsubara modes with
typical mass  $\sim \pi T$
are integrated out leaving an effective 3d theory. Therefore
this approach is also referred to as dimensional reduction. For
small enough couplings futher reduction is possible by
integrating out the static $A_0$ field (temporal component of
the gauge field) with a mass $\sim g T$.}. 
In the past few years considerable progress has been
made in understanding  the electroweak phase transition based on
this approach \cite{jako94,farakos94,kajantie96a,kajantie96b,kajantie97ew} 
(see Ref.\cite{rubakov96} for a review). Dimensional 
reduction was also successfully applied to the deconfined
phase of QCD \cite{reisz92,bielcoll,bielcollsu3,kark94}.
The dimensionally reduced QCD was used for non-perturbative
definition of the chromoelectric screening \cite{arnold95,kajantie97a}. 
However, more recent 
investigations revealed some problems in application of the
dimensional reduction in QCD \cite{kajantie97b}.
     
        The present thesis is the comprehensive summary of my work
devoted to the  investigation of field theoretical screening phenomena
by resummation of perturbative series and lattice
Monte-Carlo technique. Also I have examined  the consistency and
precision of dimensional reduction. The organisation of the presented
thesis is the following. In section 2 I am going to review the present
status of chromoelectric and chromomagnetic screening and their
applications to high temperature non-Abelian theories. Section 3 devoted
to the self-consistent determination of the screening masses of hot SU(N)
gauge theories. In section 4 the validity of dimensional
reduction is carefully examined and the screening masses are
defined by non-perturbative (lattice Monte-Carlo) technique. 
In section 5 the screening masses of SU(2) Higgs model will be 
examined and a new procedure for improving upon the standard
of dimensional reduction will be proposed.

Sections 3, 4 and 5 contain the results obtained by the author
in collaboration with F. Karsch, M. Oevers, A. Patk\'os and Zs. Sz\'ep.           
        
\newpage
\section{Review of non-Abelian Screening}
In this section a detailed review of non-Abelian screening
phenomenon will be given. The first subsection deals with
definition of non-Abelian Debye screening at leading order of
perturbation theory. Also a nice physical
picture for the non-Abelian chromoelectric screening will be
given there. In the following subsections we will discuss infrared 
problems of finite temperature gauge theories and will introduce
the idea of dimensional reduction. Finally, the problem of
chromoelectric screening beyond the leading order of perturbation
theory will be reviewed.

\subsection{Chromoelectric Screening at Leading Order}

\subsubsection{Chromoelectric Screening in Linear Response
Theory}
The phenomenon of chromoelectric screening could be easily
understood as the field induced by static colour source placed into
the plasma. First let us study the response  of the plasma to
an arbitrary weak external field. The classical source
which induces this field is $j^{\mu}_{cl}$. The perturbation
induced by this source is 
\be
V=\int d^3 x j^{\mu}_{cl} \hat A_{\mu}(x)
\ee
The classical field induced by this extrernal source is
\cite{lebellac}
\be
A_{\mu}(x)=<\hat A_{\mu}(x)>=-i \int d^3 x' D^R_{\mu \nu}(x-x') 
j^{\nu}_{cl}(x),
\ee
where $D^R_{\mu \nu}(x-x')$ is the retarded gluon propagator.
In the momentum representation the retarded gluon propagator 
could be written as \cite{lebellac}
\be
D^R_{\mu \nu}={i\over K^2-\Pi_T(K)}P^T_{\mu \nu}+{i\over
K^2-\Pi_L(K)} P^L_{\mu \nu}-i\xi {K_{\mu} K_{\nu}\over K^4},
\label{decomprop}
\ee
where the following projectors were introduced
\be
P^T_{\mu \nu}(K)=\delta_{\mu}^i (\delta_{ij}-{k_i k_j\over k^2}) \delta_{\nu}^j,
\ee
\be
P^L_{\mu \nu}(K)=(\delta_{\mu \nu}-{K_{\mu} K_{\nu}\over
K^2})-P^T_{\mu \nu}.
\ee
In the case of static chromoelectric source $J^{\mu}(x)=Q
\delta^{\mu}_0 \delta^3({\bf x})$ the induced potential
is 
\be
\Phi(r)=Q \int {d^3 k\over {(2 \pi)}^3} {e^{i \bf k x}\over
k^2+\Pi_L(k_0=0,{\bf k})}=\nonumber
\ee
\be
={Q\over 2 \pi^2} \int^{\infty}_{-\infty}{e^{ikr}-e^{-ikr}\over
2 i r} {k dk \over k^2+\Pi_L(0,k)}.
\ee
At leading order in the high temperature limit 
\be
\Pi_L(k_0=0,k)=\Pi_{00}(k_0=0,k)={1\over 3} (N+N_f/2) g^2 T^2=m_{D0}^2.
\label{phiint}
\ee
for $SU(N)$ and $N_f$ fermion flavour.
Now Eq. (\ref{phiint}) could be easily eveluated by closing the
contour in the upper and lower half complex k-plane and picking up
contribution from simple poles at $k=\pm i m_{D0}$, which yields
\be
\Phi(r)={Q\over 4 \pi r} e^{-m_{D0}r}.
\ee
Thus we can see that the field induced by static colour charge in
the plasma is described by screend Coulomb potential. This phenomenon
is called chromoelectric or Debye screening \footnote{Analogous phenomenon
is well known for the electron plasma} and the
corresponding inverse screening length is called the Debye mass. 

\subsubsection{Debye screening from classical kinetic theory}
The leading order Debye screening mass is determined by the high
temperature limit of the one-loop self energy diagram. It was shown
that the contribution of such diagrams, the so-called hard
thermal loop can be described by an effective kinetic theory 
\cite{blaziot94}.
Moreover a simple classical kinetic description could be given for
the associated phenomena \cite{kelly94}.
The idea of the classical kinetic description is the following.

The large momentum modes ($p \gtrsim T$) of the gauge and fermion
fields are described as
classical particles with some one-particle distributions. If
there are no external fields all particles have equilibrium distributions.
If one applies a weak external field to the plasma, the one-particle
distributions change and induced currents appear which are proportional
to the change of the on-particle distributions. The induced currents create
some classical field which is the response of the plasma to external
perturbation. This idea should be worked out self-consistently.
For this it is necessary to write down the equations of motion
for the set of dynamical variables which in non-Abelian case are $x^{\mu}$,
$p^{\mu}$ and $Q^a$, where $Q^a$ is the classical colour charge. Unlike
in Abelian theory here the colour charge $Q^a$ is also subject to
dynamical evolution
if particles interact with external field. The corresponding equations
of motion are \cite{wong70}
\ba
&&
m {d x^{\mu}\over d \tau}=p^{\mu},\\
&&
m {d p^{\mu}\over d \tau}=g Q^a F^{\mu \nu}_a p_{\nu},\\
&&
m {d Q^a\over d \tau}=-g f^{abc} p^{\mu} A_{\mu}^b Q^c.
\ea
Now we can write the collisionless relativistic Boltzmann equation
for the one-particle distribution $f_i(x,p,Q)$ for the i-th type of
particle:
\be
p^{\mu} \left( {\partial \over \partial x^{\mu}}-
g Q_a F^a_{\mu \nu} {\partial \over \partial p_{\nu}}-g f_{abc}
A^b_{\mu} Q^c {\partial \over \partial Q_a} \right)f_i(x,p,Q)=0.
\label{boltz}
\ee
If the external field and the gauge coupling are small
the distribution $f$ can be expanded in powers of $g$
\be
f=f^{(0)}+g f^{(1)}+...,
\ee
where $f^{(0)}$ is the thermal equilibrium distribution function.
The induced current which arises due to the departure from equilibrium is 
\be
J^{\mu}_a=g^2 \sum_i \int dP dQ p^{\mu} f_i^{(1)}(x,p,Q) Q_a,
\ee
where summation stands for all species of particles which are present
in the plasma. From Eq. (\ref{boltz}) one finds \cite{kelly94}
\be
f^{(1)}=Q_a \biggr[ A_0^a(x)-
{p_{\mu} \over p\cdot \partial}\partial_0 A_{\mu}^a(x)
\biggr] {d f_i^{(0)}\over d p_0}
\ee
and the induced current can be written as
\be
J^{\mu}_a=g^2 \sum_i \int dP dQ p^{\mu} Q_a Q_b 
\biggr(A_0^b(x)-{p^{\nu}\over p\cdot
\partial}\partial_0 A_{\nu}^b(x)\biggr) {d f^{(0)}_i\over d p_0}.
\ee
If one takes into account that
\be
dP={d^4 p\over {(2 \pi)}^4} 2 \theta(p_0) \delta(p^2-m^2) 
\ee
and
\be
\int dQ Q^a Q^b=\cases{N \delta^{ab}, {\rm for~~~gluons}\cr
{1\over 2} \delta^{ab}, {\rm for~~~quarks}}
\ee
one arrives at the following expression for the induced current
\be
J^{\mu}_a(x)=m^2_{D0} \int{d \Omega\over 4 \pi} v^{\mu}i\biggr (
{v_{\nu}
\over v \cdot \partial } \partial_0 A^{\nu}_a(x)-A^0_a(x)\biggr),
\ee 
where $v=(1,\hat {\bf p}),~~\hat {\bf p}={\bf p}/|{\bf p}|$
and $m_{D0}^2=1/3 g^2 T^2 (N+1/2 N_f)$ for $N_f$ fermion flavours.
For static fields the above expression simply reduces to 
\be
J^{\mu}_a(x)=-\m0^2 A^0_a({\bx}) \delta^{\mu}_0.
\ee
If this expression is inserted to the corresponding field equation
the screened Coulomb potential is easily recovered.

\subsubsection{Polyakov Loop Correlator and the Debye Screening}

Quantities which characterize the phases of a gauge theory are
the free energies of static configurations of quarks and antiquarks.
Following Ref. \cite{mclerran81} let us introduce operators,
$\psi_a^{\dagger}(\br_i,t)$ and $\psi_a(\br_i,t)$, which
create and anihilate static quarks with colour $a$ at position
$\br_i$ and time $t$, along with their conjugates $\psi_a^{\dagger c}$ and
$\psi^c_a$ for antiquarks.
These static fields satisfy equal time anticommutation relations
\be
[\psi_a(\br_i,t),\psi_b^{\dagger}(\br_j,t)]_{+}=\delta_{ij} \delta_{ab}
\label{com1}
\ee
The evolution equation for the static quark fields is
\be
\left({1\over i}{\partial\over \partial t}-\vec{\tau}\cdot 
\vec{A_0}\left(\br_i,t\right)\right)\psi(\br_i,t)=0, 
\ee
where $\vec{\tau}\cdot\vec{A_0}=\tau^a A_0^a$ with $\tau^a$ being the
generators of SU(N) gauge group.
The above equation may be formally integrated to yield
\be
\psi({\bf r}_i,t)= T \exp\left(i \int_0^t \vec{\tau} \cdot 
\vec{A_0}\left(\br_i,t\right)\right)\psi(\br_i,0)
\label{psit}
\ee
The free energy of a configuration of $N_q$ quarks and 
$N_{\bar q}$ at positions $\br_1,....,\br_{N_q}$ and 
$\br_1',......,\br_{N_{\bar q}}'$
\be
\exp(-\beta F(\br_1,.....,\br_{N_q},\br'_1,.....,\br'_{N_{\bar q}})=
{1\over N^{N_q+N_{\bar q}}} \sum_s <s|e^{-\beta H}|s>,
\ee
with summation over all states with heavy quarks at $\br_1,....,\br_{N_q}$
and antiquarks at $\br_1',......,\br_{N_{\bar q}}'$, $\beta=T^{-1}$ and
$N$ being the dimension of the SU(N) algebra.
The above expression could be rewriten as
\ba
&&
\exp(-\beta F_{N_q, N_{\bar q}})=\nonumber\\
&&
={1\over N^{N_q+N_{\bar q}} } \sum_{s'} <s'| \sum_{(ab)}
\psi_{a_1}(\br_1,0)....\psi_{a_{N_q}}(\br_{N_q},0)
\psi_{b_1}^c(\br_1',0)....\psi_{b_{N_{\bar q}}}^c(\br_{N_{\bar
q}}',0)\nonumber\\
&&
\times e^{-\beta H}
\psi_{a_1}^{\dagger}(\br_1,0)....
\psi_{a_{N_q}}^{\dagger}(\br_{N_q},0)
\psi_{b_1}^{\dagger c}(\br_1',0)....
\psi_{b_{N_{\bar q}}}^{\dagger c}(\br_{N_{\bar
q}}',0)|s'>,
\label{f1}
\ea
where now the sum is over all states with no heavy quarks.
Since $e^{-\beta H}$ generates Euclidian time translations, 
i.e. $e^{\beta H} \hat O(t) e^{-\beta H}=
\hat O(t+\beta)$ for any operator, (\ref{f1}) becomes
\ba
&&
\exp(-\beta F_{N_q, N_{\bar q}})=\nonumber\\
&&
={1\over N^{N_q+N_{\bar q}} } \sum_{s'} <s'|\sum_{(ab)} e^{-\beta H}
\psi_{a_1}(\br_1,\beta)\psi^{\dagger}_{a_1}(\br_1,0).....
\psi_{a_{N_q}}(\br_{N_q},\beta)
\psi^{\dagger}_{a_{N_q}}(\br_{N_q},0)\nonumber\\
&&
\times \psi_{b_1}(\br_1',\beta)\psi^{\dagger}_{b_1}(\br_1',0).....
\psi_{b_{N_{\bar q}}}(\br_{N_{\bar q}}',\beta)
\psi^{\dagger}_{b_{N_{\bar q}}}(\br_{N_{\bar q}}',0)|s'>
\label{f2}
\ea
Using (\ref{psit}) and (\ref{com1}) and introducing the Polyakov
loop \footnote{This object is also referred to as Wilson line.}
\be
L(\br)={1\over N} \tr P\exp\left(i\int_0^{\beta}dt \vec{\tau}\cdot
\vec{A_0}(\br,t)\right),
\ee
(\ref{f2}) becomes
\be
\exp(-\beta F_{N_q, N_{\bar q}})=
Tr\biggr[e^{-\beta H} L(\br_1)....L(\br_{N_q})L^{\dagger}(\br_1')
....L^{\dagger}(\br_{N_{\bar q}}')\biggr]
\ee
Let us consider the simplest case of an quark-antiquark pair at
distance $\br$ from each other. In this case the aditional
free energy due to presence of this pair is simply given by
the Polyakov loop correlator
\be
\exp(-\beta F_{q\bar q})=<L(\br)L^{\dagger}(0)>,
\ee
where $<..>$ denotes thermal average over the gluon configurations.
The Polyakov loop has following nice properties. First, it is
gauge invariant because the gauge field is periodic in time
$A_{\mu}(\br,0)=A_{\mu}(\br,\beta)$, second, its definition on lattice
is straightforward
\be
L(\br)={1\over N} \tr \prod_{i=1}^{N_t}U^0(\br,t),
\ee
where $N_t$ is the extension of the lattice in temporal dierction.
In the confined phase ($T<T_c$ with $T_c$ being the temperature
of the deconfinement phase transition) the large distance 
behavior of the Polyakov loop correlator
is the following \cite{kuti81}
\be
<L(\br)L^{\dagger}(0)>=\exp(-\sigma(T)|\br|/T)
\ee
implying confinement. In the deconfined phase at very high
temperature (small coupling) the dominant contribution to the Polyakov
loop correlators is given by the exchange of two screened electric
gluons (see Figure 1).
Calculation of the corresponding diagram yields
\begin{figure}
\epsfxsize=6cm
\epsfysize=4cm
\centerline{\epsffile{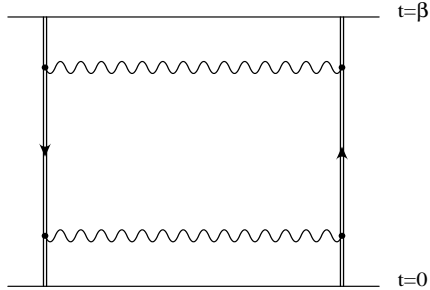}}
\caption{Two gluon contribution to the Polyakov loop correlation
function}
\end{figure}

\be
<L(\br)L^{\dagger}(0)>=\exp(-V(|\br|)/T), 
\ee
with 
\be
V(r)={N^2-1\over 8 N^2} {\left({g^2 \over 4 \pi r}\right)}^2 
e^{-2 m_{D_0}r}. 
\label{vav}
\ee
Thus the screening mass determined from the Polyakov loop correlators
is twice the Debye mass, $M_D=2 m_{D0}$.
From definition of $V(r) \equiv F_{q \bar q}$ it is obvious that it is
a thermal average over $q \bar q$ potentails in the possible colour
channels. Let us consider the simplest case of $SU(2)$ gauge group.
In this case the free energy of quark-antiquark pair (which is often 
referred to as colour averaged potential) can be written as
\be
e^{-\beta V(r)}={1\over 4} (e^{-\beta V_1(r)}+3 e^{-\beta V_3(r)})
\label{tav}
\ee
where $V_1(V_3)$ denotes the $q\bar q$ potential in the singlet
(triplet) channel for two isospin $1\over 2$ particles. These
potentials can be calculated perturbatively and contrary to $V(r)$ they
are dominated by one gluon exchange. The corresponding calculation
yields \cite{mclerran81}
\be
V_i(r)=c_i g^2 {1\over 4 \pi r} e^{-m_{D_0} r},
\ee
with $c_1=-3/4$ and $c_3=1/4$. At high temperature $g^2$ is small and
one may expand the 
exponentials in (\ref{tav}) and verify that the ${\cal O}(g^2)$ terms
cancel between singlet and triplet terms yielding Eq. (\ref{vav})
for $V$.

\subsection{Infrared divergencies Associated with Static non-Abelian
Magnetic Fields}

The application of the naive perturbation theory at finite temperature
is obstructed by infrared divergencies of thermal field theory. 
In scalar field theory and in abelian gauge theory these IR problems
can be cured by appropriate 
resummations which take into account the presence of
the screening scale $g T$ (see \cite{kapusta,lebellac} and also
discussion in section 2.5).  In non-Abelian gauge theories, however,
there exists also another sort of IR problems associated with static 
magnetic fields \cite{linde80}. 
\begin{figure}
\epsfxsize=6cm
\epsfysize=2cm
\centerline{\epsffile{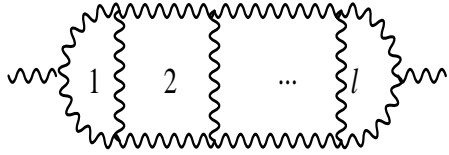}}
\caption{Divergent $l$-loop contribution to the 2-point function
}
\end{figure}

Static magnetic fields are not 
screened at leading order of perturbation theory: 
the IR limit of the 1-loop transverse self-energy vanishes  
in all gauges. 

Let us consider a generic $l$-loop contribution 
to the self-energy which is shown in Figure 2.
To evaluate the leading infrared divergent contribution we can assume
that all propagators in Figure 2 are static, i.e. consider only
contribution from zero Matsubara modes. In this case neglecting
all tensorial structures the contribution of these diagram is
\be
\underbrace{g^{2l}}_{vertex}~~\underbrace{{\left(\int d^3 p\right)}^l}_{
loop~integral}~~\underbrace{p^{2l}}_{vertex}
~~\underbrace{{(p^2+m^2)}^{-3l+1}}_{propagators}.
\ee
In the above expression we have introduced a magnetic mass since one
might expect that such a mass scale is generated by higher loop
contribution. Using simple power counting arguments one can see that
for $l=2$ the corresponding contribution is
\be
g^4 T^2 \ln{T\over m}
\ee
and for $l\ge 3$ one has
\be
g^4 T^2 {\left({g^2 T\over m}\right)}^{l-2}
\label{3ll}
\ee
Since no magnetic mass is generated at 1-loop level it is natural to
assume $m \sim g^2 T$. In this case as one can see from Eq. \ref{3ll},
infinite number of diagrams contribute to $g^4$ order self-energy.
If $m$ is proportional to the power of $g$, higher than $2$ 
no weak coupling expansion
could be obtained. However, as we will see in the next section the
magnetic mass turns out to be of order $g^2 T$.
Different approaches for determination of the numerical value of the
magnetic mass will be discussed in sections 3 and 4.

An  analysis of the higher loop diagrams similar to that presented
above shows that ${\cal O}(g^6)$ contribution to the
weak coupling expansion of the free energy is non-perturbative.
Let us emphasize that the Linde's conjecture does not imply that
no weak coupling expansion is possible in finite temperature
non-abelian gauge theory. It just states that beyond some order
the coefficients of the weak coupling expansion cannot be evaluated
by calculating finite numbers of Feynman diagrams. In fact the coefficient
of the term proportional to the weak coupling expansion of the
free energy was calculated by lattice Monte-Carlo technique \cite{karsch96}.
Linde's conjecture also implies as we will see in 
subsection 2.3 that the next-to-leading contribution to the
Debye mass is non-perturbative.

\subsection{Dimensional reduction}

Finite temparature field theories exist in a volume with one compact
dimension, the imaginary time, and the appropriate 
boundary conditions \cite{kapusta,lebellac}.
The extension of the compact dimension is $\beta=T^{-1}$. At
high temparature this dimension becomes arbitrary small.
Therefore the degrees of freedom in time direction are
essentially frozen. This situation can be also viewed differently.
Consider the Yang-Mills action at finite temperature
\be
S=\int_0^{\beta} dt \int d^3 x {1\over4} F_{\mu \nu}^a F_{\mu \nu}^a
\ee
The 4-dimensional gauge field $A_{\mu}^a(t,\bx)$ can be written
in terms of an infinite set of 3-dimensional fields using
Matsubara decomposition
\be
A_{\mu}^a(t,\bx)=\sqrt{T}\sum_{n=-\infty}^{\infty}e^{i\omega_n t}
A^a_{\mu n}(\bx), ~~~\omega_n=2 \pi T
\ee
From this expression it is clear that all modes with $n \ne 0$
have a mass $2\pi T n$. In $T \rightarrow \infty$ limit these modes become
infinitely heavy and according to conventional wisdom based on
Appelquist-Carrazone theorem these will decouple \cite{appelquist75}.
Thus the Yang-Mils action reduces to an action containing only
static fields of the following form
\be
S=\int d^3x\biggr({1\over 4} F^a_{ij} F_{ij}^a+\half{(D_i A_0^a)}^2\biggr),
\ee
with
\ba
&&
D_i A_0^a=\partial_i A_0^a-g_3 f^{abc} A_0^b A_i^c\\
&&
F_{ij}^a=\partial_i A_j^a-\partial_j A_i^b+g_3 f^{abc} A_i^b A_j^c\\
&&
g^2_3=g^2 T.
\ea
This is the action of the 3 dimensional gauge theory which interacts
with an Euclidian scalar and isovector field $A_0$. 
This matter field originates from the
time component of the original Yang-Mills field.
In more carefull analysis where non-static modes are integrated out 
at 1-loop level it has been shown that a mass term and quartic self-interaction
is generated for the $A_0$ field \cite{landsman89}. Thus the
effective action describing the dynamics of zero modes assumes the
form
\be
S=\int d^3x \biggr({1\over 4} F^a_{ij} F_{ij}^a+\half {(D_i
A_0^a)}^2
+\half m_{D0}^2 A_0^a A_0^a+\lambda_A {(A_0^a A_0^a)}^2 \biggr),
\label{l3adj}
\ee
where 
\be
m_{D0}^2={1\over 3} g^2 T^2 (N+\half N_f)
\label{md0}
\ee
is the leading order Debye mass  and
\be
\lambda_A=(6+N-N_f) {g^4 T\over 96 \pi^2}
\ee 
with $N_f$ being the number of fermion flavours \cite{landsman89,kajantie97b}.
In Ref. \cite{kajantie97b} integration of non-static modes was
performed at 2-loop level and it turned out that the 2-loop corrections
to the parameters $g^2_3$, $m_{D0}$ and $\lambda_A$ are small.
Higher dimensional and non-local operators are also generated by
1-loop integration of non-static modes, however, we have
neglected this operators because their coefficients vanish in
$T\rightarrow \infty$ limit. The impact of higher dimensional operators
was analyzed in Ref. \cite{jako94} and non-local operators were studied
in Ref. \cite{jako97}. Let us note that naive decoupling of 
non-static modes does not take place because the effective mass
and quartic vertex generated by integration of 
non-static modes  does not vanish in $T \rightarrow \infty$ limit. 
If the temperature is sufficiently high, $g_3^2 \ll m_{D0}$ i.e.
the $A_0$ field becomes very heavy and can be also integrated out.
The effective theory obtained by integrating out the $A_0$ field
is the 3 dimensional Yang-Mills theory 
\be
S_{YM_3}=\int d^3 x {1\over 4} F_{ij}^a F_{ij}^a
\ee
This theory is superrenormalizable and therefore the only dimensionfull
scale in the theory is $g_3^2$. This is the reason why the
magnetic mass is expected to be proportional to $g_3^2=g^2 T$.
Both the 3 dimensional Yang-Mills and the adjoint SU(N) Higgs
models are confining theories, thus we can see that the infrared
problems of finite temperature non-abelian gauge theory are
related to the physics of 3d confinement.

\subsection{The Non-Abelian Debye Screening beyond Leading Order}

\subsubsection{Non-Abelian Debye Screening in One-Loop Resummed
Perturbation Theory and the Magnetic Mass}

The next-to-leading correction to $\Pi_L$ is of order $g^3$ and arises
from summation of ring diagrams \cite{kapusta}. The summation of
ring diagrams simply amounts to replacing the bare propagator with
the 1-loop resummed propagator and evaluating the 1-loop diagram
with the resummed propagator. The natural candidate for next-to-leading
Debye mass would be $\Pi_L(k_0=0,k\rightarrow 0)$. However, this
quantity is gauge dependent  \cite{toimela85}.  We have seen in
section 2.1 that the Debye screening arises from poles $k=\pm i m_{D0}$ 
in the gluon propagator. Therefore the natural extension of the Debye
screening mass is 
\be
m_{D}^2=\Pi_L(k_0=0,k^2=-m_{D}^2).
\label{defmd}
\ee     
This definition
will lead to gauge invariant result because the pole of the propagator
was proven to be gauge independent at each order of perturbation
theory \cite{kobes90}.
$\Pi_L(k_0=0,k)$ was
calculated in 1-loop resummed perturbation theory in the high
temperature limit in Ref. \cite{rebhan94}
and it turns out that $k^2=-m_D^2$ is a starting point of branch
cut rather than a pole. This fact would mean a non-exponential anomalous
screening. If one introduces the magnetic mass $m_T$, i.e.
replaces the static massless magnetic propagators with massive ones the
next-to-leading result for $\Pi_L$ in $R_{\xi}$ gauge 
reads \cite{rebhan94}       
\ba
&&
\Pi_L(k_0,k)={g^2 T N\over 4 \pi}\biggr[-m_D-m_T\nonumber\\
&&
+{2 m_D^2-2
k^2-m_T^2\over k}
\arctan{k\over m_D+m_T}+\nonumber\\
&&
(k^2+m_D^2)\biggr({k^2+m_D^2\over m_T^2 k} \biggr(\arctan{k\over m_D+
\sqrt{\xi} m_T}-\arctan{k\over m_D+m_T}\biggr)+\nonumber\\
&&
\biggr(\sqrt{\xi}-1
\biggr){1\over m_T}\biggr)\biggr]
\ea
As one can see from this expression $\Pi_L$ has gauge invariant poles
at $k=\pm i m_D$ and the starting point of the branch cut has been 
shifted to $k=i(m_D+m_T)$.
Thus the exponential nature of the screening is restored. 
As far as one is interested in obtaining the weak coupling
expansion of the Debye mass, $m_D$ should be replaced
by $m_{D0}$ on the rhs. of equation (\ref{defmd}). Then the next-to-leading
correction for the Debye mass reads     
\be
m_{D1}^2-m_{D0}^2={N g^2 T\over 2 \pi} m_{D0} \biggr[
(1-{m_T^2\over 4 m_{D0}^2})\ln{2 m_{D0}+m_T\over
m_T}-\half-{m_T\over 2 m_{D0}}\biggr].   
\ee
Thus we can see that the value of the Debye mass at next-to-leading
order is strongly influenced by the magnetic mass. An alternative possibility
which will be considered in section 3 is a self-consistent determination
of $m_D$ from the equation  (\ref{defmd}), which is equivalent diagrammatically
to a summation of "super-daisy" diagrams \cite{dolan74}.          
Analogously to the Debye mass the magnetic screening mass can be defined as
\be
m_T^2=\Pi_T(k_0=0,k^2=-m_T^2),
\label{defmt}
\ee
where $\Pi_T$ is defined by equation (\ref{decomprop}).
Though the pole of the transverse gluon propagator is called magnetic
screening mass its relation to the screening of static magnetic fields
is not obvious. The magnetic screening could be defined in terms of the
free energy of monopole-antimonopole pair
\footnote{The divergence of the static non-Abelian 
magnetic field is non-zero because the the corresponding non-Abelian
field equation is ${\bf \nabla} \cdot {\bf B}^a=g f^{abc} {\bf A}^b \cdot
{\bf B}^c$. Therefore there may exist non-Abelian field configurations
which act as magnetic monopoles}.
However, it cannot be stated
that for the weak coupling regime this energy is dominated by exchange of
a static transverse gluon. Because of the Linde's conjecture all multi
gluon exchange have to be considered. 

The magnetic screening can be defined by studying the large distance 
behaviour of the field created by a 
magnetic monopole. This was realized on lattice investigations 
\cite{billoire81,degrand82} by calculating the magnetic flux leaving
the spatial 3d cube and also by semiclassical approxiomation by
B\'{\i}r\'o and M\"uller \cite{biro93}. In the lattice investigation
the corresponding inverse screening length was found to be $0.24g^2T$
\cite{billoire81} and $0.27(3)g^2 T$ in Ref. \cite{degrand82}. The
semiclassical approach of Ref. \cite{biro93} yields the result $0.25g^2
T$ which is consistent with the previous ones. From the arguments given
above it is clear, however, that this inverse screening length cannot
be directly related to the magnetic mass defined from the propagator.

\subsubsection{Non-Perturbative definitions of the Debye mass}

We are interested in definitions of the Debye mass which do not rely
on perturbation theory and can be easily implemented in lattice Monte-Carlo
simulations. A natural way to determine the Debye screening mass
non-perturbatively is to measure the gluon propagator on lattice.
Such measurement were done  in Refs. \cite{heller95,heller98} in
finite temperature 4d SU(2) gauge theory and also in 3d effective  
theory in Ref. \cite{karsch98k}. We will discuss the non-perturbative
determination of the propagators in section 4 in detail.

Let us consider the Polyakov loop correlator which is a
manifestly gauge invariant quantity. 
At leading order the Polyakov loop correlator is determined by
the exchange of two electrostatic gluons as shown in Fugure 2 (see
subsection 2.1.3) leading to exponential screening
with the inverse screening length equal to $2 m_{D0}$. 

Beyond leading order, however, Polyakov loops can also couple to
magnetostatic gluons as shown in Figure 3. 
\begin{figure}
\epsfxsize=6cm
\epsfysize=4cm
\centerline{\epsffile{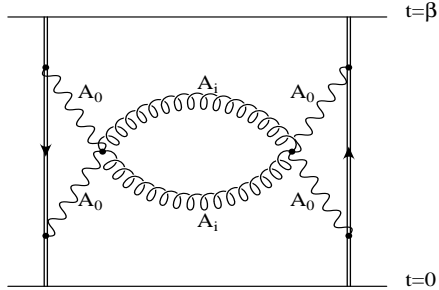}}
\caption{Contribution of magnetic gluons to the Polyakov loop correlation
function}
\end{figure}

At high temparatures different mass scales which are present in
the plasma are separated $2 \pi T \gg g T \gg g^2 T$ due to assymptotic
freedom. Then the infrared problems associated with static
magnetic fields are related to the physics of 3 dimensional confinement.
Because of the 3 dimensional confinement Polyakov loops cannot
exchange a pair of magnetic gluons, the pair will instead form
a glueball with a mass of order $g^2_3=g^2 T$. Thus the large distance
behaviour of the Polyakov loop correlators will be determined by
the lightest glueball state of the 3 dimensional Yang-Mills theory
\cite{nadkarni86,braaten94}. Thus the true large distance behaviour
of Polyakov loop correlator at very high temperature has nothing
to do with the physics of the Debye screening.

For vectorially coupled theories (like QCD at zero chemical
potential but not the electroweak theory) 
an alternative non-perturbative definition 
of the Debye mass was suggested by Arnold and Yaffe \cite{arnold95}
This definition relies on the fact that in vectorially-coupled
theories at finite temperature there is a symmetry, namely the Euclidian time
reflection symmetry that  excludes the unwanted exchange of
pairs of magnetostatic gluons. The crucial property is that the temporal
component of the gauge field $A_0$ is odd under this symmetry
while the spatial gauge field $A_i$ is even. Thus
non-perturbatively the Debye mass can be associated with a correlation
function of some gauge invariant operator which is odd under Euclidian
time reflection and which has the largest correlation length. 
Some gauge invariant time-reflection odd operators are listed in
Ref. \cite{arnold95}. As far as one is interested in the leading
non-perturbative correction i.e. the coefficient of the $g^2 \ln
g$ and $g^2$ terms in
\be
m_D=m_{D0}+c_1 g^2 \ln g+c_2 g^2 +...
\ee
it is not necessary to calculate correlation function in the full
4 dimensional theory. One can instead consider the effective theory,
construct the corresponding operators and measure their correlation
function there. Various realisation of this strategy have been
recently discussed in the literature \cite{kajantie97a,laine98,laine99a,
korthals}.

\subsection{Screened Perturbation Theory}
   One of the main motivation for studying the screening masses is
their role in the physical expression of different 
thermodynamic quantities like thermodynamic potentials. The
weak coupling expansion of the QCD free energy does show very bad
convergence properties
\cite{shuryak78,kapusta79,kalashnikov81,toimela83,arnold95zhai,kastening95}.
Only in the $TeV$ temperature range can one find a satisfactory
numerical convergence rate \cite{nieto97}. 

Recently, 
in the case of
scalar $\phi^4$ theory, where  conventional perturbative expansion
encounters similar convergence problems, several proposals 
have been made for new types of 
perturbative expansion which shows  better convergence
\cite{karsch97k, reinbach97, drummond98}. The outcome of the resummed
perturbative calculation for the free energy density may depend on the
correct choice of the screening mass \cite{karsch97k,drummond98}.

        In Ref. \cite{karsch97k} it was shown that the poor convergence of
perturbative expansion can be improved if the series expansion
is reorganized in
loop expansion with screened propagators; i.e. instead of expanding
around the free energy density of a massless ideal gas, 
loop expansion is performed starting from a
massive ideal gas \footnote{The propagators become massive 
by the mechanism of Debye 
screening generated by the thermal fluctuation in analogous way as
in gauge theories.}.
Following Ref. \cite{karsch97k} consider an $O(N)$ symmetric scalar
field theory with the following Lagrangian
\begin{eqnarray}
&
L=L_0+L_{int},\nonumber\\
&
L_0={1\over 2}({(\partial \phi_i)}^2+m^2 \phi^2_i),\nonumber\\
&
L_{int}=-{1\over 2} m^2 \phi_i^2+{g^2\over 24 N} {(\phi_i^2)}^2+
{g^2\over 24 N} (Z_2-1){(\phi_i^2)}^2.
\end{eqnarray}
Following ref. \cite{frenkel92,parwani92} we have introduced and substracted
a mass term with $m\sim {\cal O} (g)$ in order to reorganize the
perturbative
expansion.
The coupling constant renormalisation factor is given by \cite{amit}
\begin{equation}
Z_2=1+{3 g^2\over {(4 \pi)}^2} {N+8\over 18 \epsilon} +
{\cal
O}(g^4),
\end{equation}
and we have neglected the field renormalisation factor $Z_1$, which is
unity
up to ${\cal O}(g^4)$.
\begin{figure}
\vskip-3truecm
\epsfxsize=10cm
\epsfysize=7cm
\centerline{\epsffile{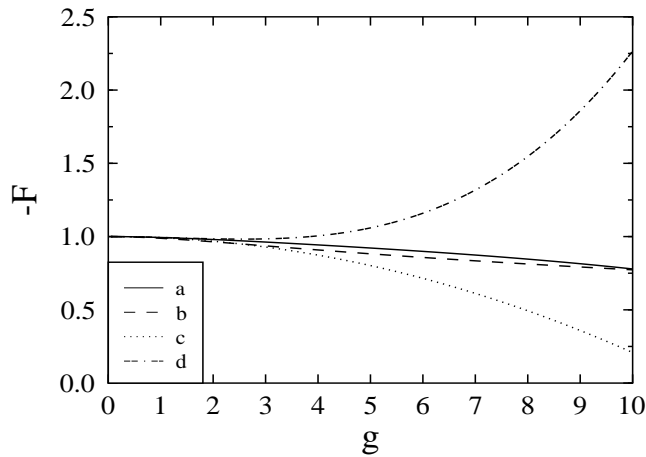}}
\caption{Shown is the free-energy density of the N=1 scalar
field theory
as a function of the scalar self-coupling g in the units of the
free energy density of a massless ideal gas ($\pi^2 T^4\over 90$)
(from Ref. \cite{karsch97k}).
The curves
represent 2-loop (a) and 1-loop (b) results of the screened loop
expansion as well as the
${\cal O}(g^2)$ (c) and  ${\cal O}(g^3)$ (d) results of the
conventional perturbative expansion.}
\label{fig:1}
\end{figure}

The free energy can be easily evaluated with massive propagators up
to 2-loop, but in the limit of large $N$ also the 3-loop contribution can
be evaluated. The details of the calculation can be found in Ref.
\cite{karsch97k}. To evaluate the free energy numerically the
screening mass has to be specified. It turns out that the final result
is sensitive to the actual value of the screening mass
\cite{karsch97kproc}. The most natural way of the determination of the
screening mass is the use of a 1-loop gap equation
\cite{dolan74,espinosa93,buchmuller94fodor}, which for our case reads
\footnote{In the limit of large $N$ this equation is exact and
therefore its solution can also be used in 3-loop calculations.
The gap equation should be properly renormalized before it is solved
numerically. Therefore a divergent and scale dependent counterterm was
added; see discussion in Ref. \cite{drummond98} on this issue.} 
\begin{equation}
m^2={g^2 (N+2) \over 6 N}(I(m)+{m^2\over 16 \pi^2}({1\over \epsilon}+
\ln{{\bar \mu}^2\over T^2})),
\label{gapscal}
\end{equation}
with
\be
I(m)=\sum_{n=-\infty}^{\infty} \int {d^3 p\over {(2 \pi)}^3}
{1\over p^2+\omega_n^2+m^2},~~~\omega_n=2 \pi T n.
\ee
The numerical results for the free energy density in the case of the $N=1$
component theory as a function of the self-coupling calculated in the screened
perturbation theory is shown in Figure 4.
Also shown are
the results obtained with the conventional perturbative expansion up
to
${\cal O}(g^2)$ and ${\cal O}(g^3)$ in the scalar self-coupling.
The Figure clearly shows that the difference between the 1-loop and
2-loop
results of the screened perturbative expansion
is small and essentially stays constant over the entire range of
couplings. In the conventional perturbative approach, on the other
hand,
different orders in the expansion differ largely from each other.
For large $N$ as it was mentioned above also the 3-loop contribution 
can be calculated. The corresponding results are shown in Figure 5
for $N=10$. As one can see from the figure the convergence of the
perturbative expansion is substantionally improved as compared to
the usual perturbative expansion. 

Let us notice that an attempt to 
generalize this approach to the case of QCD have been recently reported
in Ref. \cite{braaten99}
\begin{figure}
\vskip-2truecm
\epsfxsize=10cm 
\epsfysize=7cm
\centerline{\epsffile{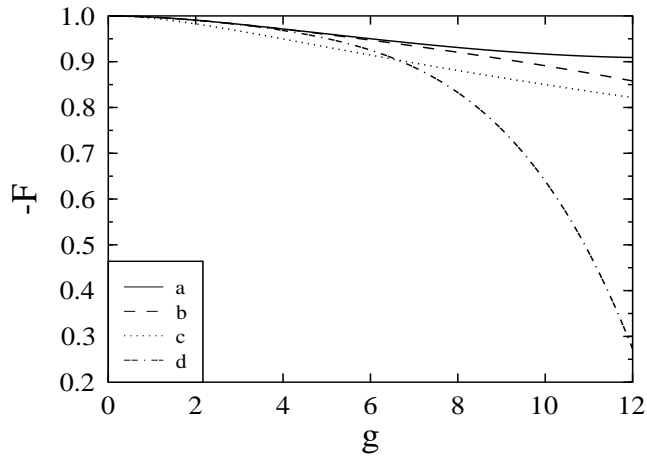}}
\caption{The free energy density of the scalar field theory in
the large-$N$ limit as
a function of the scalar self-coupling $g$ in units of the free
energy
density of a massless ideal gas (from Ref \cite{karsch97k}). 
Shown are results for $N=10$ (see
text).
The curves represent 3-loop (a) 2-loop (b) and 1-loop (c) results of
the screened loop expansion as well as the
${\cal O} (g^4)$ (d) results of the conventional
perturbative expansion.}
\label{fig:2}
\end{figure}

The authors were able to find a compact formula for the leading order 
massive gas contribution to the free energy density of the hot gluon
gas. In distinctions to phenomenological approaches \cite{levai98},
their expression also takes into account the contribution of the 
longitudinal fluctuations and also the effect of Landau-damping
appears in addition to the quasiparticle piece of the free energy. 
The screening mass of the gluons remains a free parameter, which
emphasizes the importance of its independent determination.

\newpage
\section{Coupled Gap Equation for the Screening Masses in Hot
SU(N) Gauge Theory}
As it was discussed in section 2.4 a natural way to calculate the Debye
mass is to solve the ( gap ) equation $m^2_D=\Pi_L(k_0=0,k^2=-m_D^2)$ in a
self-consistent way. This definition is gauge invariant but requires
the introduction of the magnetic mass which can be associated with
a dynamically generated mass gap of pure 3d gauge theory if the
temperature is high enough. The magnetic mass was calculated 
using self-consitent approach in the effective 3d gauged Higgs
and also in 3d pure gauge theories 
\cite{buchmuller94fodor, espinosa93, buchmuller95, alexanian95, jackiw97}.
In the case of the  
Electroweak Theory (or SU(2) Higgs model ) the
gauge coupling is small for the temperature range of interest 
and a hiearchy of the different mass scales holds
$2\pi T>>g T>>g^2 T$. Due to this fact different screening masses
could be determined separately in the corresponding effective theories.
In hot SU(N) theory for the temperature range of interest the above
hierachy of scales fails to hold. Therefore we
should investigate a coupled set of gap equations for all the screened 
modes and determine the corresponding screening lengths simultanously. 

The most straightforward way to do this would be to
derive gap equations in the full four-dimensional theory. However, it is
not known how to generalize the by-now well-established
three-dimensional gauge invariant resummation \cite{buchmuller95,
philipsen94,alexanian95,jackiw97} to 
four dimensions. There are gauge transformations which might mix
static and non-static modes, therefore the resummation of the
static modes only,
which was suggested in \cite{arnold93} violates gauge invariance.
Since screening masses are static quantities it is natural to
calculate them in the framework of an effective three-dimensional
theory which is, however, valid only up to scales $k\lesssim g T$. It was
shown that the accuracy of the description of such theories is improved if in 
the action of the effective theory besides superrenormalizable
operators one also takes into account higher dimensional and
non-local operators \cite{jako96,jako97}. 

When deriving the gap equation in the
framework of three-dimensional adjoint Higgs model it is important  to
address the question whether the symmetric or the broken phase is the
physical one. Dimensional reduction is valid if $A_0 \ll \sqrt{T}$.
However, in the broken phase the expectation value of the $A_0$ field
is $A_0 \sim {1\over g}$. Therefore for small $g$ 
the broken symmetry phase cannot be the physical one.
We will return to this question in section
4, but here we will assume no $A_0$ background.

In this section we shall assume only the separation of the static and 
non-static scales in the pure SU(N) gauge theory, that is we assume
that $g(T)<<2\pi$, and derive different versions of the coupled gap
equations in the corresponding effective theory, the 3d adjoint Higgs
model. Presentation in this section relies on Ref. \cite{patkos98k}.

\subsection{Gauge non-invariant resummation scheme}

This scheme for the evaluation of the magnetic mass was first
suggested in \cite{philipsen94} and for the Debye mass in
\cite{rebhan94}. It was noticed in Ref. \cite{philipsen94} that magnetic mass 
obtained in this scheme is gauge dependent and therefore cannot be regarded as 
physically meaningfull. 
However, even in gauge invariant resummation schemes the 
value of the magnetic mass cannot be defined unambigously, 
because its approxiomate value depends 
on the specific resummation scheme \cite{jackiw97}. Therefore it
is of a certain  interest 
to calculate the magnetic mass even in a gauge non-invariant
scheme just to compare the amount of gauge dependence with the ambiguity of 
different gauge invariant approaches, when evaluated to leading order
is some approxiomation scheme.

The three-dimensional effective Lagrangian relevant for 1-loop
calculations can be written with $R_\xi$ gauge as (see Eq.
(\ref{l3adj}))
\ba
&
L={1\over 4}F^a_{ij} F^a_{ij}+{1\over 2}\left({(D_i A_0)}^2+m_D^2 A_0^a A_0^a
\right)+
\nonumber\\
&
{1\over 2} m_T^2 A^a_i A^a_i+(\partial_i \bar c^a D_i c^a+
m_G^2 \bar c^a c^a)+
{1\over 2 \xi} {(\partial_i A_i)}^2+L_{ct},\nonumber\\
&
L_{ct}={1\over 2}(m_{D0}^2-m_D^2)A_0^a A_0^a-{1\over 2} m_T^2 A_i^a A_i^a-
m_G^2 \bar c^a c^a
\ea

Here we have added and substracted a  mass term for $A_i$, $A_0$ and the
ghost fields ($c$) 
with their exact screening masses. $m_{D0}^2$ is the tree-level 
(from the point of view of the effective theory) Debye mass for $A_0$, which
was generated during the procedure of the dimensional reduction and
defined by Eq. (\ref{md0}).
The propagators can be read from the
quadratic part of the Lagrangian and can be found in the Appendix A, where one
also finds some details of the evaluation of the relevant Feynman diagrams 
contributing to the different 2-point functions. It should be noticed
when performing the resummation of the pure gauge sector with the unique
mass term  $m_T$, the longitudinal and transverse gluons
acquire different masses, which are, however related by
$m_L=\sqrt{\xi} m_T$ ($m_T$ is the transverse and $m_L$ is the
longitudinal mass). It is also possible to perform the resummation by
introducing independent masses for the longitudinal and transverse gluons,
but then the corresponding gap equations will have only complex
solutions.
The gauge boson self-energy can be decomposed as
\be 
\Pi_{ij}(k)=(\delta_{ij}-{k_{i} k_{j}\over k^2})
\Pi_T(k,m_T,m_D,m_G)+{k_{i} k_{j}\over k^2} \Pi_L(k,m_T,m_D,m_G).
\ee
In the Appendix A we give the expression of $\Pi_{ij}$ in terms of a few
fun\-da\-men\-tal three-dimensio\-nal loop-integrals. These integrals are 
easily evaluated and an explicit but very cumbersome functional form can be 
written for the longitudinal and transversal projections of the polarisation 
matrix. The self-energy of the ghost fields is also given in Appendix A.

The self energy for $A_0$ was first calculated in \cite{rebhan94}:
\bea
&
\Pi_{00}(k,m_D,m_T)=m^{2}_{D0}+{g^2 N\over 4 \pi} \biggl[-m_D-m_T+
{2 (m_D^2-k^2-m_T^2/2)\over k} \arctan{k\over m_D+m_T}\nonumber\\
&
+(k^2+m_D^2) \biggl({k^2+m_D^2\over m_T^2 k} (\arctan{k\over
m_D+\sqrt{\xi} m_T}-\arctan{k\over m_D+m_T})+(\sqrt{\xi}-1){1\over
m_T}\biggr)\biggr].
\label{reb}
\eea
\begin{figure}
\epsfbox[-90 0 80 180]{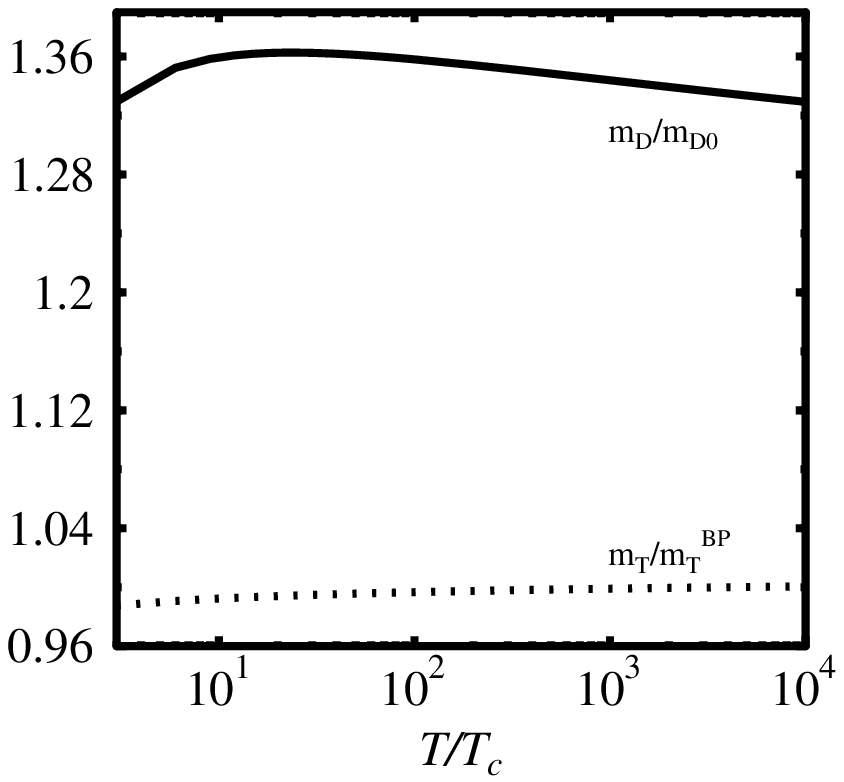}
\caption{The temperature dependence of $m_D/m_{D0}$ (solid line) and
$m_T/m_T^{BP}$ for Feynman ($\xi=1$) gauge, $m_{D0}$ is  the leading 
order result for the Debye mass and $m_T^{BP}$ is the value of the
magnetic mass obtained by Buchm\"uller and Philipsen for pure $SU(2)$
gauge theory.}
\vskip 2truecm
\epsfbox[-90 0 80 180]{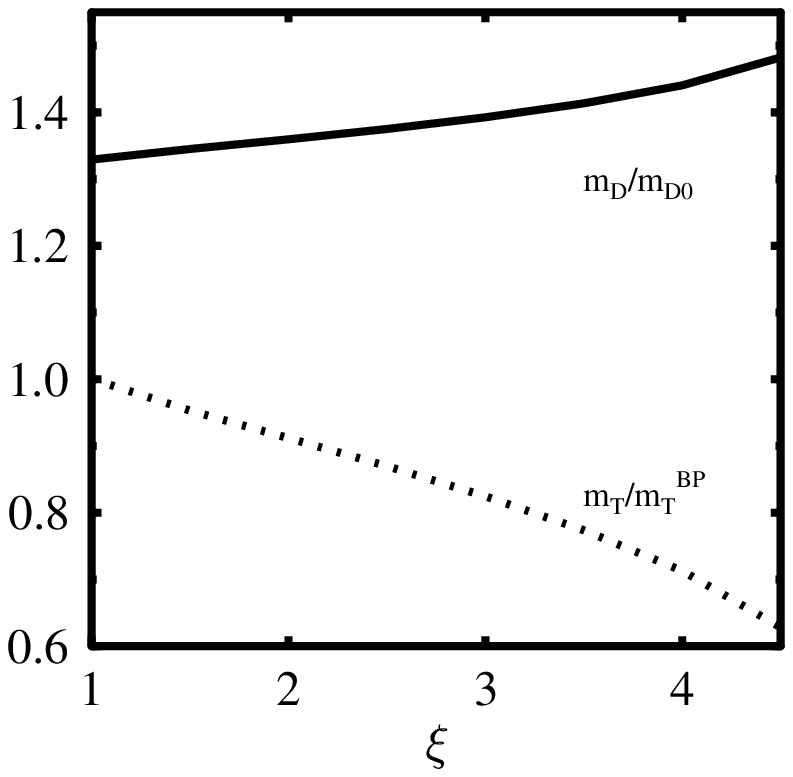}
\caption{The dependence of $m_D/m_{D0}$ (solid line)
and $m_T/m_T^{BP}$ on  gauge parameter $\xi$ at $T=10^4 T_c$, $m_{D0}$ 
is the leading order result for the Debye mass and $m_T^{BP}$ is the value 
of the magnetic mass obtained by Buchm\"uller and Philipsen for pure $SU(2)$
gauge theory.} 
\end{figure}

The above self-energies are functions of the magnitude of momentum
but also depend parametrically on the screening masses.
The on-shell gap equations now can be written as
\begin{eqnarray}
\nonumber
m_T^2 &=&\Pi_T(k^2=-m_T^2,m_T,m_D,m_G), \\
m_D^2 &=&\Pi_{00}(k^2=-m_D^2,m_D,m_T), \\\nonumber
m_G^2 &=&\Pi_G(k^2=-m_G^2,m_T,m_G). 
\end{eqnarray}
On the mass-shell $\Pi_{00}(k^2=-m_D^2,m_D,m_T)$ is gauge parameter
independent, but  $\Pi_T(k^2=-m_T^2,m_T,m_D,m_G)$ and $\Pi_G(k^2=-
m_G^2,m_T,m_G)$ do depend on the gauge fixing parameter, therefore the
masses obtained from this coupled set of gap equations are gauge dependent. 

In the following numerical investigations we shall consider the case
of the SU(2) gauge group. The 4-dimensional coupling constant is taken
at scale $\mu=2 \pi T$, where $\mu$ is the $\overline{MS}$ scale and
1-loop relation for the gauge parameter of the effective theory is used. To set 
the temperature scale we use the relation $T_c/{\Lambda_{\overline{MS}}}=1.06$
obtained from numerical simulation of the finite temperature SU(2) gauge 
theory \cite{heller98}.

The temperature dependence of $m_D$ in the Feynman gauge  is plotted in 
Figure 6. As one can see from the plot $m_D$ receives $30 \%$ positive
correction compared to the leading order result, while the
magnetic mass stays very
close to the value calculated by Buchm\"uller and Philipsen
\cite{buchmuller95}, given
below in Eq.(\ref{egy6}). Since the masses are gauge dependent in this
approach, it is important to investigate the dependence of the screening masses 
on the gauge parameter $\xi$. It turns out that one gets real values for the 
masses from the gap equations only if $\xi\in[1,5)$. The dependence of the 
screening masses on $\xi$ in this range is shown in Figure 7. One can see, that
the $\xi$-dependence of $m_T$ in this interval is 40 \%, while for $m_D$ it
remains in the 10 \% range.

\subsection{Gauge Invariant Approach}

Gauge invariant approaches  for the magnetic mass
generation in three-dimensional pure SU(N) gauge theory  were
suggested by Buchm\"uller and Philipsen (BP) \cite{buchmuller95} and by
Alexanian and Nair (AN) \cite{alexanian95}. 
The basic idea of both approaches is to add and substract a gauge
invariant mass generating term to the original Lagragian:
\be
L={1\over 4} F_{ij}^a F_{ij}^a+\delta L-l \delta L,
\ee
where $l$ is the (formal) loop expansion parameter. In the final result
one should set $l=1$ of course. In the scheme of BP $\delta L$ is
just the Lagrangian of the gauged non-linear sigma model. In the
scheme of AN the mass generating term relevant for 1-loop 
calculation assumes the form \cite{alexanian95}
\be
\delta L  = {1\over 2} m_T^2 A_i (\delta_{ij}-{\partial_i \partial_j\over
\partial^2}) A_j+ f^{abc} V_{ijk} A_i^a A_j^b A_k^c
\label{dlan}, 
\ee
where the explicit expression
for $V_{ijk}$ can be find in Ref. \cite{alexanian95}.
Till now only these two gauge invariant
schemes are known to provide real values for the magnetic mass
\cite{jackiw97}.
 
In these approaches the gauge boson self-energy is automatically
transverse and there is no need to project the transverse part from the 
polarisation tensor. The corresponding expression for the on-shell
self-energy reads
\be
\Pi_T(k=i m_T,m_T)=C m_T,
\ee
where 
\be
\label{egy6}
C=\cases{{g^2 N\over 8 \pi} [{21\over 4} \ln 3-1],AN ,\cr
{g^2 N\over 8 \pi} [{63\over 16} \ln 3-{3\over 4}],BP.}
\ee
In gauge invariant approaches the ghost field remains massless and
there is no gap equation corresponding to it.

Since we are interested in calculating the screening masses in the 
three-dimensional SU(N) adjoint Higgs model, $\Pi_T(k,m_T)$ should be
supplemented by the corresponding contribution coming from $A_0$
fields. This contribution is calculated from diagrams d) and e) of the
Appendix to be
\be
\delta \Pi_{ij}^{A_0}(k,m_D)=
{g^2 N\over 4 \pi} \left(-{m_D\over 2}+{k^2+4 m_D^2\over 4 k}
\arctan{k\over 2 m_D}\right) \left(\delta_{ij}-{k_{i} k_{j}\over k^2}\right).
\label{dpi_a0}
\ee
It is transverse and gauge parameter independent, it also does
not depend on the specific resummation scheme applied to the
magnetostatic sector. It should be also noticed that it starts to contribute to 
the gap equation  at ${\cal O}(g^5)$ level in the
weak coupling regime, thus preserving the magnetic mass scale to
be of order $g^2 T$. This is the reason why no "hierarchy" problem arises in
this case, at least for moderate g values.

The self energy of $A_0$ depends on the specific resummation scheme. 
To calculate this one has to fix a specific gauge. In Ref.
\cite{jackiw97} it was shown that it is possible to integrate out 
the sigma fields and the resulting Lagrangian describes massive
gauge fields and has no gauge-fixing and ghost terms. 
Therefore to calculate the self-energy of $A_0$ in BP scheme 
one should evaluate diagrams $f$ and $g$ in appendix A with massive
vector boson propagators. The corresponding calculation leads to  
\ba
&
\Pi_{00}(k,m_D,m_T)=m^{2}_{D0}+{g^2 N\over 4 \pi} \biggl[-m_D-m_T+
{2 (m_D^2-k^2-m_T^2/2)\over k} \arctan{k\over
m_T+m_D}-\nonumber\\
&
{(k^2+m_D^2)\over m_T^2} \biggl(-m_T+{k^2+m_D^2\over k} \arctan{k\over m_T+m_D}
\biggr)\biggr].
\ea
This expression is different from the expression of $\Pi_{00}$ calculated in 
the gauge non-invariant approach (see eq. (\ref{reb}) ) but its
analytic properties and on-shell value is the same as of (\ref{reb}).
For the resummation scheme of AN  it is important
to find a convenient gauge fixing term which reads
\be
L_{g.f.}^{AN}={1\over 2} \partial_i A_i^a 
\biggl(1-m_T^2{1\over \partial^2}\biggr) \partial_j A_j^a
\label{dlangf}
\ee
With this gauge fixing the expression for the self-energy  of $A_0$
coincides with (\ref{reb}) if it is evaluated at $\xi=1$.
The coupled set of gap equations now can be written as
\bea
\nonumber
m_T^2&=&C m_T+\delta \Pi^{A_0}(k=i m_T,m_D), \\
m_D^2&=&\Pi_{00}(k=i m_D,m_D,m_T).
\eea
The temperature dependence of $m_D$ obtained from this coupled set of gap 
equations is shown in Figure 8 for both schemes, where we have again normalized the 
Debye mass by the leading order result, $m_{D0}$. The temperature dependence of
the magnetic  mass is shown in Figure 9, where we have normalized $m_T$ by the 
value of the magnetic mass obtained for pure three-dimensional SU(2) theory, 
in the BP (AN) gauge invariant calculations \cite{buchmuller95,alexanian95}. 
As one can see the contribution of $A_0$ 
to the magnetic mass is between 1 and 3\%. 
\begin{figure}
\epsfbox[-90 0 80 160]{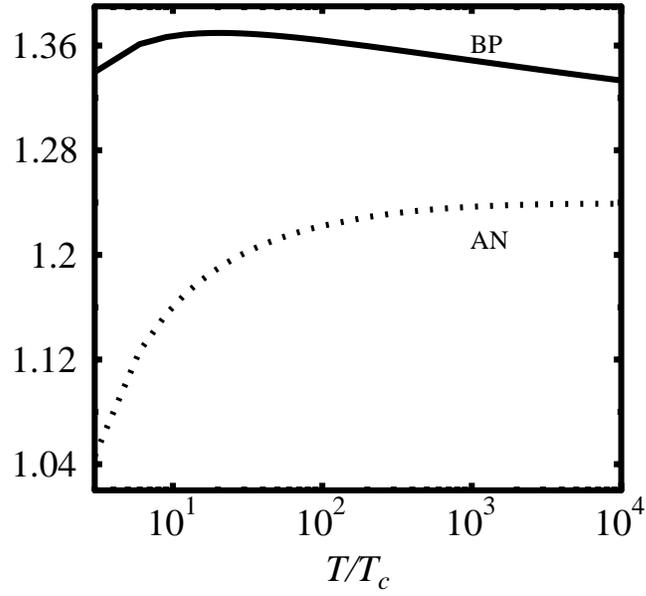}
\caption{The temperature dependence of the scaled Debye mass for
BP resummation scheme (solid) and for the AN resummation scheme
(dashed). The scaling factor is $m_{D0}$.}
\end{figure}

\begin{figure}
\vskip 2truecm
\epsfbox[-90 0 80 160]{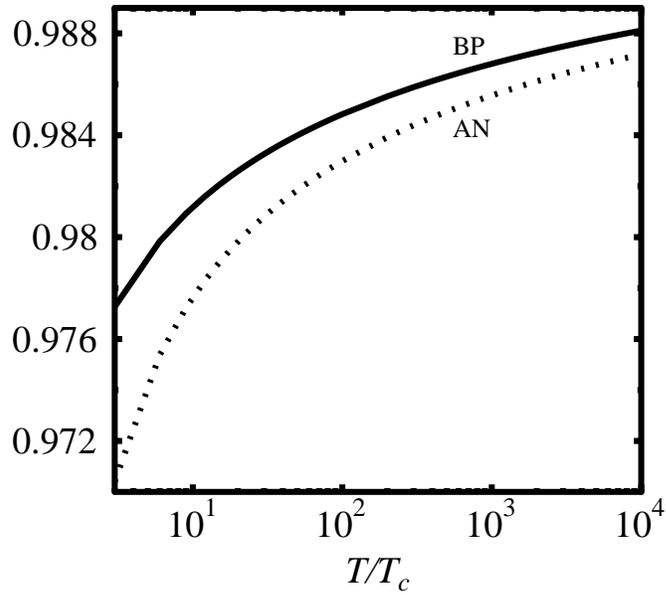}
\caption{The temperature dependence of the scaled magnetic  mass for
BP resummation scheme (solid) and for the AN resummation scheme
(dashed). The scaling factors are $m_{T}^{BP}$ and $m_{T}^{AN}$,
respectively.}
\end{figure}

From Figures 8 and 9 it is also seen that the temperature dependence of the 
screening masses is very similar to the temperature dependence of the
respective leading order results.

\subsection{Contribution of non-Local Operators to the Gap\\ Equation}
In the previous section the gap equations were derived for an effective
local superrenormalizable theory. In this case the effect of non-static modes
in the 2-point function were represented by the thermal mass for $A_0$ and
by the field renormalization factors which relate 3d fields to the
corresponding 4d ones. It was shown in \cite{farakos94,jako94a} 
that when one performs the procedure of the dimensional
reduction in $R_{\xi}$ gauges  the parameters of the superrenormalizable
effective theory are gauge independent, only the expressions of 3d
fields in terms of 4d ones depend on the gauge parameter. However, this
does not hold for higher dimensional and non-local operators, which are
generally gauge dependent. At 1-loop level the only diagrams contributing
non-locally to the gap equation are those which have two non-static line
inside the loop, diagrams with one static and one non-static line inside
the loop are forbiden because of 4-momentum conservation. Therefore at
1-loop level the non-locality scale $(2\pi T)^{-1}$ is much smaller than the
relevant length scales.

In general the non-static contribution to the static 2-point function 
$\Pi_{\mu \nu}(k_0=0,k)$ can be written as
\bea
&
\Delta \Pi^{ns}_{\mu \nu}(k_0=0,k)=
\delta^0_{\mu} \delta_{\nu}^0 \Pi^{ns}_{00}(k)+\delta_{\mu}^i \delta_{\nu}^j (\delta_{ij}-
{k_i k_j\over k^2}) \Pi(k)
\nonumber, \\
&
\Pi_{00}^{ns}(k)=m_{D0}^2+a_1(\mu,\xi)k^2+T^2 \sum_{n=2}^{\infty}a_n(\xi)
 {({k^2\over 
2 \pi T})}^{2 n},\nonumber\\
&
\Pi^{ns}(k)=b_1(\mu,\xi) k^2+T^2 \sum_{n=2}^{\infty}b_n(\xi) {({k^2\over 
2 \pi T})}^{2 n},
\label{nonloc}
\eea
where the coefficents $a_n$ and $b_n$ can be calculated for arbitrary $n$.
The first two terms in the expressions of $\Pi_{00}$ and first term of
$\Pi (k)$ are already included into the 3d 
effective theory as part of the tree level mass and the
definition of the 3d fields in
term of 4d fields. The last two sums will contribute to the 3d effective
lagrangian as quadratic non-local operators. There are also higher dimensional
operators as well as non-local 3- and 4-point verticies in the effective
lagrangian, however, since we restrict our interest to 1-loop gap equations 
these are not important for us. Their contribution  would correspond to 2 or 
higher loop contribution in the full four-dimensional theory.
 
We have estimated by direct numerical evaluation of the infinite sums
in (\ref{nonloc}) the contribution of non-local operators to be less
than 1\% in the temperature range $T=\left(3-10^4\right)T_c$, thus neither 
their contribution nor their gauge dependence is essential.

\subsection{What Have We Learnt from the Coupled Gap Equations ?}
In this section an attempt have been made to extract the electric and
magnetic screening masses from the coupled set of gap equations of the 
three-dimensional SU(N) adjoint Higgs model considered as an effective theory 
of QCD. The screening masses have been studied using gauge non-invariant as
well as gauge invariant resummation schemes. In the gauge non-invariant
formalism we have observed rather strong gauge parameter dependence,
therefore the results extracted from it are not very informative.
It is still interesting to note that in Feynman gauge ($\xi=1$) the
results for the magnetic mass
are rather close to those obtained from the gauge invariant
resummation scheme of Buchm\"uller and Philipsen. In gauge invariant treatments 
we have compared two different resummation schemes, that of Buchm\"uller and 
Philipsen (BP) and one proposed by Alexanian and Nair (AN). 
Qualitatively these 
two resummations lead to similar results, but in the BP scheme one has
smaller magnetic mass and larger Debye mass than in the AN scheme.

Let us summarize our view on the interaction of the electric $A_0$ and the
magnetic $A_i$ fields. In both schemes one can see that the 
dynamics of $A_0$ is
largely influenced by the magnetic sector, however, no similar feedback on the 
magnetic sector is seen. The magnetic masses calculated from the coupled gap 
equations provide screening masses which are 1\% smaller than evaluated in the 
pure gauge theory. This fact suggests that 
the influence of the adjoint scalar field
is similar to that of the fundamental Higgs field, because the magnetic mass 
calculated in the symmetric phase of 3d SU(2) Higgs theory by the
gap equation technique is roughly the same as in pure gauge theory 
\cite{buchmuller95}. 

Finally we compare our results with recent Monte-Carlo data for the screening 
masses obtained in 4d finite temperature SU(2) gauge theory \cite{heller98}. 
The data on the magnetic mass found from this simulation in the temperature 
range $T=(10-10^4)T_c$ can be fitted well by the formula $m_T=0.456(6) g^2 T$. 
This value of the magnetic mass is considerably larger than what one obtains 
from the magnetic gap equation and 3d simulations, where the results are 
approximately $m_T=0.28 g^2 T$ for the BP scheme, $0.38g^2 T$ for the
AN scheme and $0.35(1) g^2 T$ from 3d simulations.
The corresponding results for the 3d adjoint Higgs model are
also close to these values.
 
The data from Monte-Carlo simulation for the  Debye mass in the above mentioned 
temperature interval can be fitted using the following leading order-like  
ansatz $\sqrt{1.69(2)} g(T) T$ \cite{heller98}, which means that the Debye mass 
is  roughly $1.6 m_{D0}$, where $m_{D0}$ is the leading order result.
The gap equations at the same time, as one can see from Figure 8 give  
$(1.2-1.3) m_{D0}$, depending on the resummation scheme. While there is no 
quantitative agreement between masses measured in Monte-Carlo simulation and 
those obtained from the gap equation, 
the temperature dependence of these masses 
in the temperature interval $T=(10-10^4)T_c$ seems to follow the temperature 
dependence of the leading order result.

Finally let us discuss the consistency of the gap equation approach.
The validity of 1-loop gap equation for the magnetic mass
in pure gauge theory was critically questioned in
Refs. \cite{jackiw96,cornwall98}. We have seen that the two
different resummation schemes lead to somewhat different results.
If the gap equation aproach is valid the difference between the
values of the magnetic masses calculated in two schemes 
$m_T^{AN}-m_T^{BP}$ should be of the same order in magnitude as
the 2-loop correction to the magnetic mass \cite{jackiw97}.
A recent 2-loop analysis of the magnetic mass yields $m_T=0.34 g_3^2$
\cite{eberlein98}. Thus the gap equation approach seems to be viable.
Moreover in Ref. \cite{nair98}
a strong coupling expansion for the magnetic mass was carried out
with the result $0.32g_3^2$. In 
next section we will discuss the determination of the magnetic mass
using lattice Monte-Carlo technique.
\newpage
\section{ Screening Masses of Hot SU(2) Gauge Theory from 
Monte-Carlo Simulation of the Lattice 3d Adjoint Higgs Model}
Since screening masses are static quantities it is expected that they
can be determined in
a 3d effective theory, the 3d SU(N) adjoint Higgs model, provided
the temperature is high enough. However, in the case of the hot SU(N) 
gauge theory one may
worry whether the standard arguments of  dimensional reduction
apply. First of all the coupling constant is large $g \sim 1$  
for the physically interesting temperature range and thus the requirement
$g T << \pi T$ is not really satisfied. 
Another problem is the identification of the physical phase of
the 3d adjoint
Higgs model, i.e. the phase which corresponds to the high temperature
phase of SU(N) gauge theory \footnote{In the vicinity of the
deconfinement phase
transition the gauge coupling is very large and the dimensional
reduction is certainly not valid. Therefore the two phases of the
3d adjoint Higgs model has nothing to do with the phases of 4d
SU(N) gauge theory}. 
The 3d adjoint Higgs model is known
to have two phases, the symmetric (confinement) phase and the broken
(Higgs) phase \cite{nadkarni90, hart97, kajantie97b} 
separated by a line of strong $1^{st}$ order transition for small Higgs
couplings. The perturbative
calculation of the effective
potential \cite{kajantie97b,polonyi90} 
suggests
that the Higgs phase of the reduced theory corresponds to the deconfined
phase of the 4d SU(N) theory. This conclusion seems to be supported
by the 2-loop level dimensional reduction and the analysis of
the phase diagram performed in Ref. \cite{kajantie97b}. Though 
an earlier study \cite{kark94} indicated that the symmetric phase of 
the 3d adjoint Higgs model is the physical one, a more carefull
analysis shows that this conclusion may be altered by
higher order corrections generated in the process of
dimensional reduction (see appendix B.1).
The fact that the broken phase turns out to be the physical one 
is contrary to naive expectations
(see the discussion in the previous
section).
This circumstance and the ${\cal O}(1)$ value of the gauge coupling
raises serios doubts about the validity of the dimensional
reduction. Therefore a non-perturbative study is nedeed to
clarify the applicability of the 3d effective theory. 

The aim of this section is to clarify whether the screening masses,
defined as  poles of the corresponding lattice propagators
can be determined in the effective theory for the simplest
case of the SU(2) pure gauge theory. For this model 
precise 4d data on  screening
masses  exist for a huge temperature range \cite{heller98}.
Inclusion of 
fermions into the model is straightforward \footnote{Fermionic
fields do not appear in the effective theory, their role is only to 
modify 
the parameters of the effective theory.} and generalization to the
SU(3) case will not introduce any novel feature.

Gauge invariant definitions of the Debye mass as it was discussed in
section 2 rely
on the 3d effective theory \cite{arnold95,kajantie97a,kajantie97b}.
Although these definitions are the best
in the sense that they are explicitly gauge invariant, little is known
about the corresponding
correlators 
in the full 4d theory. 
Gauge invariant correlators yield larger masses than the pole masses
\cite{kajantie97a, kajantie97b}
and further studies are required to establish the connection between them.
We 
will try to relate these masses in  
section 4.5 of the thesis in the spirit of the so-called additive model
\cite{buchmuller97}.

\subsection{The 3d SU(2) Adjoint Higgs Model on Lattice}
The lattice action for the 3d adjoint Higgs model used in the present paper
has the form
\bea
&&
S=\beta \sum_P \half Tr U_P + 
\beta \sum_{\bx,\hat i} \half \tilde A_0(\bx) U_i(\bx) \tilde A_0(\bx+\hat i)
U_i^{\dagger}(\bx) + \nonumber\\
&&
\sum_{\bx} \left[-\beta\left(3+\half h\right) \half \tr \tilde A_0^2(\bx) + 
\beta x { \left( \half \tr
\tilde A_0^2(\bx)\right)}^2 \right],
\label{act}
\eea
where $U_P$ is the plaquette, $U_i$ are the usual link variables and
the adjoint Higgs field is parameterized, as in \cite{kajantie97b} 
by anti-hermitian matrices $\tilde A_0=i \sum_a \sigma^a
A_0^a$ ($\sigma^a$ {are the usual Pauli matricies}). Furthermore 
$\beta$ is the lattice gauge coupling, $x$ parameterizes the 
quartic self coupling of the Higgs field and $h$ denotes the
bare Higgs mass squared. The details of the derivation of the lattice
action can be found in appendix B.1. 

In principle the  parameters appearing in
eq.~(\ref{act}) can be related to the parameters of the original 4d theory
via the procedure of dimensional reduction
\cite{kajantie97b} 
which is essentially perturbative and discussed in appendix B.1.
The obvious problem with this approach, as it was already discussed
before, is that the physical phase turns out to be the broken phase.
This problem may signal  that higher dimensional and
non-local operators have to be kept in the effective action
to ensure the consistency of the whole approach
\footnote{An attempt to caclulate the higher dimensional operators
in the effective 3d action was recently done in Ref. \cite{bronoff98}.}. 
Certainly it is difficult to take into account all possible 
(higher dimensional and non-local )operators in a lattice
investigation.

There are two ways out
from this unfortunate situation. 
One can assume that the complicated action containing all kinds of
higher dimensional and/or non-local operators can be mapped onto
the effective action (\ref{act}) which is the action of some 
coarse-grained theory
with some finite UV cutoff and does not necessarily correspond to the
naive continuum 3d adjoint Higgs model. Then parameters appearing in
(\ref{act}) are found 
by
matching some quantities which are equally well calculable both in the full
4d  lattice theory and in the effective 3d lattice theory, 
i.e. by a {\em non-perturbative matching}.

Another possibility which was suggested in Ref. {\cite{kajantie97b} is
based on observation that the parameters of the 3d adjoint Higgs model 
which correspond to the high temperature SU(2) gauge theory lie
in the region of the parameter space which is close to the transition
line and where for any finite volume available in numerical
simulations the symmetric phase turns out to be metastable (see
appendix B.1 for a detailed discussion on this issue). 
For practical purposes,
e.g. measurements, the metastable phase behaves as if it was stable
and thus parameters appearing in (\ref{act}) can be fixed at their
values given by standard dimensional reduction. One proceeds futher by
arguing that if  higher dimensional and/or non-local operators
were taken into account the transition line would be shifted so that
the parameters corresponding to 4d physics would move to the region where
the symmetric phase is absolutely stable but, 
the inclusion of
these operators would have little impact on the values
of the quantities we are interested in. 

Here we will consider both possibilities. Comparison between
corresponding 3d and 4d data, hopefully enables us to clarify which
scenario takes place. 

Let us start to discuss the first strategy for fixing the parameters
$\beta, ~h, ~x$, namely the non-perturbative matching. We will turn
to the discussion of the second strategy in the next subsection.
The deconfined high temperature phase of the 4d SU(2) gauge theory
corresponds to some surface in the parameter space ($\beta,~h,~x$),
the {\em surface of 4d physics}. For fixed value of $\beta$ the 4d
physics is described by a line on this surface, which will be refered
to as the {\em line of 4d physics}. 
One of the aim of numerical simulations is the determination of the
surface of 4d physics.
In general this would require a matching analysis in a 3d parameter space
($\beta, ~x,~h$), which is clearly a difficult task. We thus will follow
a more moderate approach and fix two of the three parameters
namely $\beta$
and $x$, to the values obtained from the perturbative 
procedure of dimensional reduction. The values of these parameters
at 2-loop level are \cite{kajantie97b}
\vskip0.3truecm
\ba
&&
\beta={4\over g_3^2 a}, \nonumber \\[0.1cm]
&&
g_3^2=g^2\left(\mu\right) T \left[1+{g^2\left(\mu\right)\over 16 \pi^2}
\left(L+{2\over3}\right)\right], \\
&&
x={g^2\left(\mu\right)\over 3 \pi^2}\left[1+
{g^2\left(\mu\right)\over 16 \pi^2} \left(L+4\right)\right],\\
&&
L={44\over 3} \ln{\mu\over 7.0555 T},
\ea
with $a$ 
and $T$ denoting the lattice spacing and temperature, respectively.
The coupling constant of the 4d theory $g^2(\mu)$ is defined through 
the 2-loop formula
\be
g^{-2}(\mu)={11\over 12 \pi^2} \ln{\mu\over \Lambda_{\overline{MS}}}+
{17\over 44 \pi^2}\ln\left[2\ln{\mu\over
\Lambda_{\overline{MS}}}\right].
\label{4dg}
\ee
The parameter $h$ will be left free to allow non-perturbative matching.
In order to be able to compare the results of the 3d simulation with 
the corresponding calculations
in the 4d theory it is necessary to fix the renormalization and 
the temperature
scale. We choose the renormalization scale to be $\mu=2 \pi T$, which ensures
that corrections to the leading order results for the parameters
$g_3^2$ and $x$ 
of the effective theory are small. Furthermore we use the 
relation $T_c=1.06 \Lambda_{\overline{MS}}$ from \cite{heller98}. 
Now the temperature scale is fixed 
completely and the physical temperature may be varied by varying the 
parameter $x$.  The lattice spacing was chosen according to the criterium
$a<<m^{-1}<<N a$, where $N$ is the extension of the lattice and $m$ is the
mass we want to measure. 

\subsection{Determination of the Propagators and the Screening Masses
on Lattice}
The main goal of this section is to study the propagators  of
scalar and vector (gauge) fields. For this purpose one has 
to fix a specific gauge,
which is chosen to be the Landau gauge
$\partial_{\mu} A_{\mu}=0$. The choice of this gauge is
motivated by the fact that the 4d Landau gauge condition
used in \cite{heller95,heller98} for static field configuration is
equivalent to the 3d Landau gauge condition. 
On the lattice this condition is realized by
maximizing the quantity \cite{mandula87}:
\be
\tr \left[ \sum_{\bx,i} \left ( U_i(\bx)+U^{\dagger}_i(\bx) \right) \right]
\label{max}
\ee
The gauge fixing is performed using the overrelaxation algorithm
\cite{mandula90}, which 
in our case is as efficient as the combined overrelaxation and $FFT$ algorithm 
used in \cite{heller95,heller98}.
The vector field is defined in terms of the link variables as
\be
\tilde A_i(\bx)={1\over 2 i} (U_i(\bx)-U^{\dagger}_i(\bx))
\ee
We are interested in extracting the electric (Debye) $m_D$ and the magnetic
$m_T$ screening masses from the long distance behaviour of the scalar and
vector propagators defined as
\bea
&&
G_D(z)=< \tr \tilde A_0(z)\tilde A_0^{\dagger}(0) >\sim \exp(- m_D z),
\label{gdz}\\
&&
G_T(z)= {1\over 2} ( G_1(z) + G_2(z)) \sim \exp(- m_T z),
\label{gtz}
\eea
with  
\bea
&&
G_i(z)=< \tr \tilde A_i(z) \tilde A_i(0) >,\nonumber\\
&&
\tilde A_\mu(z)=\sum_{x,y} \tilde A_\mu(x,y,z),\quad \mu=0,~1,~2
\eea
Note that due to the Landau gauge condition $G_3(z)$ should be constant. This
fact can be used to test the precision and validity of the gauge fixing 
procedure. In our case this condition is satisfied with an accuracy of 
$0.001\%$.

Besides the scalar and vector propagators of the adjoint Higgs model 
we also calculate the
gauge invariant scalar correlators and analyze the propagators
also in the limit
of a 3d pure gauge theory.

To extract the masses from the correlation functions we have used the
general fitting ansatz
\be
A \left[ {\exp\left(-m z\right) \over {z}^b}+ {\exp \left(- m \left (N_z-z
\right)\right)\over {\left(N_z-z\right)}^b} \right].\quad  
\label{fitI}
\ee
Previous investigations of gauge boson and quark propagators in Landau gauge
\cite{heller95,bernard91,dimm95} have shown 
that effective masses extracted from
the correlation functions rise with increasing Euclidean time separation
and eventually reach a plateau. This is contrary to local masses extracted 
from gauge invariant correlation functions which approach a plateau from 
above and is a direct consequence of a non-positive definite 
transfer matrix in 
Landau gauge. A fit with $b\ne 0$  
allows to extract stable masses already at shorter distances.  
Most of our numerical studies have been performed on lattices of
size $32^2\times 64$ \footnote{
Additional calculations also have been performed on a
$16^2\times 32$ lattice and it have been 
checked explicitly that results for the electric
mass show no volume dependence. Furthermore some calculations have
been performed on a $32^2\times 96$ lattice to check that the lattice size
used for our calculations was sufficient for the determination of the
magnetic mass.}.
From previous studies of gluon propagators in the 4-dimensional SU(2)
gauge theory we know that such large lattices are needed to observe a plateau in
local masses \cite{heller98}. In particular, this is the case for 
the rather small magnetic screening mass which leads to a rather slow decay
of the correlation functions. We thus use the same spatial lattice size as in 
those studies.

We have used correlated (Michael-McKerrell) fits with eigenvalue smearing
\cite{michael95}. 
Our fits have been constrained to the region where local masses
\footnote{The definition of the local masses are given in
Appendix B.2 where also futher details about the fitting
procedure can be found} show a 
plateau (typically $z\sim 15$ for the magnetic mass and $z\sim 5$ for 
the electric mass). In this region fits with $b=0$ and $b\ne 0$ yield 
mutually consistent results within statistical errors. 
From fits with $b\ne 0$ we 
find best fits with $b < 0$, in accordance with the 
behaviour of local masses discussed above. 
Although these good quality fits with $b <0$  start at shorter distances, 
the magnitude of $b$ is not well determined
within our present statistical accuracy. In the following we thus will
quote results from fits with $b=0$.

\subsection{ Numerical Results for the Propagators of the 3d SU(2) 
Higgs Model}
One of the aim of the numerical simulations is to clarify
how far the screening mass can be determined in the framework
of the effective 3d theory and to find the most suitable choice
of $h(x,\beta)$.

As it was already noticed before
it is expected that the symmetric phase of the 3d model corresponds
to the physics of the high temperature 4d SU(2) theory.
Nevetherless let us investigate the propagators in the broken
phase by fixing the parameter $h$ to its value given by perturbative
dimensional reduction $h_{4d}^p(x)$.
In the broken phase the simulations were done for two sets of parameters:
$\beta=16,~~x=0.03,~~h=-0.2181$ and $\beta=8~~,x=0.09,~~h=-0.5159$.
The propagators obtained by us in the broken phase show 
a behaviour which is very different from
that in the symmetric phase and that in the 4d case studied in Refs.
\cite{heller95, heller98}. The magnetic mass extracted from the 
gauge field propagators is $0.104(20)g_3^2$ for the first set of 
parameters and $0.094(8)g_3^2$ for the second set of parameters.
It thus is a factor 4 to 5 smaller than the corresponding 4d result.
Moreover, the propagator of the $A_0$ field does not seem to
show a simple exponential behaviour, this fact actually is in qualitative 
agreement with the results of analytic calculations of Ref. \cite{rebhan94}.
Taken together these facts suggest that the broken phase does not
correspond to the high $T$ physics of the 4d system.

Now let us discuss the procedure of choosing the parameter
$h$ in the symmetric phase. The first step of this procedure is
to chose for each value of $x$ a number of trial values for $h$
for some fixed $\beta$ and measuring the propagators for these
parameters. 
The value of $\beta$ was chosen to be
$\beta=16$. The second step might be an extrapolation of the screening masses
measured for different values of $h$ for each $x$ to the value of the screening
masses obtained in 4d simulation and determination of the corresponding
value of $h$.
The guiding principle in choosing the trial values of $h$ was the fact
that
these  values should be close to the value corresponding to the
transition. This is based on observation that the perturbative line
of 4d physics lies close to the transition line. We suspect the same
is true for the physically relevant $h$ values with the exception that
they are located in the symmetric phase. 
Therefore first we have to determine the transition line $h_{c}(x)$. 
The transition line
as function of $x$ in the infinite volume limit was found in
\cite{kajantie97b} in terms of the renormalized mass parameter
$y$ (see appendix B.1 for definition of $y$). The transition
line in terms of $y$ turns out to be independent of $\beta$.
Then using Eqs. (\ref{fityc}) and (\ref{hx}) one can calculate $h_{c}(x)$.
The usage of the infinite volume result for the transition
line seems to be justified 
because most of our simulation were done  on a
$32^2 \times 64$ lattice. 
The two sets of $h(x)$ values, which appear on Figure 10,
were chosen so that the renormalized mass
parameter $y$ always 
stays $10 \%$ and $25 \%$ away from the transition line.
The values of the parameters in the symmetric phase are shown 
in Table 1 where the two sets of $h$ values are 
denoted as (I) and (II) and also the values
of $h_{c}$ corresponding to the transition line are given.
Theses trial values together with perturbative line
of 4d physics are also depicted in Figure 10.
To check the suitability of trial values shown in Table ~1 simulation 
for $x=0.07$ and value of $h=-0.2179$ which lies deeply in 
the symmetric phase was done. The corresponding value of the Debye
mass was found to be $2.41(11)$. This value should be compared with
the value of the Debye mass obtained $m_D/T = 1.85$ obtained in
4d case \cite{heller98} for $T/T_c= 12.57$ which corresponds to
$x=0.07$. Clearly this choice of $h$ is not suitable and one
has to choose smaller $h$ values to reproduce 4d data.
The analysis described above motivated our choice of trial values for
$h$ at other values of $x$ as given in Table~1.

\vskip0.5truecm
\begin{center}
\begin{tabular}{|l|l|l|}
\hline
$Temperature~scale~~$ &$~~~~~~~~~~~~~~~~~~~~~~~~~~~~~~h~~~~~~~~~~~~~~~~~~~~~~~~~~~~~$ \\
\end{tabular}
\begin{tabular}{|l|l|l|l|l|}
\hline 
$~~~~x~~~~$  &$~~~~T/T_c~~~~$   &$~~~~~~~I~~~~~~~$  &$~~~~~~~II~~~~~~~$ 
&$~~~~transition~~~~$\\
\hline
$~~0.09~~$  &$~~~4.433$   &$~~-0.2652$   &$~~-0.2622$   &$~~-0.2672(4)$\\
$~~0.07~~$  &$~~~12.57$   &$~~-0.2528$   &$~~-0.2490$   &$~~-0.2553(5)$\\
$~~0.05~~$  &$~~~86.36$   &$~~-0.2365$   &$~~-0.2314$   &$~~-0.2399(6)$\\
$~~0.03~~$  &$~~~8761$   &$~~-0.2085$   &$~~-0.2006$   &$~~-0.2138(9)$\\
\hline
\end{tabular}
\vskip0.3truecm
Table 1: {The two sets of the
bare mass squared used in the simulation and those
which correspond to the transition line for $\beta=16$}
\end{center}

\begin{figure}
\vspace{-1.5cm}
\epsfysize=9cm
\epsfxsize=11cm
\centerline{\epsffile{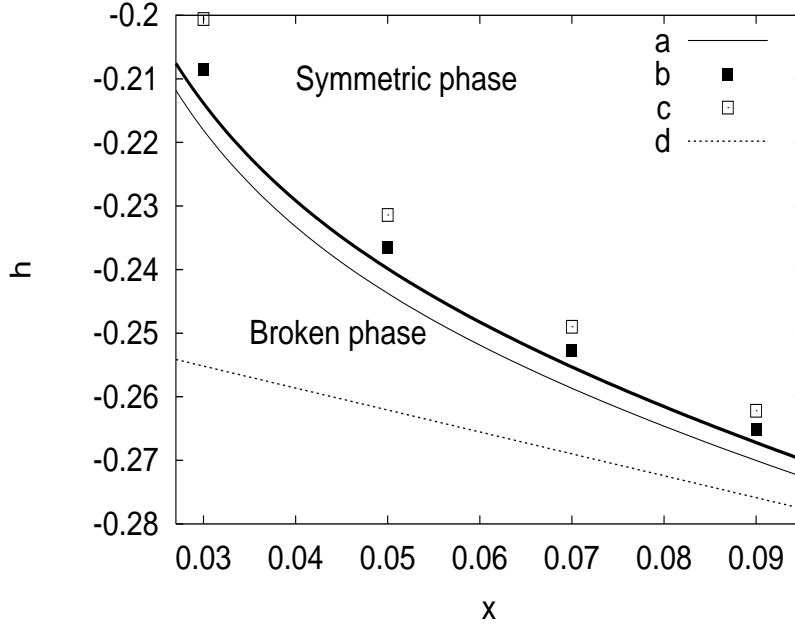}}
\vspace{-1cm}
\caption{The  phase diagram of the 3d SU(2) adjoint Higgs model 
and bare mass parameters $h$ used 
in our analysis (squares) for $\beta=16$: perturbative line (a) and
two sets $I$ (b) and $II$ (c). For a discussion of their
choice see text. 
The thick solid line is the transition line. 
The dashed line (d) indicates the values of $h$ below which no
metastable branch exists and was estimated based on results of
Ref. \cite{kajantie97b}.}
\end{figure}

Yet another possiblity to be discussed is the simulation in
the metastable symmetric phase along the line of perturbative
4d physics. As was already pointed out due to the fact that the
transition line is strongly first order for all 
physically relevant values of $x$ metastable symmetric branch 
exists below the transition line (see Figure 10).
In fact the perturbative line of 4d physics lies in the region
where this metastable branch exists. If one prepares the system
to be in this metastable phase by starting a Monte-Carlo simulation
with disordered random configuration it will not tunel to the
other broken phase corresponding to the broken phase \footnote{
The tuneling between two branches is supressed by a factor which
is the exponential of the volume for all standard 
Monte-Carlo updating algorithms.}.
Simulations in the metastable phase were done for parameters shown
in Table~1.
\begin{figure}
\epsfysize=8cm
\epsfxsize=11cm
\centerline{\epsffile{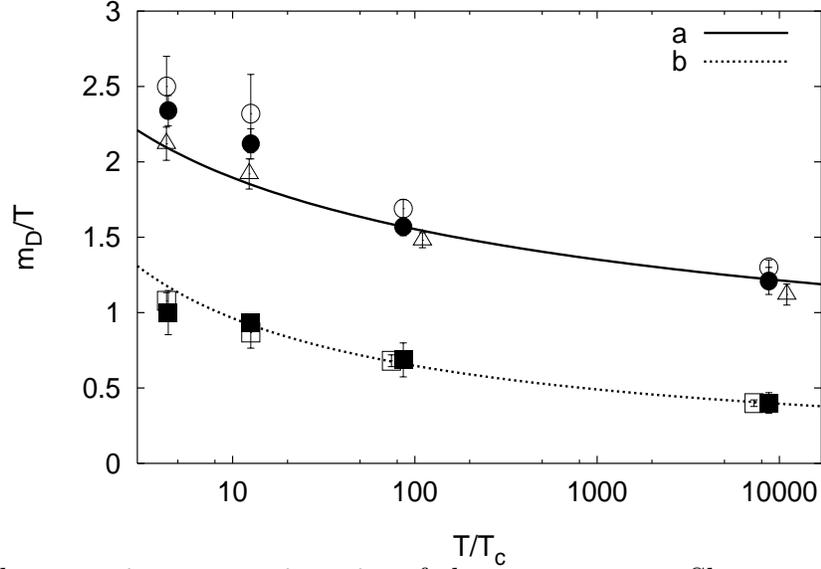}}
\vspace{-0.7cm}
\caption{The screening masses in units of the temperature. 
Shown are the Debye mass
$m_D$ for the first (filled circles) and the second (open circles)
set of $h$, and
the magnetic mass  $m_T$ for the first (filled squares)
and the second (open squares) set of $h$.
Also shown there are the values of the Debye mass measured in
metastable phase along the perturbative line of 4d physics
(open triangles).
The line (a) and line (b) represent the fit for the temperature
dependence of the Debye and the magnetic mass from 4d
simulations
from \cite{heller98}. Some data points at the temperature $T \sim 90
T_c$ and $ \sim 9000 T_c$ have been shifted in the temperature scale for
better visualization.}
\end{figure}

The temperature dependence of the screening masses 
obtained in the symmetric phase for two sets of  
the parameters, as well for the perturbative line of 4d physics $h_{4d}^p(x)$.
in the metastable phase is shown in Figure~11. 
Also given there is the result of the 4d simulations \cite{heller98}, 
$m_D^2/T^2 = Ag^2(T)$,
with $A=1.70(2)$ for the electric mass and $m_T/T= C g^2(T)$, with
$C=0.456(6)$ for the magnetic mass. 
As one can from Figure 11 the magnetic mass agrees quite well with
the 4d data for both sets of $h$ values given in Table~1 and it 
is not sensitive to the actual value of $h$. The magnetic mass 
calculated for the perturbative line of 4d physics yields roughly
the same result. The electric mass shows more pronounced
dependence on $h$ for temperatures smaller than $15 T_c$ and as
one can see from Figure 11 only the values of the electric mass
calculated for $h_{4d}^p$ are consistent with the 4d data. For larger
temperatures the dependence of the electric mass on the actual
value of $h$ weakens but the second set of $h$ values, which 
lies more deeply in the symmetric phase, seems to overestimate 
the value of $m_D$. 

Based on these observations we can conclude:
the the interpolation procedure for finding the true value of
$h_{4d}(x)$ is not necessary and one can assume that the correct
choice is $h_{4d}^p(x)$. This choice for the line of 4d physics
reproduces the screening masses with about $5\%$ 
accuracy in the entire temperature range. This is exactly the
precision which is reasonable to assume for the 3d effective theory. 
Nevertheless it is important to notice that an accepatable description
of the screening masses can be also achived with the choice of $h$ 
corresponding to the symmetric phase.

\subsection{Magnetic Mass in 3d SU(2) Pure Gauge Theory}
The magnetic mass found in the previous section seems to scale with
the 3d gauge coupling $g_3^2\sim g^2(T)T$. A similar behaviour 
was found for the square root of the spatial string tension 
in the deconfined phase of the finite temperature SU(2) gauge
theory \cite{bali93}. Moreover, the value of the spatial string tension
is $\sqrt{\sigma_s}=0.369(14)g^2(T)T$, which is very close the the
value of the string tension calculated in the 3d pure SU(2) gauge
theory:
$\sqrt{\sigma_3}=0.3340(25)g_3^2$ \cite{teper92}.
This behaviour can be understood 
for very high temperature (small coupling) because in this case,
as it was discussed in subsection 2.3, 
the heavy $A_0$ field  with mass
$\sim g T$ can also be integrated out and the $IR$ behaviour of
high temperature 
{\bf $SU(N)$ }
gauge theory is described by a 3d pure gauge theory in
which the only dimensionfull scale is $g_3^2 \sim g^2 T$.
Unfortunately for the realistic temperature range (which is also
studied here ) the above argument fail to 
hold because the coupling is large. 
A non-perturbative study is therefore needed to establish the 
relation between
the magnetic mass found in the finite temperature SU(2) theory and 
the magnetic mass  
of the 3d pure gauge theory.

We have measured the Landau gauge propagators for the 3d SU(2) gauge theory
and from its large distance behaviour extracted the magnetic mass.
The results for different values of $\beta$ are listed in Table~2.
\vskip0.3truecm
\begin{center}
\begin{tabular}{|l|l|l|l|l|l|}
\hline
$~~~\beta~~~~$  & $~~~m_T/g_3^2~~~$ & $~~~\chi^2/d.o.f~~~$\\
\hline
$12.00$ & $0.48\pm0.036$ & $~~~~~0.580 $\\
\hline
$16.00$ & $0.42\pm0.070$ & $~~~~~0.497 $\\
\hline
$20.00$ & $0.44\pm0.068$ & $~~~~~1.439 $\\
\hline
\end{tabular}
\vskip0.5truecm
Table 2: {The results of the fit for the magnetic mass in 3d pure
gauge theory. Simulations for $\beta=20$ were done on $40^2\times 96$
lattice}
\end{center}

Using the data from Table~2 one finds $m_T=0.46(3)g_3^2$.
This value  is in good agreement with  the 3d adjoint Higgs
model result. The magnetic mass thus  is rather insensitive to the dynamics
of the $A_0$ field. 
This finding is 
in accordance with the gap equation study of the 
adjoint Higgs model described in the previous section.
Gauge boson propagators were also studied in the 3d SU(2) Higgs model
in Ref. \cite{karsch96}. It was found that in the symmetric phase the
magnetic mass extracted from the propagator is insensitive to the
values of the scalar couplings and its value is $0.35(1)g_3^2$.
The magnetic mass in the 3d pure gauge theory was also measured 
there and its value was found to a bit larger but close  $0.35(1)g_3^2$
(see Figure 3 in Ref. \cite{karsch96a}). This value is considerably
smaller than the value found by us. This fact is probably due to small
lattices used in Ref. \cite{karsch96a} and to the different method for
extracting the screening masses. These issues are discussed in
appendix B.2. 

The decoupling of static magnetic sector is also seen in
study of gauge invariant correlators. In Refs.
\cite{kajantie97b,philipsen99} the mass of the lowest lying gluball
was measured in the 3d SU(2) adjoint Higgs model and its value was found
rather insensitive to the values of scalar couplings and close to
the corresponding value of the 3d pure SU(2) gauge theory.

\subsection{Gauge Invariant Correlators and the Additive Constituent 
Gluon Model}
In this subsection I will establish a simple
relation between gauge invariant and gauge dependent 
correlators.
Gauge invariant correlators for the SU(2) adjoint 
Higgs model were measured in
Refs. \cite{kajantie97b,philipsen99}. 
The correlators of the following gauge invariant operators
were studied: $\tr A_0^2,~h_{ij}=\tr A_0 F_{ij}$ and $G=F_{ij}^a
F_{ij}^a$ \footnote{In Ref.\cite{kajantie97b} the correlation
function of the oparator $h_i=\epsilon_{ijk} \tr A_0 F_{jk}$
was studied instead of $h_{ij}$. But the masses extracted from this 
correlation function are close to those extarcted from $h_{ij}$}.

Let us first consider
the correlation function of the scalar operator $Tr A_0^2$ 
whose large distance
behaviour  gives the mass of the $A_0-A_0$ bound state: $m(A_0)$.
The mass of this
bound state is also expected to determine the exponential falloff of the 
Polyakov loop correlator \cite{kajantie97b}. 
Let us discuss the relation of $m(A_0)$ and the Debye mass defined through
the propagator.
If the coupling constant is small enough then
$m(A_0) \sim 2 m_{D0}$. In our case this 
perturbative relation is not 
satisfied. However, following the analysis of Ref. \cite{buchmuller97}
the mass of the 
$A_0-A_0$ bound state can be represented by the sum of two constituent
adjoint scalar masses, which is $m_D$.  

Next we discuss the gauge invariant 
correlator contaning gauge field operators. In Ref. \cite{buchmuller97}
it was suggested that the screening masses extracted from these
correlators can be related to the screening masses extracted from
propagators using the notion of the constituent gluon. According to this
idea one can associate with each covariant derivative in the
corresponding operator a constituent gluon with the mass equal to
the propagator pole mass $m_T$. It was shown in Ref.
\cite{buchmuller97} that using this simple idea the spectrum of 3d 
SU(2) Higgs model can be qualitatively well interpreted. 
The mass of the lowest lying glueball state was measured in
Ref. \cite{philipsen99} 
\footnote{Earlier measurements of 
Ref. \cite{kajantie97b} overestimated the mass of this state.}
and  was found to be rather independent of the
couplings $h$ and $x$ of the $A_0$ field, the value of this mass
is  $m_G \sim 1.67(2) g_3^2$.
In terms of the constituent model this state can  be viewed as a 
loose bound state 
of four constituent gluons \cite{buchmuller97} with a constituent mass equal 
to $m_T$, thus $m_G \sim 4 m_T$. If one uses the value of magnetic mass 
in pure gauge theory found in Monte-Carlo simulations one finds 
$m_G=1.84(12)g_3^2$. For the value of the magnetic mass given by
Nair \cite{nair98} one gets $m_G \sim 1.28g_3^2$.

An attempt 
for a gauge invariant definition of the constituent gluon mass was given
in Ref. \cite{laine98}. The approach is
based on the expactation value of the following operator 
\be
G_{Fijkl}(x,y)=< F_{ij}^a(x) U_{ab}(x,y) F_{kl}^b(y)>,
\ee
where
\ba
&&
U_{ab}(x,y)=2 \tr T^a U(x,y) T^b U^{\dagger}(x,y),\nonumber\\
&&
U(x,y)=P \exp\biggl(ig \int_x^y dx_i A_i^a T^a\biggr)
\ea
with $T^a$ being SU(N) generators. For large separation one has
$G_F \sim \exp(-M_s |x-y|)$. According to the additive constituent 
gluon model
$M_s=2 m_T$. The numerical value
was found to be $M_s \sim
0.73g_3^2$. Based on the additive constituent gluon model and the
numerical value of the magnetic mass in pure SU(2) gauge theory one
would expect $M_s \sim 0.92(6)g_3^2$. On the other hand with the value
of the magnetic mass obtained by Nair \cite{nair98} one expects 
$M_s=0.64 g_3^2$.

Finally we consider the correlation function of $h_{ij}$. This
correlation function was used for non-perturbative definition
of the Debye mass \cite{kajantie97a,laine98,laine99a}. In terms 
of the additive constituent gluon model the mass extracted from
this correlation function is $m_h=m_D+2 m_T$.
The masses predicted by the constituent model compared with the results
of direct measurements are shown in Table~3.
\vskip0.3truecm
\begin{center}
\begin{tabular}{|l|l|l|l|}
\hline
$~~~~parameters~~~~$  &$~~~~~~~~~~~~~~~~~m(A_0)/g_3^2~~~~~~~~~~~~$
&$~~~~~~~~~~~~~~~m_h/g_3^2~~~~~~~~~~~~~$\\
\hline
$\beta~~~\vline~~~~x~~~~\vline~~~~h~~~~~$ &$~measured~$ \vline $~constituent
~model~$ &$~measured$ \vline $~~constituent~model~$\\
\hline
$9~~~~~0.10~~~-0.4764$  &$~~~1.01(2)^a~~~~~~~~~~~~1.75(18)$
&$1.39(6)^a~~~~~~1.79(15)~~[~1.51(1)~]$\\
$16~~~~0.09~~~-0.2622$  &$~~~1.54(15)~~~~~~~~~~~~1.78(6)$
&$~~~~~~~~~~~~~~~~~~~~~~~~~~~$\\
$16~~~~0.05~~~-0.2314$  &$~~~2.28(20)~~~~~~~~~~~~2.38(16)$
&$~~~~~~~~~~~~~~~~~~~~~~~~~~~$\\
$9~~~~~0.04~~~-0.2883$  &$~~~2.41(2)^a~~~~~~~~~~~~2.74(2)$
&$2.16(20)^a~~~~~~2.29(7)~~[~2.01(1)~]$\\
$32~~~~0.04~~~-0.1247$  &$~~~2.20(11)^b~~~~~~~~~~~2.42(1)$
&$2.42(15)^b~~~~~~2.12(7)~~[~1.85(1)~]$\\
$24~~~~0.03~~~-0.1475$  &$~~~3.03(65)~~~~~~~~~~~~3.28(30)$
&$~~~~~~~~~~~~~~~~~~~~~~~~~~$\\
\hline
\end{tabular}
\vskip0.3truecm
Table~3: {The masses of the adjoint scalar and heavy-light bound states 
in units of $g_3^2$ compared with
the predictions of the aditive constituent gluon model. Some bound state
masses were taken from Ref. \cite{philipsen99} (a) and Ref. \cite{kajantie97b}
 (b), other values were taken from Ref. \cite{karsch98k}.
In square brackets we indicate the predictions of the constituent
gluon model based the value of the magnetic mass estimated by Nair
\cite{nair98}.}
\end{center}
\vskip0.3truecm
As one can see from the Table except for the largest value of $x$
which corresponds to the temperature $T \sim 3T_c$
the simple additive model can describe
the spectrum of gauge invariant screening masses with the accuracy 
of $15-20 \%$. This is the accuracy which is reasonable to expect from
such simple bound state picture which does not take into account binding energy.

The success of the additive model in describing the mass of the
$A_0-A_0$ bound state has important physical consequence concerning 
the physics of the screening defined from the Polyakov loop
correlator. In section 2.4 it was argued that the true large distance
behaviour of the Polyakov loop correlator is determined by the lightest
3d gluball state. However, as one can see from Table 3
for $x \le 0.5$ the the masses extracted from the $\tr A_0^2$
correlator are larger
than the 3d gluball mass and can be well described by the additive
model. This suggests that the dominant contribution to the Polyakov
loop correlator comes from the exchange of two constituent $A_0$ field
in a similar manner as it happens in the leading order of perturbation
theory.

\subsection{Gauge Dependence of the Screening Masses}

Before closing this section we briefly want to address the question of a 
possible gauge dependence of the screening masses extracted by us.
The pole of the propagator was proven to be gauge invariant in
perturbation theory \cite{kobes90,kronfeld98}. The statment
on the propagator pole masses extracted from lattice simulation 
is less clear. Here one faces the numerical problem
to isolate the asymptotic large distance behaviour of the correlation
function from possible short distance (powerlike) corrections which are
gauge dependent \cite{bernard91}.  
In Ref. \cite{dimm95} quark-propagators were studied 
in axial, Coulomb and Landau
gauges and the effective masses extracted at quite short
physical distances 
were found to depend on the gauge. In Ref. \cite{bernard91} the
quark and gluon propagators were investigated in so-called
$\lambda$-gauges. Masses extracted from the propagators using
exponential fits like Eq.~(\ref{fitI}) also show a mild dependence
on the gauge parameter.
To study the gauge dependence of our results, following, \cite{bernard91}
we have introduced $\lambda$-gauges which in our case are defined by the
condition
\be
\lambda \partial_3 A_3+\partial_2 A_2+\partial_1 A_1=0.
\ee
The propagators have been measured in $\lambda$ gauges and the masses were
extracted from the propagators using the functional form given in 
Eq.~(\ref{fitI}).

Some preliminary results on the gauge dependence of the screening masses
were already published in \cite{karsch99}. However the conclusion 
presented there are somewhat different from the present situation due to
the low statistics of the first analysis.

For the numerical analysis the following values of the parameter $\lambda$
have been chosen : $\lambda=0.5,~1.0,~2.0,8.0$. Simulations
were performed on a $32^2\times 96$ lattice at $x=0.03$ and for two values
of $\beta$ and $h$.
Values of the Debye
mass measured for different $\lambda$ and two values of $\beta$ are
summarized in Table 4.
\vskip0.5truecm
\begin{center}
\begin{tabular}{|l|l|l|l|l|}
\hline
$\lambda$ &$0.5$ &$1.0$ &$2.0$ &$8.0$\\
\hline
$\beta=16,~h=-0.2085$ &$1.28(6)$ &$1.21(9)$ &$1.15(5)$ &$1.08(6)$\\
$\beta=24,~h=-0.1510$ &$1.26(7)$ &$1.31(9)$ &$1.20(5)$ &$1.11(3)$\\
\hline
\end{tabular}
\vskip0.3truecm
Table 4: The Debye mass in units of the temperature for
$x=0.03$ and calculated for different $\lambda$
parameters.
\end{center}
\begin{figure}
\epsfxsize=10cm
\epsfysize=7cm
\centerline{\epsffile{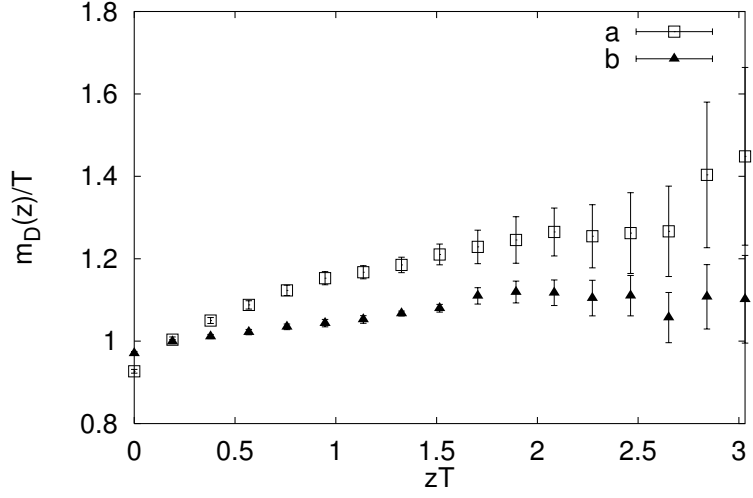}}
\caption{The local electric masses for $\beta=16$, $x=0.03$ and
$h=-0.1510$
in units of $g_3^2$ measured 
on $32^2 \times 96$ lattice for different values of the gauge parameter 
$\lambda$. Shown are the local electric mass for $\lambda=0.5$ (a)
and for $\lambda=8.0$ (b).
}
\label{maloclam}
\end{figure}

\begin{figure}
\epsfxsize=10cm
\epsfysize=7cm
\centerline{\epsffile{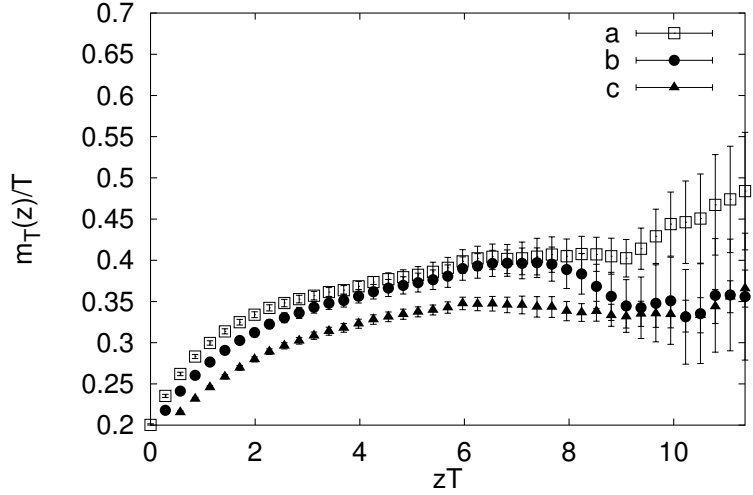}}
\caption{The local magnetic masses for $\beta=16$, $x=0.03$ and $y=0.9279$
in units of $g_3^2$ measured 
on $32^2 \times 96$ lattice for different values of the gauge parameter 
$\lambda$. Shown are the local magnetic mass for $\lambda=0.5$ (a), for
$\lambda=1.0$ (b) and for $\lambda=2.0$ (c).
}
\end{figure}

As one can see from the results the Debye mass shows a weak dependence
on $\lambda$. Its value is decreasing as $\lambda$ increases but for
$\lambda \ge 2$ it seems to approach a constant value which is independent
of $\lambda$. This dependence on $\lambda$ is similar to those which
was found in
Ref. \cite{bernard91}. In Figure \ref{maloclam} we show the local
electric masses for $\beta=24$ and $h=-0.1510$ for the largest
($\lambda=8$) and the smallest ($\lambda=0.5$) values of the 
$\lambda$ parameter.

The local magnetic masses measured at $\beta=16$, $x=0.03$ and
$h=-0.2085$
for $\lambda=0.5,~1.0,~2.0$ are shown in Figure 13. 
Similary to the electric mass the magnetic mass also show a mild
dependence on $\lambda$. 

Futher studies are required to clarify the
source of the non-perturbative gauge dependence of the screening
masses.

\newpage
\section{Coupled Gap Equation for SU(2) Higgs Model}

As it was already mentioned in the Introduction, considerable progress  
has been achived 
in understanding the thermodynamics of the electroweak phase transition
in the past 5 years.    
For the electroweak theory the separation of different mass scales
holds ($2 \pi T \gg gT \gg g^2 T$)\footnote{The renormalized
gauge coupling for the electroweak theory has a moderate value 
$g \sim 0.66$. In the following
discussion we will ignore the presence of the U(1) sector and fermions
because they are not essential from the point of view of basic properties
of the electroweak phase transition.} and the
superheavy modes (i.e. the non-zero Matsubara modes with typical
mass $\sim 2 \pi T$) together with the heavy $A_0$ field 
(with a mass $\sim g T$)
are integrated. The thermodynamics is described by an effective
theory, the 3d SU(2) Higgs model \cite{jako94,farakos94,kajantie96a,jako97}. 
The properties of the phase transition and the screening masses were
studied in great detail using lattice Monte-Carlo simulations
of the reduced model\cite{kajantie96b,karsch96a}. 
The application of the gap equation approach
\cite{buchmuller94fodor,buchmuller95} led to similar results. 
Both approaches predict that the line of first order transitions
ends for some Higgs mass $m_H=m_H^c$ and the value of the critical mass
$m_H^c$ was found to be the same within $20\%$. 

Recent 4d Monte-Carlo 
simulations of the finite temperature SU(2) Higgs model 
\cite{aoki99,csikor99} provide good 
non-perturbative tests for the validity of dimensional reduction.
A detailed discussion of relating 4d and 3d results
was published very recently in Ref. \cite{laine99b}.

The purpose of the investigations presented in this section is twofold.
The first aim is to provide 
some non-perturbative evidence for the decoupling of the $A_0$ field from the
gauge + Higgs dynamics in the vicinity of the phase transition. 
The coupled set of gap equations for the screening masses will be
derived and solved numerically in 
the 3d fundamental + adjoint Higgs model. This model emerges when the 
non-static modes are integrated out in the full finite temperature Higgs 
system. Its predictions for the screening masses will be compared 
with those obtained by Buchm\"uller and Philipsen (BP)
\cite{buchmuller95} in the 3d Higgs model using the same technique.
This model is obtained from the 3d fundamental + adjoint Higgs model by
integrating out the heavy $A_0$ field.
The main result of 
this comparison is a proposition for a non-perturbative and non-linear
mapping between the two models ensuring
quantitative agreement between the screening masses in a wide temperature
range on both sides of the transition. This high quality evidence for 
the decoupling of the $A_0$ field at the actual finite mass ratios presumes, 
however,
the knowledge of the "exact" value of the Debye screening mass, since for the 
proposed mapping its non-perturbatively determined value turns out to be 
essential.

The second aim of the present investigations is
to examine the symmetric phase in more detail. 
There the Higgs and the Debye screening masses are both of the same order 
of  magnitude $\sim gT$ and thus in that regime there is no {\it a priori} 
reason for the $A_0$ field to 
decouple. This circumstance makes the quantitative relation of the screening 
masses calculated in the 3d fundamental + adjoint Higgs model particularly 
interesting in the high-T phase. Here we are going to apply two different 
resummation techniques and check to what extent 
persists a non-perturbative mass hierarchy in this part of the spectra.
Investigations presented in this Section 
are based on Ref. \cite{petreczky99}.

\subsection{The Extended Gap Equations}
The Lagrangian of the three dimensional SU(2) fundamental $+$ adjoint Higgs 
model is \cite{buchmuller95,farakos94}
\bea\label{l3d}
L^{3D}&=&Tr\bigl[\frac{1}{2}F_{ij}F_{ij}+(D_i\Phi)^+(D_i\Phi)+\mu^2\Phi^+\Phi+
2\lambda (\Phi^+\Phi)^2\bigr]\nonumber\\
&&
+\frac{1}{2}(D_i\vec{A}_0)^2+
\frac{1}{2}\mu_D^2{\vec{A}_0}^2+{\lambda_A\over 4}(\vec{A}_0^2)^2+
2c\vec{A}_0^2Tr\Phi^+\Phi,
\eea
where
\bea
\Phi=\frac{1}{2}(\sigma 1+i\vec{\pi}\vec{\tau}), \quad 
D_i\Phi=(\partial_i-igW_i)\Phi,\quad W_i=\frac{1}{2}\vec{\tau}\vec{W}_i.
\eea
The relations between the parameters of the 3d theory  and those of the
4d theory are derived at 1-loop level perturbatively \cite{farakos94}:
\bea
&
g^2=g^2_{4d}T,\qquad\quad
\lambda=\left(\lambda_{4d}+\frac{3}{128\pi^2}g^4_{4d}\right)T,\qquad\quad
\lambda_A=\frac{17}{48\pi^2}g^4_{4d}T,\nonumber\\
&
c=\frac{1}{8}g^2_{4d}T,\qquad\quad
\mu_D^2=\frac{5}{6}g^2_{4d}T^2,\qquad\quad
\mu^2=\left(\frac{3}{16}g^2_{4d}+\frac{1}{2}\lambda_{4d}\right)T^2-
\frac{1}{2}\mu^2_{4d}.
\eea 
If the integration over the $A_0$ adjoint Higgs field is performed we obtain
the model investigated in \cite{buchmuller95} with parameters $\bar g, \bar\lambda,
\bar\mu$. These couplings of the reduced theory are related to the
parameters of the 3d fundamental $+$ adjoint Higgs theory through the
following relations:
\be
\label{eq:relpar1}
\bar g^2=g^2\left(1-\frac{g^2}{24\pi \mu_D}\right),\qquad
\bar \lambda=\lambda-\frac{3c^2}{2\pi\mu_D},\qquad
\bar\mu^2=\mu^2-\frac{3c\mu_D}{2\pi}.
\label{pertmap}
\ee   
In particular, we note that the $\bar\mu$ scale serves as the temperature 
scale of the fully reduced system, while $\mu$ is the scale for the
system containing both the fundamental and the adjoint scalars. The two are 
related perturbatively by a constant shift.

In order to perform the actual calculations in the broken phase it is 
necessary to shift 
the Higgs field, $\sigma \rightarrow v + \sigma'$. After this shift  
and the gauge-fixing (the gauge fixing parameter is denoted by
$\xi$) the Lagrangian including the ghost terms assumes the form
\bea\label{lexplit}
L &=& {1\over 4}\vec{F}_{\mu\nu}\vec{F}_{\mu\nu} + {1\over 2\xi}
(\partial_{\mu}\vec{W}_{\mu})^2 + {1\over 2}m_{T0}^2\vec{W}_{\mu}^2 \nn\\
&& + {1\over 2}(\partial_{\mu}\s')^2 + {1\over 2} M_0^2 \s'^2
   +{1\over 2}(\partial_{\mu}\vec{\pi})^2  
   +\xi{1\over 2}m_{T0}^2\vec{\pi}^2\nn\\
&& +{g^2\over 4}v\s'\vec{W}_{\mu}^2 + {g\over 2}\vec{W}_{\mu}\cdot(
    \vec{\pi}\partial_{\mu}\s'-\s'\partial_{\mu}\vec{\pi})
   + {g\over 2}(\vec{W}_{\mu}\times\vec{\pi})\cdot\partial_{\mu}\vec{\pi}\nn\\
&& + {g^2\over 8}\vec{W}_{\mu}^2(\s'^2+\vec{\pi}^2)
   + \lambda v\s'(\s'^2+\vec{\pi}^2)+{\lambda\over 4}(\s'^2+\vec{\pi}^2)^2\nn\\
&& + {1\over 2 } {(D_i \vec{A_0})}^2+{1\over 2} m_{D0}^2 {\vec{A_0}}^2
   + {\lambda_A\over 4} {({\vec{A_0}}^2)}^2+2 c v \s' {\vec{A_0}}^2
   + c {\vec{A_0}}^2 ({\s'}^2+{\vec{\pi}}^2)\nn\\
&& + \partial_{\mu}\vec{c^*}\partial_{\mu}\vec{c}
   + \xi m_{T0}^2 \vec{c^*}\vec{c}\nn\\
&& + g\partial_{\mu}\vec{c^*}\cdot(\vec{W}_{\mu}\times\vec{c})
   + \xi{g^2\over 4}v\s'\vec{c^*}\vec{c}
   + \xi{g^2\over 4}v\vec{c^*}\cdot(\vec{\pi}\times\vec{c})
   + {1\over 2}\mu^2 v^2 + {1\over 4}\lambda v^4\nn\\
&& + {1\over 2}(\mu^2+\lambda v^2)(\s'^2+\vec{\pi}^2)
   + v(\mu^2 +\lambda v^2)\s'\ ,
\label{shiftL}
\eea
where the following notations were introduced for the tree-level masses: 
$m_{T0}^2={1\over4} g^2 v^2$ (the vector boson mass), $M_0^2=\mu^2+3 
\lambda v^2$ (the Higgs mass) and $m_{D0}^2=\mu_D^2+2 c v^2$ (the Debye mass). 
The last two terms of (\ref{shiftL}) 
arise from the Higgs potential after the shift in the Higgs 
field $\s$. For $\mu^2 < 0$, they vanish if one expands around the classical 
minimum $v^2 = -\mu^2/\lambda$. These terms have to be 
kept \cite{buchmuller95}, when the equation for the vacuum
expectation value is to be deduced.

In order to obtain the coupled gap equations one replaces the tree-level masses
by the exact masses
\be
m_{T0}^2\rightarrow m_T^2+\delta m_T^2,~~~~~ M_0^2 \rightarrow
M^2+\delta M^2,~~~~~m_{D0}^2\rightarrow m_D^2+\delta m_D^2,
\ee
and treats the differences 
$\delta m_T^2=m_{T0}^2-m_T^2,~~\delta M^2=M_0^2-M^2,
~~\delta m_D^2=m_{D0}^2-m_D^2$
as counterterms. The exact Goldstone and ghost masses are both equal to
$\sqrt{\xi} m_T$, where $m_T$ is the exact gauge boson mass.
The gauge invariance of the self-energies of the Higgs and gauge
bosons is ensured by introducing appropriate vertex resummations. 
Their explicit formulae can be found in \cite{buchmuller95}. In the present 
extended model, a resummation of the Higgs-$A_0$
vertex would be also necessary if the gauge invariance of the $A_0$
self-energy is to be ensured. Then the only source of the gauge 
dependence which would remain is the equation for the vacuum expectation 
value $v$.

All these resummations are equivalent to work with 
the following gauge invariant Lagrangian: 
\bea
L^{3D}_I & = & \frac{1}{4}\vec{F}_{ij}\vec{F}_{ij}+
Tr\left((D_i\Phi)^+D_i\Phi-\frac{1}{2}M^2\Phi^+\Phi\right)+
\frac{1}{2}(m_D^2-\frac{8cm_T^2}{g^2})\vec{A}_0^2\nonumber\\
&&
+\frac{g^2M^2}{4m_T^2}Tr(\Phi^+\Phi)^2+
2c\vec{A}_0^2Tr\Phi^+\Phi.
\eea
In this Lagrangian one shifts the Higgs field around its classical 
minimum $\s\rightarrow \s' + {2 m_T \over g}$ and adds the corresponding gauge 
fixing and ghost terms \cite{buchmuller95}. Shortly, we shall argue that the 
$A_0$-Higgs vertex resummation arising from the replacement of $v$ by 
$2m_T/g$ when the scalar field is shifted in the last term of the above 
Lagrangian destroys the mass-hiearchy between the heavy 
$A_0$ and the light gauge and Higgs fields. Therefore we have to 
give up the full gauge independence of the resummation scheme. 
The numerical solution to be presented below shows that the gauge dependence
of the $A_0-\Phi$ vertex
in our resummation scheme introduces only a minor additional gauge dependence 
beyond that of the equation for the
vacuum expectation value \cite{buchmuller95} appearing below 
in Eq. (\ref{eq:gapeq4}). 

The coupled set of gap equations is constructed from that of Ref.
\cite{buchmuller95} by adding the contributions due to the presence of the 
adjoint Higgs field. The self-energy contributions 
for the 3d adjoint Higgs model were already calculated in section 3.
Below we list only the additional contributions to the 
self-energies, which all contain at least one $A_0-\Phi$ vertex (the
corresponding diagrams are listed in the Appendix B). Let us emphasize
once again that no resummation of the $A_0-\Phi$ vertex was applied.

The additional contribution to the self-energy of the $A_0$ 
field coming from Higgs, Goldstone, gauge and ghost fields (diagrams {\bf a-i}) 
is
\bea
\delta\Pi_{A_0}^{H,G,gh} & = & -{4 c v^2(\mu^2+\lambda v^2)\over M^2}+
{3 c g v\over \pi} \left( {M\over 4 m_T}+{m_T^2\over M^2} \right)-{c M
\over 2 \pi}
+{3 \sqrt{\xi} \over 4 \pi} (g v-2 m_T)\nonumber\\
&&
+{4 c^2 v^2\over \pi}\left[
{3\over 2} {m_D\over M^2}-
\frac{1}{p}\arctan\frac{p}{m_D+M}\right].
\eea
There is also an additional contribution  
to the gauge boson self-energy  coming from the adjoint Higgs field 
(diagram {\bf m}):
\be
\delta\Pi_T^H(p,m_T,M,m_D)=\frac{3 c g}{2 \pi}\frac{m_T v}{M^2}m_D.
\ee
The contribution of $\vec{A}_0$ to the Higgs self-energy
(diagrams {\bf j-l}) is the following:
\bea
\delta\Pi_H^{A_0}(p,m_T,m_D)=-\frac{3 m_D
c}{2 \pi}-\frac{6 c^2 v^2 }{\pi}
\frac{1}{p}\arctan\frac{p}{2m_D}+{9 gcv \over 4\pi m_T} m_D.
\eea

Making use also of the pieces of the self-energies calculated
in section 3 and also in \cite{buchmuller95} we write
down a set of coupled on-shell gap equations for the screening masses of the
magnetic gauge bosons, fundamental Higgs and adjoint $A_0$ fields in the 
form:
\bea
\label{eq:gapeq2}
m_T^2&=&\Pi_T(p^2=-m_T^2,m,M)+\delta\Pi_T^{A_0}(p^2=-m_T^2,m_D)\nonumber\\
&&
+\delta\Pi_T^{H}
(p^2=-m_T^2,m_T,M,m_D)\\
\label{eq:gapeq3}
M^2&=&\Sigma(p^2=-M^2,m_T,M)+\delta\Pi_H^{A_0}(p^2=-M^2,m_T,m_D),\\
\label{eq:gapeq1}
m_D^2&=&\Pi_{00}(p^2=-m_D^2,m_T,m_D)+\delta\Pi_{A_0}^{H,G,gh}(p^2=-m_D^2,
m_T,M,m_D),
\eea
where $\Pi_T$ and $\Sigma$ are defined by  Eqs. (17),~(18) of Ref. 
\cite{buchmuller95}. $\delta \Pi_T^{A_0}$ and $\Pi_{00}$ are defined by Eqs. 
(\ref{dpi_a0}) and (\ref{reb}).

If on the right hand side of the third equation one inserts the tree
level masses, the next-to-leading order result of Ref.\cite{rebhan94h} 
is recovered
for the Debye mass in the SU(2) Higgs model. 

\begin{figure}[t]
\epsfysize=7cm
\epsfxsize=10cm
\centerline{\epsffile{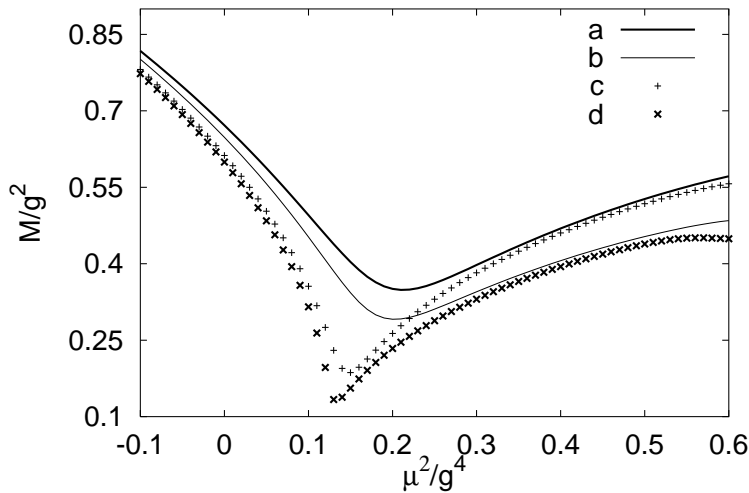}}
\caption{\small
The Higgs mass in units of $g^2$ as function of $\mu^2/g^4$ calculated at
$\lambda/g^2=1/8$ using the gauge-invariant
$A_0-\Phi$ vertex resummation version of the gap equations. Shown are the
Higgs mass derived in the full static theory in the Landau- (a) and in the
Feynman-gauge (b) and the Higgs mass, and in the $A_0$-reduced theory in the
Landau-gauge (c) and in the Feynman-gauge (d). The $\mu^2$-shift indicated
by  eq.(\ref{pertmap}) was applied.}
\label{Mbpinvar}
\end{figure}

The equation for the vacuum expectation value completes the set of the above 
three equations:
\be
v(\mu^2+\lambda v^2)=\frac{3}{16\pi}g\left(4m_T^2+
\sqrt{\xi} M^2+\frac{M^3}{m_T}
\right)+\frac{3c}{2 \pi}v m_D.
\label{eq:gapeq4}
\ee
It is important to notice that this equation can be rewritten as
\be\label{vac2}
v(\mu_{eff}^2+\lambda v^2)=\frac{3}{16\pi}g\left(4 m_T^2+\sqrt{\xi} 
M^2+\frac{M^3}{m_T}\right),
\ee
with 
\be
\mu^2_{eff}=\mu^2-\frac{3c}{2 \pi}m_D.
\label{muscale}
\ee 
This equation is formally identical to the equation of BP for the vacuum
expectation value \cite{buchmuller95}. On the basis of this observation, 
we expect that the main effect of the $A_0$ integration is the above shift 
in the $\mu^2$-scale. Since $m_D$ is itself a non-trivial function of $\mu$
this non-perturbative mapping is also nonlinear.

A very similar set of equations could be derived for the
case of the gauge invariant resummation of the $A_0-\Phi$ vertex. 
For instance, one would write in the last term on the right hand side
of (\ref{eq:gapeq4}) $2m_T/g$ on the place of $v$, which would
prevent the absorption of this term into a redefinition of the temperature
scale. This was the reason when we solved the gap equations first for 
the case of gauge invariant $A_0-\Phi$ vertex resummation, that we have found
substantial corrections to the BP results, beyond the shift of the 
$\mu^2$-scale indicated in eq. (\ref{pertmap}). This situation is illustrated
in Figure \ref{Mbpinvar}.

This makes it clear that the decoupling of the $A_0$-field for the physical
value of the parameter $c$ would not work, the hierarchy of the screening
masses would be destroyed under such vertex resummation scheme. 
Since our goal is to propose such
a mapping onto the fundamental Higgs-model, which already for finite
$A_0$-mass shows evidence
for the asymptotic Appelquist-Carazzone theorem \cite{appelquist75}, we have
renounced from the gauge invariant resummation of the $A_0-\Phi$ vertex.

\subsection{Numerical Results}

The main goal of the present investigation is to analyze non-perturbatively
in quantitative terms the decoupling of the $A_0$ field. We are interested
in those effects which go beyond the perturbative mapping implied by 
eq.(\ref{pertmap}). Therefore,
we will first compare the predictions for the Higgs and gauge boson masses
from the coupled gap equations  (\ref{eq:gapeq2})-(\ref{eq:gapeq4}) 
of the 3d fundamental + adjoint Higgs model with
those obtained in the $A_0$-reduced theory, the 3d Higgs model
\cite{buchmuller95}. The
corresponding Higgs masses are shown in Figure \ref{Mbpnoninvar}
using two different gauges. 
The results obtained in the $A_0$-reduced theory are displayed after the shift
required by eq.(\ref{pertmap}) is performed.
The difference between the full and the reduced theory is still
visible in the vicinity of the crossover. In this region the relative 
difference between the prediction of the full and the reduced theory
is about $20 \%$. 
\begin{figure}
\epsfysize=7cm
\epsfxsize=10cm
\centerline{\epsffile{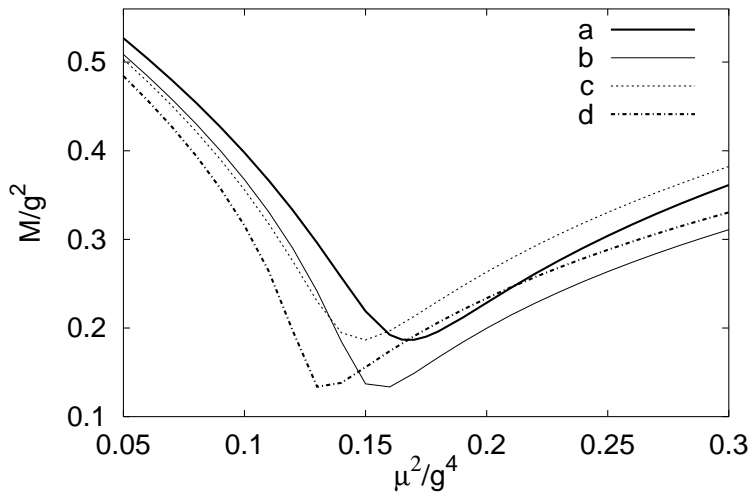}}
\caption{\small The Higgs boson masses at $\lambda/g^2=1/8$ (crossover region) 
in units of $g^2$ as function of $\mu^2/g^4$ in the 3d fundamental + adjoint 
Higgs model and in the 3d SU(2) Higgs ($A_0$-reduced) theory. Shown are the 
Higgs mass in the full static theory in the $\xi=0$ (Landau) gauge (a) and in 
the $\xi=1$ (Feynman) gauge (b), and the Higgs boson mass in the $A_0$-reduced
theory in the $\xi=0$ gauge (c) and in the the $\xi=1$ gauge (d).}
\label{Mbpnoninvar}
\end{figure}

Our proposal to resolve this relatively 
large deviation is to introduce a more complicated relationship between
the couplings. Having gained intuition from eq.(\ref{muscale}), 
we have plotted the mass-predictions for the Higgs-field
derived from our full set of equations against the results of BP calculated
for couplings taken from (\ref{pertmap}) with a replacement $\mu_D\rightarrow
m_D$:
\be 
g_{eff}^2=g^2(1-{g^2\over 24\pi m_D}),\qquad
\lambda_{eff}=\lambda-{3c^2\over 2\pi m_D},\qquad 
\mu_{eff}^2=\mu^2-{3c\over 2\pi}m_D.
\label{nonpertmap}
\ee  
The non-trivial nature of this replacement becomes clear from Figure
\ref{mdbp}
where the $\mu^2$-dependence of $m_D$ is displayed. Clearly, its
non-trivial $\mu^2$-dependence is most expressed in the neighbourhood 
of the phase transformation (crossover) point $\mu^2/g^4 \in (0.1-0.2)$. 
The application of this
mapping to the data obtained from the model containing both the fundamental 
and the adjoint representation leads to a perfect agreement of the two data 
sets for large values of $\lambda /g^2$. For smaller values of $\lambda /g^2$ 
(1/32,1/64) the mapping (\ref{nonpertmap}) works very well in the symmetric 
phase, but in the broken phase (\ref{pertmap}) seems to be the better choice.
\begin{figure}
\epsfxsize=10cm
\epsfysize=7cm
\centerline{\epsffile{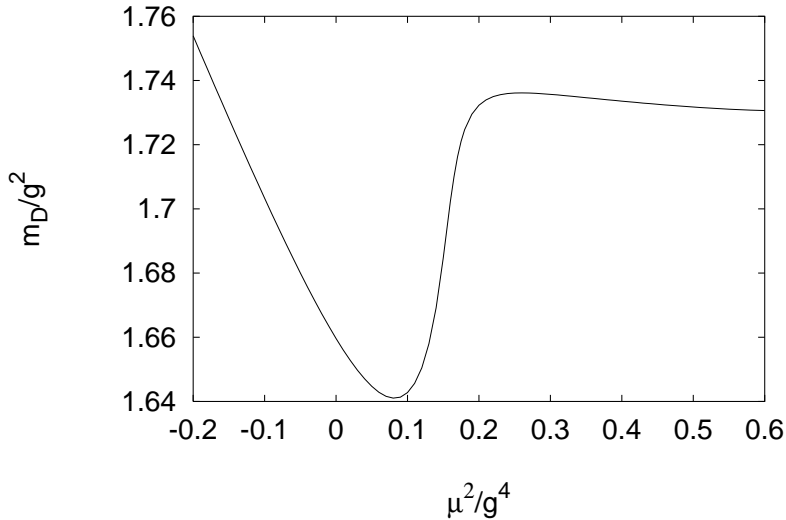}}
\caption{\small
The $\mu^2$ dependence of the Debye mass for
$\frac{\lambda}{g^2} =1/8$.}
\label{mdbp}
\end{figure}

We suspect, that the tree level piece in $m_D$ arising from the Higgs-effect,
should not be included into the correction of (\ref{pertmap}), since it is 
itself a tree-level effect. Therefore we propose the following replacement in
(\ref{nonpertmap}):
\be
m_D\rightarrow \sqrt{m_D^2-2cv^2}.
\label{nonpertmap_new}
\ee
In Figure \ref{Mbpnonpertmap} it is obvious that a very good 
agreement could be obtained with this
mapping between the Higgs mass predictions of the one-loop
gap equations of the full static and the $A_0$-reduced theory for $\lambda /g^2
=1/32$. The quality of the agreement on both sides of the phase transition
is good, signalling that the influence of the ``mini-Higgs'' effect in the 
symmetric phase is negligible. Therefore it is not surprising that for
$\lambda /g^2=1/8$ the same quality of agreement is obtained like before.

It is important to notice that there is a strong
gauge parameter dependence in the symmetric phase and in the vicinity
of the crossover. The variations due to the change in the gauge 
are equal  in the full and in the 
reduced theory, which indicates that the additional gauge dependence,
introduced by the gauge non-invariant resummation of the $A_0$ field is
negligible. The mapping (\ref{nonpertmap_new}) performs equally well
in Landau- and Feynman-gauges.

\begin{figure}
\epsfxsize=10cm
\epsfysize=7cm
\centerline{\epsffile{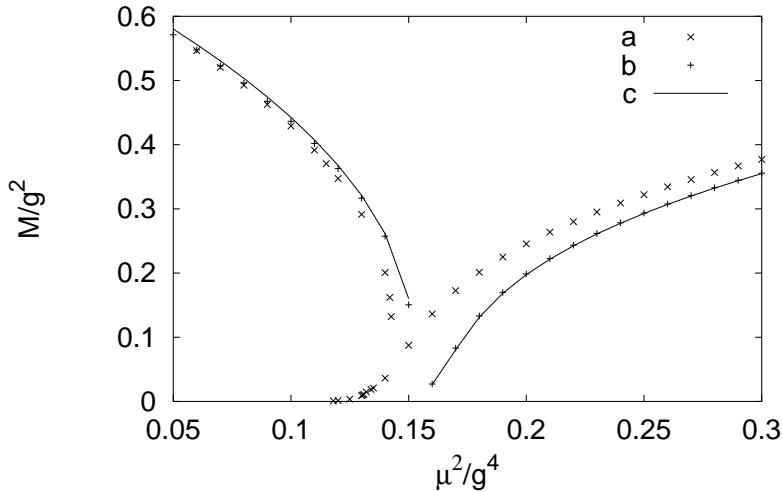}}
\caption{\small
The Higgs boson mass in units of $g^2$ as function of $\mu^2/g^4$
calculated at $\lambda/g^2=1/32$ in the Landau gauge 
in the full static theory and in the
$A_0$-reduced theory. Shown are the Higgs mass in the reduced theory obtained
by perturbative reduction (a), in the reduced theory obtained by 
non-perurbative
matching (cf. eqs. (\ref{nonpertmap}), (\ref{nonpertmap_new})) (b)  and
in full static theory (c).}
\label{Mbpnonpertmap}
\end{figure}

Other quantities which are worth of considering for the comparison 
of the full 3d and the reduced theories are $\lambda_c/g^2$, the endpoint of 
the first order transition line and $\mu_{+}/g^2$, the mass parameter above 
which the broken phase is no longer metastable. 
The  values of $\mu_{+}^2/g^4$ for different scalar couplings and different
gauges in the full and in the reduced theory are summarized in Table 1.
Here the mapping (\ref{nonpertmap}) could be implemented only
by extrapolating from smaller $\mu^2/g^4$, since the
end-points of metastability do not correspond to each other, and in some
cases $m_D$ could not be determined from the gap equations. Also here
for larger values of $\lambda/g^2$ the application of (\ref{nonpertmap}) led
to an improved agreement between the end-point $\mu_+^2/g^4$ values, while for 
$\lambda /g^2 =1/48,1/64$ the mapping (\ref{pertmap}) works better. 
In the table
we have displayed $\mu_+^2/g^4$ values of the $A_0$-reduced theory shifted 
perturbatively and with help of the best performing non-perturbative
mapping (\ref{nonpertmap_new}). For both gauges the latter agrees with the
$\mu_+^2/g^4$-values of the full static theory very well.
\begin{center}
\begin{tabular}{|l|l|l|l|}
\hline
$~~~~\lambda/g^2~~~$  &$~~~~~~~~~~~~~A~~~~~~~~~~$
& $~~~~~~~~~~~~~B~~~~~~~~~~$ & $~~~~~~~~~~~~~C~~~~~~~~~~$\\
\cline{2-4}
$~~~~~~~~~~~~~~$ & $~~~\xi=0~~~$ \vline $~~~\xi=1~~~$
& $~~~\xi=0~~~$ \vline $~~~\xi=1~~~$ 
& $~~~\xi=0~~~~$\vline $~~~~\xi=1~~~$
\\
\hline
$~~~1/32~~~$  &$~~~0.1516~~~~~~~0.1423~~~$
&$~~~0.1426~~~~~~0.1341~~~$
&$~~~0.1499~~~~~~~0.1405~~~$\\
$~~~1/48~~~$  &$~~~0.1647~~~~~~~0.1558~~~$
&$~~~0.1627~~~~~~0.1541~~~$
&$~~~0.1637~~~~~~~0.1546~~~$\\
$~~~1/64~~~$  &$~~~0.1841~~~~~~~0.1750~~~$
&$~~~0.1881~~~~~~0.1808~~~$
&$~~~0.1875~~~~~~~0.1792~~~$\\
\hline
\end{tabular}
\vskip0.3truecm
Table.~1: {{\small Values of $\mu_{+}^{2}/g^4$ in the full static theory (A), 
in the perturbatively reduced theory (B) and in the reduced theory
obtained using non-perturbative matching described in the text (C). 
Calculations were done 
in the Landau ($\xi =0$) and in the Feynman ($\xi=1$) gauges.}}
\end{center}

The endpoint of the $1^{st}$ order line in the Landau gauge in the 
3d Higgs theory was found at $\lambda_c/g^2=0.058$. The
corresponding critical scalar coupling in the full 3d theory is within
the 1\% range. In Feynman gauge we find $\lambda_c/g^2=0.078$ for the
$A_0$-reduced theory and the
corresponding value for the full 3d theory lies again very close to it. 
Thus the $A_0$ field has almost no effect on the position of the endpoint.
The strong gauge dependence of $\lambda_c$ indicates, however, that
higher order corrections to this quantity are important.

The gauge dependence of the screening masses is even more pronounced deep 
in the symmetric phase
($\mu^2/g^4>0.3$). For example the value of the gauge boson mass is roughly
$0.28 g^2$ in the symmetric phase for the Landau gauge. The
corresponding value in the Feynman gauge is about $0.22 g^2$. The gauge
dependence of the gauge boson mass is somewhat weaker at the 2-loop
level \cite{eberlein98}. It should be also noticed that the gauge boson
mass depends weakly on the parameters of the scalar sector
($\mu,~\mu_D,~\lambda,~\lambda_A$). This fact was also noticed in
previous investigations \cite{buchmuller95,patkos98k}.

\subsection{Screening Masses in the Symmetric Phase with 
a Gauge Invariant Resummation Scheme}

The main motivation for the present investigation was to gain  insight
into the decoupling of the dynamics of the fundamental and 
the adjoint Higgs fields. The degree of the decoupling is 
expected to depend on the mass ratio of the fundamental
and adjoint Higgs fields. In the symmetric phase both
masses are of the same order in magnitude (eg. $\sim gT$). 
Therefore the hierachy of the $A_0$ and Higgs masses 
can only be present  due to numerical prefactors. The persistence 
of the perturbatively calculated ratio should be checked in any 
non-perturbative approach.

As we have seen in the previous section the gauge dependence in the
symmetric phase is too strong in the applied schemes to give a stable 
estimate for the mass ratio of the fundamental and the adjoint Higgs fields.
A reliable non-perturbative estimate for the Higgs mass deep in the symmetric
phase (defined through the
pole of the propagator) is even more interesting because it was not
measured so far on lattice. Therefore, in this section
we will investigate a coupled set of gap equations in the symmetric phase
which is based on the gauge invariant resummation scheme of 
Alexanian and Nair (AN) \cite{alexanian95}. 
In this approach one can avoid any vacuum 
expectation value for the Higgs field in
the symmetric phase and because of this fact this approach is gauge
invariant.

To derive the gap equations in this scheme one has to add and substract
the mass generating term defined by Eq. (\ref{dlan}) and
also add the corresponding gauge fixing term defined by Eq.
(\ref{dlangf}). Among diagrams shown in Appendix B those which
contain gauge boson propagators should be reevaluated.
Straightforward calculations lead to the following equations:
\bea
&&
m_T^2=C g^2 m_T + {g^2 m_T\over 4 \pi} \biggl(2 f(m_D/m_T)+ 
f(M/m_T)\biggr),\nonumber\\
&&
M^2=\mu^2+{1\over 4 \pi} \biggl({3\over 4} g^2 M F(M/m_T)-6 \lambda M-6 c m_D
\biggr),\nonumber\\
&&
m_D^2=\mu_D^2+{1\over 4 \pi} \biggl( 2 g^2 m_D F(m_D/m_T)-5 \lambda_A m_D-
8 c M\biggr),
\label{aneq}
\eea
where $C={1\over 4\pi} (21/4 \ln3-1)$ \cite{alexanian95} and the
following functions were introduced
\bea
&&
f(z)=-{1\over 2} z+\bigl(z^2-{1\over 4}\bigr) {\rm arctanh}{1\over 2 z},\\
&&
F(z)=-1-{1\over z}+\bigl( 4 z - {1\over z} \bigr) \ln(1+ 2 z).
\eea

\begin{figure}
\epsfxsize=10cm
\epsfysize=7cm
\centerline{\epsffile{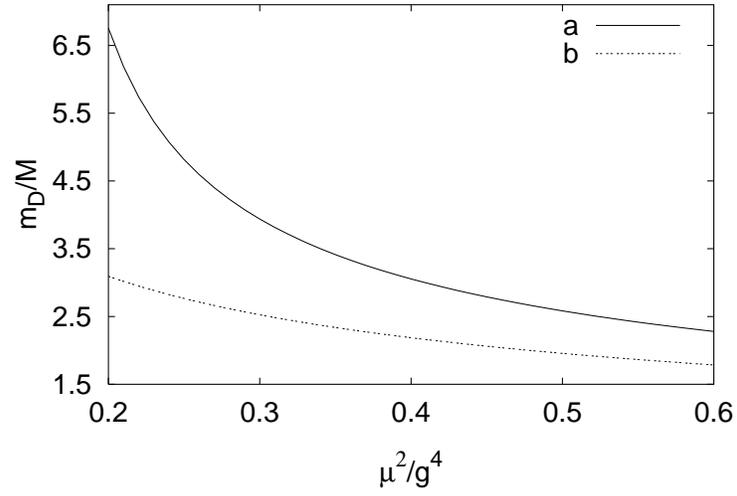}}
\caption{\small
The ratio of the Debye and the fundamental Higgs masses for
${\lambda\over g^2}=1/8$ calculated from gap eqs. (\ref{aneq}) (a) and
the leading order result (b).}
\label{mdpMan}
\end{figure}

\begin{figure}
\epsfxsize=10cm
\epsfysize=7cm
\centerline{\epsffile{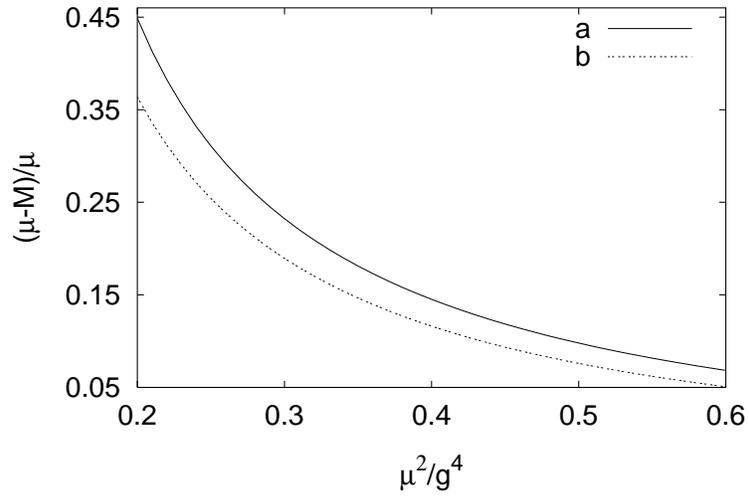}}
\caption{\small
The non-perturbative correction to the Higgs mass as
the function of $\mu$ calculated from the full static (a)
and from the $A_0$-reduced theory (b).}
\label{ManNP}
\end{figure}

Let us first discuss the ratio of the $A_0$ and the fundamental Higgs
masses. In Figure \ref{mdpMan} this ratio is shown as calculated from the eqs.
(\ref{aneq}) and compared with the corresponding perturbative value.
The $\mu$ interval in this plot corresponds to the temperature range 
relavant for the 
electroweak theory $T<1 TeV$.
We have also analyzed the $\mu$-dependence of the fundamental
Higgs mass alone in the full static and in the $A_0$-reduced model.
For $\mu^2/g^4$ in the interval $(0.2-0.3)$ the result of the gauge invariant
approach agrees fairly well with the masses obtained in the BP-scheme.
In Figure \ref{ManNP} the difference between the Higgs masses calculated 
from the coupled set of gap equations (\ref{aneq}) and the leading
order perturbative result ($M_0$) is shown. As one can see, 
the non-perturbative correction to the Higgs mass is the largest for
small $\mu$ and is decreasing as $\mu$ increases reaching the percent level
for large enough $\mu$.

The relative difference between the full static and the $A_0$-reduced theory,
however, is slowly increasing as $\mu$ increases and the hierarchy between 
the $A_0$ and the Higgs masses becomes less pronounced as $\mu$
gets larger.
The relative difference between the Higgs masses calculated in the 
full static and in the $A_0$-reduced theory varies between $20\%$ for 
$\mu^2/g^4=0.2$ and $35\%$ for
$\mu^2/g^4=0.6$.

It is also important to notice that the $A_0$ field is 
not sensitive to the dynamics of the Higgs field. In particular it
turns out that $m_D$ depends weakly on $\mu$ and $\lambda$ in the
symmetric phase and its value is close to the corresponding value 
calculated in 3d adjoint Higgs model. Let us notice that the magnetic
mass in this resummation scheme also seems to be insensitive to the
dynamics of scalars, therefore the magnetic and electric screening masses
are close to their values determined in the pure SU(2) gauge model.

\newpage
\section{Conclusions}

In my Thesis I have studied the problem of screening in hot 
non-Abelian plasma. A detailed review of the  phenomenon of chromoelectric
screening was given. It was shown that the chromoelectric screening
is a well understood phenonenon in the leading order of perturbation
theory and it is quite similar to the well known Debye screening
phenomenon in non-relativistic electron plasma. Beyond the leading
order, however, it turns out to be very complicated and at present
it is even not fully clear how to define it properly. In my
numerical investigations
Debye screening phenomenon was defined through the long distance
(small momentum) behaviour of the static longitudinal propagator. 
It has been shown that such a definition is 
free of infrared singularities in higher orders of perturbation 
theory
only if one
introduces the so-called magnetic screening mass, a concept which
was introduced long ago by Linde to cure infrared divergencies of
finite temperature gauge theories \cite{linde80}. 

In recent years the problem of magnetic mass was actively studied
using self-consistent resummation of perturbative series.
The introduction of
the magnetic mass allows a self-consistent 
determination of the Debye screening mass (the inverse
screening length) through coupled gap equations, 
which was discussed in section 3. It was argued
that such determination so far is only possible in the framework of the 
dimensionally reduced effective theory. Thus one has to address the 
question how far the dimensionally reduced effective theory faithfully
represents the high temperature limit of gauge theories. This question
is far from being trivial since the separation of different mass scales
$2 \pi T \gg gT \gg g^2 T$ 
which is necessary for dimensional reduction is not satisfied for
any realistic temperature range 
in the theory which corresponds to quantum chromodynamics
because the gauge coupling constant $g$
is large. This observation suggested that the electric (Debye) and the
magnetic screening masses should be determined simultaneously from the
coupled set of gap equation. In section 3 where the screening masses
were studied through coupled gap equation the precision of dimensional
reduction was tested also by examining the effect of non-local operators and
it turned out that their contribution is small thus justifying the 
approach based on dimensional reduction. 

In section 4 the screening
masses were studied using lattice Monte-Carlo technique in the 3d
effective theory corresponding to the high temperature
phase of SU(2) gauge theory, the 3d adjoint Higgs model. Comparison of the
screening masses obatined in these simulations with the corresponding
masses obtained recently in the full 4d finite temperature theory
\cite{heller98} allows to test non-perturbatively and quantitatively
the precision of the dimensionally reduced theory. It was shown that
the predictions of the 3d effective theory are quite reasonable even
at temperature $\sim 4 T_c$ and become more and more precise as the
temperature is increased. This fact gives prospects for
studying the screening phenomena in the effective theory approach. 
The predictions of the gap equation approach were compared with the
corresponding results. It turns out that the results obtained from
coupled gap equations describe correctly the temperature dependence
of the screening masses, however, so far there is no quantitaive
agreement between the predictions of the coupled gap equations and the
Monte-Carlo data. The Debye mass measured in lattice Monte-Carlo
simulations is about $25\%$ larger than what is predicted from the
coupled set of gap equations. 

In section 5 the screening masses of the finite temperature SU(2) Higgs
model were studied. Though the application of the dimensional reduction
approach here seems to be more justified (the corresponding 
gauge coupling constant is $g \sim 0.66$), the non-perturbative quantitative
test of the precision of the dimensional reduction is also important
here. The study of the screening masses in this model 
through coupled gap equations tests
the decoupling of the heavy $A_0$ field (temporal component of the
static gauge field). Detailed numerical analysis revealed high
precision decoupling of the $A_0$ field if 
in the mapping of the full static theory onto the $A_0$-reduced system
its mass
determined non-perturbatively is used.
This observation also implies that for the realistic
value of the $A_0$ mass perturbative integration over the heavy $A_0$
field has limited precision due to the coupling of the heavy $A_0$
field to the light magnetostatic sector. This coupling to the magnetostatic
sector is taken into account in the proposed non-perturbative mapping
procedure. It is very interesting that this coupling between the heavy
$A_0$ field and the light magnetostatic sector can described quite
precisely by the magnetic mass obtained from the gap equations.

From the above discussion it is clear that the concept of the magnetic
screening mass is extremely usefull since it allows higher
order definition of the
electric screening mass, which  quantitatively agrees with the
corresponding lattice predictions and also the high quality mapping
onto the effective theory mentioned above. The values of 
magnetic mass determined in different resummation schemes 
are quite close to each other. A recent 2-loop gap equation of the
magnetic mass has shown that the value of the magnetic mass does not
change dramatically 
relative to the 1-loop results 
and the corresponding value was found to be
$0.34g_3^2$ \cite{eberlein98} 
(which should be compared with the 1-loop predictions
$0.28g_3^2$ and $0.38g_3^2$). Moreover the non-perturbative analysis of
the mass gap in 3d pure gauge theory yields a very similar value
$0.32 g_3^2$ \cite{nair98}. All these facts give us confidence that the
magnetic mass is a viable concept. However, in Refs.
\cite{jackiw96,cornwall98} some inconsistencies in the 1-loop 
determination of the
magnetic mass were found. In the numerical simulations the value of the
magnetic mass was found to be $\sim 0.46 g_3^2$ which is
substantially larger than the predictions from the analytical approaches.
Thus I think, further studies are required to clarify
the phenomenon of non-Abelian magnetic mass generation.

\newpage
\section*{Appendices}
\appendix
\setcounter{equation}{0}
\renewcommand{\theequation}{\Alph{section}.\arabic{equation}}
\section{Calculation of Different Diagrams in SU(N) Adjoint
Higgs Model}

The propagators of different fields can be read off the quadratic part of
the Lagrangian. These exact propagators are listed below.
The gauge boson propagator:
\be
D_{ij}(k)=\left(\delta_{ij}-{k_i k_j\over k^2}\right){1\over k^2+m_T^2}+{k_i
k_j\over k^2}{\xi\over k^2+ m_L^2},
\ee
where $m_L=\sqrt{\xi} m_T$,
the propagator for the adjoint scalar field $A_0$:
\be
D^{A_0}(k)={1\over k^2+m_D^2},
\ee
and finally, the ghost propagator:
\be
\Sigma(k)={1\over k^2+m_G^2}.
\ee

The Feynman diagrams contributing to the 2-point functions of the relevant
fields are shown below. In these diagrams every line corresponds to a
resummed propagator. The wavy line corresponds to the gauge
bosons, the solid one to $A_0$ and the dashed one to the ghost.
The corresponding analytic expressions can be written in terms of the
following standard integrals:
\begin{eqnarray}
I(k,r,s)&=&\int\frac{d^d p}{(2\pi)^d}\frac{1}{(p+k)^{2r}p^{2s}}\nonumber\\
&=&\frac{k^{d-2(r+s)}}{(4\pi)^{d/2}}\frac{\Gamma\left(r+s-\frac{d}{2}\right)}
{\Gamma(r)\Gamma(s)}\,\frac{\Gamma\left(\frac{d}{2}-s\right)
\Gamma\left(\frac{d}{2}-r\right)}{\Gamma(d-s-r)},\\
I^{m}(k,r,s)&=&\int\frac{d^d p}{(2\pi)^d}\frac{p_{m}}{(p+k)^{2r}p^{2s}}
\nonumber\\
&=&-k^{m}\,\frac{k^{d-2(r+s)}}{(4\pi)^{d/2}}\frac{\Gamma\left(r+s-\frac{d}{2}
\right)}{\Gamma(r)\Gamma(s)}\,\frac{\Gamma\left(\frac{d}{2}+1-s\right)
\Gamma\left(\frac{d}{2}-r\right)}{\Gamma(d+1-s-r)},\\
I^{mn}(k,r,s)&=&\int\frac{d^d
p}{(2\pi)^d}\frac{p^{m}p^{n}}{(p+k)^{2r}p^{2s}}
\nonumber\\
&=&\frac{k^{d-2(r+s)}}{(4\pi)^{d/2}}\biggl[\frac{k^2}{2}\frac{\Gamma\left(r+s-1-
\frac{d}{2}\right)}{\Gamma(r)\Gamma(s)}\,\frac{\Gamma\left(\frac{d}{2}+1-s\right)
\Gamma\left(\frac{d}{2}+1-r\right)}{\Gamma(d+2-s-r)}\enspace\delta^{mn}
\nonumber\\
&&\qquad\qquad+\frac{\Gamma\left(r+s-\frac{d}{2}\right)}{\Gamma(r)\Gamma(s)}\,
\frac{\Gamma\left(\frac{d}{2}+2-s\right)\Gamma\left(\frac{d}{2}-r\right)}
{\Gamma(d+2-s-r)}\enspace k^{m}k^{n}
\biggr],\\
J(k,m_1,m_2)&=&\int\frac{d^3 p}{(2\pi)^3}\frac{1}{\left((p+k)^2+m_{1}^{2}
\right)\left(p^2+m_{2}^{2}\right)}=\frac{A}{8\pi},\\
J^{m}(k,m_1,m_2)&=&\int\frac{d^3
p}{(2\pi)^3}\frac{p_{m}}{\left((p+k)^2+m_{1}^{2}
\right)\left(p^2+m_{2}^{2}\right)}\nonumber\\
&&=\frac{k^{m}}{8\pi}\biggl[\frac{m_{1}-m_{2}}{k^2}-\frac{k^2+m_{1}^{2}
-m_{2}^{2}}{2k^2}\,A\biggr],\\
J^{mn}(k,m_1,m_2)&=&\int\frac{d^3 p}{(2\pi)^3}\frac{p^{m}p^{n}}{\left((p+k)^2+
m_{1}^{2}\right)\left(p^2+m_{2}^{2}\right)}\\
&=&-\frac{\delta^{mn}}{8\pi}\left[\frac{m_1}{2}+\frac{4k^2m_{2}^{2}+
\left(k^2+m_{1}^{2}-m_{2}^{2}\right)^2}{8k^2}\,A\right]\nonumber\\
&&-\frac{k^{m}k^{n}}{8\pi}\left[\frac{m_1}{2k^2}-\frac{4k^2m_{2}^{2}
+3\left(k^2+m_{1}^{2}-m_{2}^{2}\right)^2}{8k^4}\,A\right]\nonumber\\
&&+\frac{1}{8\pi}\frac{(m_{1}-m_2)\left(k^2+m_{1}^{2}-m_{2}^{2}\right)}{4k^2}
\left[\delta^{mn}-3\,\frac{k^{m}k^{n}}{k^2}\right],
\eea
where $A=\frac{2}{k}\arctan\left(\frac{k}{m_1+m_2}\right)$
\bea
j(m)&=&\int\frac{d^3 p}{(2\pi)^3}\frac{1}{p^2+m^2}=-\frac{m}{4\pi},\\
l^{mn}(m)&=&\int\frac{d^3 p}{(2\pi)^3}\frac{p_{m}p_{n}}{p^2+m^2}=
-\frac{\delta^{mn}}{3}m^2j(m).
\end{eqnarray}
These integrals were evaluated using dimensional regularization. 

The diagrams contributing to the 2-point functions of $A_i$ are the
following:

\vskip0.15 truecm
\hspace{2.4cm}
\epsfbox{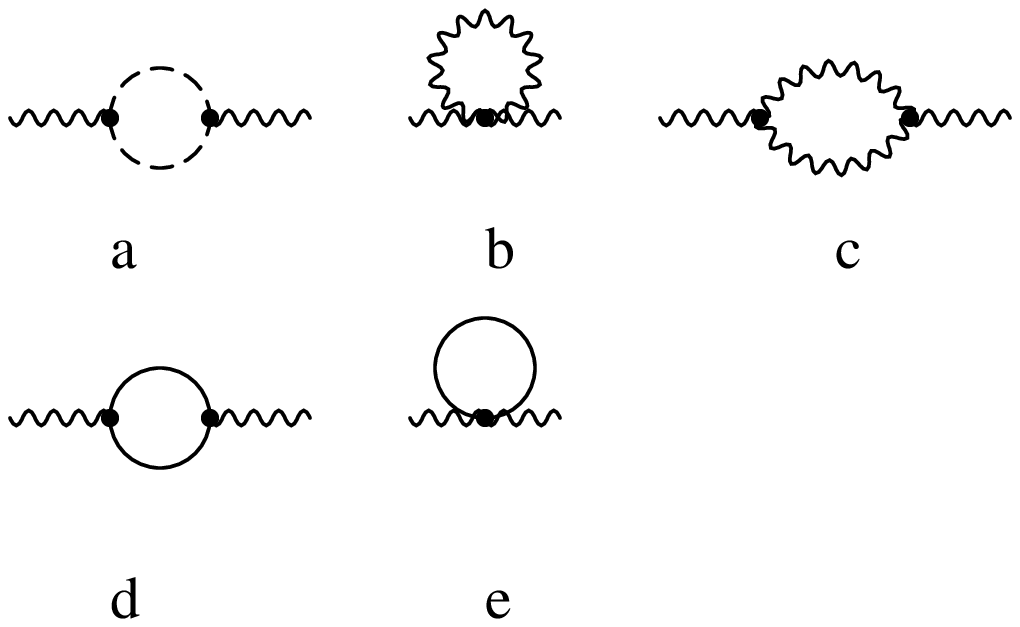}
\vskip0.1truecm
with the analytical contribution:

\begin{eqnarray}
{\Pi^{(a)}}^{mn}_{ab}(k)&=&g^2N\delta_{ab}\biggl[J_{m n}(k,m_G,m_G)
+k_{m}J_{n}(k,m_G,m_G)\biggr],\\
{\Pi^{(b)}}^{mn}_{ab}&=&g^2N\delta_{ab}\delta^{mn}\biggl[\frac{4}{3}j(m_T)+
\frac{2}{3}\xi j(m_L)\biggr],\\
{\Pi^{(c)}}^{mn}_{ab}(k)&=&-\frac{1}{2}g^2N\delta_{ab}
\biggl\{
\delta_{mn}\biggl[-\frac{\left(m^{2}_{T}+k^2\right)^2}{m_{T}^{2}}
\left(J(k,m_T,0)+J(k,0,m_T)\right)\nonumber\\
&&\quad+\frac{2k^2\left(4m^{2}_{T}+k^2\right)}{m^{2}_{T}}J(k,m_T,m_T)
-2\,\left(1+\frac{k^2}{m^{2}_{T}}\right)j(m_{T})\biggr]\nonumber\\
&&\quad+k_{m}k_{n}\biggl[-\frac{\left(m^{2}_{T}+k^2\right)^2}
{4m_{T}^{4}}J(k,m_{T},0)
+\left(\frac{k^4}{4m^{4}_{T}}-\frac{k^2}{m^{2}_{T}}-6\right)J(k,m_T,m_T)
\nonumber\\
&&\quad+\left(-\frac{k^4}{4m^{4}_{T}}+\frac{3}{2}\,\frac{k^2}{m^{2}_{T}}+
\frac{7}{4}
\right)J(k,0,m_T)+\frac{3}{2m^{2}_{T}}j(m_T)+\frac{k^4}{4m^{4}_{T}}I(k,1,1)
\biggl]\nonumber\\
&&\quad+\biggl[-\frac{k_{m}k_{n}}{m^{2}_{T}}j(m_T)+\frac{k^4}{4m^{4}_{T}}
I_{mn}(k,1,1)+\left(\frac{k^4}{4m^{4}}+4\frac{k^2}{m^{2}_{T}}+8\right)
J_{mn}(k,m_T,m_T)\nonumber\\
&&\quad-\frac{2}{m_{T}^{2}}l_{mn}(m_T)-\left(1+\frac{k^2}{m^{2}_{T}}\right)^2
\left(J_{mn}(k,0,m_T)+J_{mn}(k,m_T,0)\right)\biggr]\nonumber\\
&&\quad+\biggl[-\frac{k_{m}k_{n}}{m_{T}^{2}}j(m_T)+\frac{k^4}{m^{4}_{T}}
k_mI_{n}(k,1,1)-
\left(\frac{k^2}{m^{2}_{T}}+1\right)\left(\frac{k^2}{m^{2}_{T}}+3\right)k_m
J_{n}(k,m_T,0)\nonumber\\
&&\quad+\left(1-\frac{k^4}{m^{4}_{T}}\right)k_{m}J_{n}(k,0,m_T)+\left(
\frac{k^4}{m^{4}_{T}}+4\frac{k^2}{m^{2}_{T}}+8\right)k_{m}J_{n}(k,m_T,m_T)
\biggr]
\biggr\}\nonumber\\
&&-g^2N\delta_{ab}\xi\
\biggl\{
\delta_{mn}\biggl[\frac{\left(m_{T}^{2}+k^2\right)^2}{m^{2}_{L}}\left(
J(k,0,m_T)-J(k,m_L,m_T)\right)\nonumber\\
&&\qquad\qquad\qquad\qquad+\frac{\left(m_{T}^{2}+m_{L}^{2}+k^2\right)}
{m^{2}_{L}}j(m_L)
\biggr]\nonumber\\
&&\quad+\frac{k_{m}k_{n}}{4m_{T}^{2}m_{L}^{2}}\biggl[\left(6k^2m_{T}^{2}+
7m_{T}^{4}+
2m_{T}^{2}m_{L}^{2}-\left(m_{L}^{2}+k^2\right)^2\right)J(k,m_L,m_T)\nonumber\\
&&\qquad\qquad\qquad+m_{L}^{2}j(m_T)-7m_{T}^{2}j(m_L)+\left(k^4-6k^2m_{T}^{2}
-7m^{4}_{T}\right)J(k,0,m_T)\nonumber\\
&&\qquad\qquad\qquad-k^4I(k,1,1)+\left(m_{L}^{2}+k^2\right)^2J(k,m_L,0)\biggr]
\nonumber\\
&&\quad+\frac{1}{m_{L}^{2}m_{T}^{2}}\biggl[m_{T}^{2}k_{m}k_{n}j(m_L)
+k^4\left(J_{mn}(k,m_L,0)-I_{mn}(k,1,1)\right)\nonumber\\
&&\qquad\qquad\qquad+\left(k^2+m_{T}^{2}\right)^2\left(J_{mn}(k,0,m_T)
-J_{mn}(k,m_L,m_T)\right)+m_{T}^{2}l_{mn}(m_L)\biggr]\nonumber\\
&&\quad+\biggl[-\frac{\left(k^2+m_{T}^{2}\right)\left(k^2-m_{T}^{2}+
m_{L}^{2}\right)}
{m_{L}^{2}m_{T}^{2}}k_mJ_n(k,m_L,m_T)
-\frac{k^4}{m_{L}^{2}m_{T}^{2}}k_mI_n(k,1,1)\nonumber\\
&&\quad\,\,\,+\frac{k_mk_n}{m_{L}^{2}}j(m_L)+\frac{k^2}{m^{2}_{T}}
\left(\frac{k^2}{m^{2}_{L}}+1\right)k_mJ_n(k,m_L,0)
+\frac{k^4-m_{T}^{4}}{m_{T}^{2}m_{L}^{2}}k_mJ_n(k,0,m_T)\biggr]\biggr\}
\nonumber\\
&&-\frac{1}{2}N\delta_{ab}\xi^2
\biggl\{
\frac{k_mk_n}{4m_{L}^{4}}\biggl[-2m_{L}^{2}j(m_L)+k^4\left(I(k,1,1)+
J(k,m_L,m_L)
\right)\nonumber\\
&&\qquad\qquad\qquad-\left(k^2+m_{L}^{2}\right)^2J(k,m_L,0)-\left(k^2-m_{L}^{2}
\right)^2J(k,0,m_L)\biggr]\nonumber\\
&&\quad+\frac{k^4}{m_{L}^{4}}\biggl[I_{mn}(k,1,1)-J_{mn}(k,m_L,0)-
J_{mn}(k,0,m_L)
+J_{mn}(k,m_L,m_L)\biggr]\nonumber\\
&&\quad+\biggl[-\frac{k^2}{m^{2}_{L}}\left(\frac{k^2}{m^{2}_{L}}+1\right)
k_mJ_n(k,m_L,0)
-\frac{k^2}{m^{2}_{L}}\left(\frac{k^2}{m^{2}_{L}}-1\right)k_mJ_n(k,0,m_L)
\nonumber\\
&&\qquad\qquad\qquad\qquad+\frac{k^4}{m^{4}_{L}}\left(k_mJ_n(k,m_L,m_L)+
k_nI_m(k,1,1)\right)\biggr]\biggr\},\\
{\Pi^{(d)}}^{mn}_{ab}(k)&=&-\frac{1}{2}g^2N\delta_{ab}\biggl[
4J^{mn}(k,m_D,m_D)+4k^m J^{n}(k,m_{D},m_{D})+k^m k^n J(k,m_D,m_D)\biggr],\\
{\Pi^{(e)}}^{mn}_{ab}&=&g^2N\delta_{ab}\delta^{mn}j(m_D).
\end{eqnarray}
Diagrams a), b) and c) were calculated in \cite{kajantie96b} using the Landau
gauge ($\xi=0$). The results of these calculations coincide with ours if in the 
above formulas one sets $\xi=0$.
The diagrams contributing to the 2-point function of $A_0$ are given by
diagrams f) and g): 
\vskip0.3truecm
\hspace{4.5cm}
\epsfbox{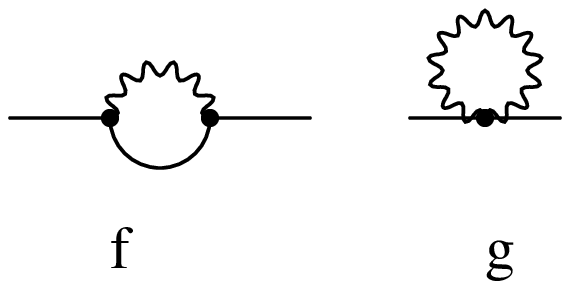}
\vskip0.3truecm
The corresponding analytical contribution is:
\begin{eqnarray}
{\Pi^{(f)}}^{\mu \nu}_{ab}(k)&=&-g^2N\delta_{ab}\delta^{0\mu}\delta^{0\nu}
\biggl[j(m_T)-\frac{k^2+m_{D}^{2}}{m^{2}_{T}}j(m_T)-\frac{\left(m_{D}^{2}+k^2
\right)^2}{m^{2}_{T}}J(k,m_D,0)\\
&&\quad+\frac{1}{m_{T}^{2}}\left(2k^2\left(m_{T}^{2}+m_{D}^{2}
\right)+k^4+
\left(m_{T}^{2}-m_{D}^{2}\right)^2\right)J(k,m_D,m_T)-j(m_D)
\biggr]\nonumber\\
&&-g^2N\delta_{ab}\delta^{0\mu}\delta^{0\nu}\frac{\xi}{m_{L}^{2}}
\biggl[
\left(m_{L}^{2}+m_{D}^{2}+k^2\right)j(m_L)+\left(m_{D}^{2}+k^2\right)^2J(k,m_D,0)
\nonumber\\
&&\qquad\qquad-\left(m_{D}^{2}+k^2\right)^2J(k,m_D,m_{L})
\biggr],\nonumber\\
{\Pi^{(g)}}^{\mu\nu}_{ab}&=&g^2N\delta_{ab}\delta^{0\mu}\delta^{0\nu}
\biggl[2j(m_T)+\xi j(m_L)\biggr].
\end{eqnarray}
Finally the single diagram contributing to the ghost 2-point function is
given by diagram h):
\vskip0.3truecm
\hspace{6cm}
\epsfbox{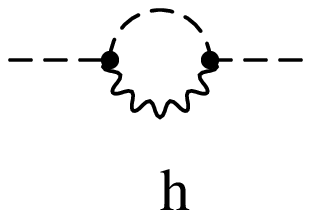}
\vskip0.3truecm
The corresponding analytical contribution is: 
\begin{eqnarray}
{\Sigma}_{ab}(k) &=&-\frac{1}{2}g^2N\delta_{ab}\biggl\{
\biggl[\frac{m_{T}^{2}-\left(k^2+m_{G}^{2}\right)}{4m_{T}^{2}}j(m_T)
-\frac{\left(k^2+m_{G}^{2}\right)^2}{4m_{T}^{2}}J(k,m_G,0)\\\nonumber
&&\qquad\qquad+\left(\frac{\left(k^2+m_{G}^{2}\right)^2}{4m^{2}_{T}}+
\frac{k^2-m_{G}^{2}}{2}+\frac{m_{T}^{2}}{4}\right)J(k,m_G,m_T)
-\frac{1}{4}j(m_G)\biggr]\\\nonumber
&&-\xi\biggl[\left(\frac{\left(k^2+m_{G}^{2}\right)^2}{4m^{2}_{L}}+
\frac{k^2+m_{G}^{2}}{2}+\frac{m_{L}^{2}}{4}\right)J(k,m_G,m_L)
-\frac{1}{4}j(m_G)\\\nonumber
&&\qquad\qquad-\frac{\left(k^2+m^{2}_{G}\right)^2}{4m^{2}_{L}}J(k,m_G,0)
-\frac{3m_{L}^{2}+k^2+m_{G}^{2}}{4m^{2}_{L}}j(m_L)
\biggr]\biggr\}.\\\nonumber
\end{eqnarray}

\section{3d Adjoint Higgs Model on Lattice}
\subsection{Phase Diagram of 3d Adjoint Higgs Model and
Dimensional Reduction}
As it was discussed in section 2 the 
high temperature limit of 4d SU(N) theory is described by 3d adjoint
Higgs model with the following action
\be
S=\int d^3x \biggr({1\over 4} F^a_{ij} F_{ij}^a+\half {(D_i
A_0^a)}^2
+\half m_{D0}^2 A_0^a A_0^a+\lambda_A {(A_0^a A_0^a)}^2 \biggr),
\label{l3adj_c}
\ee
The relations between the parameters appearing in this equation
and those of the original 4d SU(N) theory are \cite{kajantie97b}
\ba
g_3^2&=&g^2(\mu)T\Bigl[1+\gw(L+c_g)\Bigr], \la{g3g}\\
m_{D0}^2&=&\fr13\Bigl(N+\fr12N_f\Bigr)
g^2(\mu)T^2\Bigl[1+\gw(L+c_m)\Bigr],\la{mdg}\\
\lambda_A&=&(6+N-N_f){g^4(\mu)T\over24\pi^2}
\Bigl[1+2\gw(L+c_l^{(N)})\Bigr],
\la{lag}
\ea
where
\ba
L&=&\fr{22N}3\ln{\mu\over\mu_T}-{4N_f\over3}\ln{4\mu\over\mu_T},\\
c_g&=&\fr{N}3,\la{cg}\\
c_m&=&{10N^2+2N_f^2+9N_f/N\over 6N+3N_f},\\
c_l^{(2)}&=&{7/3-109N_f/96\over 1-N_f/8}+\fr23 N_f,\\
c_l^{(3)}&=&{7/2-23N_f/18\over 1-N_f/9}+\fr23 N_f \la{cl3}
\ea
with $\mu_T=4 \pi e^{-\gamma_E} T \sim 7.0555T$.
One can easily verify that the parameters $g_3^2,~m_{D0}$ and $\lambda_A$
do not depend on the renormalization scale $\mu$ at leading order.
Following Ref. \cite{kajantie97b} we introduce the following
dimensionless parameters which will be useful in further analysis
\be
x={\lambda_A\over g_3^2},~~~~y={m_{D0}^2\over g_3^4}.
\ee
The lattice action corresponding to the continium action (\ref{l3adj_c}) 
using standard discretization procedure could be written
as
\ba
S_A&=&\beta \sum_P U_P+\sum_{\bx,\hat i}2 \biggl[\tr aA_0^2(\bx)-\tr
aA_0(\bx)U_i(\bx)
A_0(\bx+\hat i)U_i^\dagger(\bx)\biggr]\nonumber\\
&&+\sum_{\bx}\biggl[(\tilde m_0a)^2\tr aA_0^2(\bx)+a\lambda_A
(\tr aA_0^2)^2\biggr], \la{sa}
\ea
where the first term is the standard Wilson action for the gauge
fields, $A_0=\sum_b \tau^b A_0^b$ with $\tau^b$ being the gararators of the
SU(N) gauge group and $a$ is the lattice spacing. Futhermore,
\be
\beta={2 N\over g_3^2 a}
\ee
and $\tilde m_0$ is the cutoff dependent 
bare mass of the $A_0$ field calculated in lattice regularization.
In the following we will consider only the SU(2) case.
In order to relate the  results of lattice calculations to the physics
of the high temperature phase of the original 4d SU(2) gauge theory, it is
necessary to find the relation between the bare mass $\tilde m_0$ and
and the renormalized mass at scale $\mu$ in $\overline{MS}$ scheme
$m_{D}^r(\mu)$.
Since the theory is superrenormalizable the only running coupling is
the mass parameter. The scale dependence of the renormalized mass is
given by \cite{kajantie97b}
\be
m_{D}^r(\mu)=m_{D0}+{5 (2 g_3^2-\lambda_A) \lambda_A \over 16 \pi^2}
\ln{g_3^2\over
\mu}.
\ee
The bare mass $\tilde m_0$ could be written as 
\be
\tilde m_0=m_{D}^r(\mu)+\delta \tilde m(\mu),
\label{tm0}
\ee
where 
\be
\delta \tilde m(\mu)=-{3.1759114\over 4 \pi a} (4 g_3^2+5 \lambda_A)-
{g_3^4\over 16 \pi^2} 
\biggl[ (20 x-10x^2) (\ln{6\over a \mu}+0.09)+8.7+11.6x\biggr]
\label{deltatm0}
\ee
is the 2-loop lattice counterterm calculated in
Ref. \cite{laine95}.
It is convenient to write the lattice action in terms of antihermitian 
matricies $\tilde A_0$ defined by $\beta \tilde A_0=2 i a^{1/2} A_0$.
Futhermore introducing the bare mass parameter $h=\tilde m_0^2 a^2$
and taking into account that 
\be
\lambda_A \tr (a A_0^2)^2=x \beta (\half \tr \tilde A_0^2)^2 
\ee
the lattice action could be written in the form
\ba
&&
S=\beta \sum_P \half \tr U_P +
\beta \sum_{\bx,\hat i} \half \tr \tilde A_0(\bx) U_i(\bx) \tilde A_0(\bx+\hat i)
U_i^{\dagger}(\bx) + \nonumber\\
&&
\sum_{\bx} \left[-\beta\left(3+\half h\right) \half \tr \tilde A_0^2(\bx) +
\beta x { \left( \half \tr
\tilde A_0^2(\bx)\right)}^2 \right].
\label{act1}
\ea
Making use of Eqs. (\ref{tm0}) and (\ref{deltatm0}) 
the bare mass parameter $h$ could be written in terms of $\beta$ and continuum
parameters $x$ and $y$ as
\be
h(x)={16\over \beta^2}y-{3.1759114(4+5 x)\over \pi \beta}-{1\over \pi^2 \beta^2}
\biggl( (20x-10x^2) (\ln{3\over 2} \beta+0.09)+8.7+11.6 x \biggr).
\label{hx}
\ee
The phase diagram of the 3d SU(2) adjoint Higgs model was established
in Ref. \cite{kajantie97b} in terms of $x$ and $y$ using multicanonical
update technique \cite{berg91}. It has a symmetric
(confinement) phase and the broken (Higgs) phase 
separated for small $x$ by a line
of $1^{st}$ order phase transition. The transition becomes weaker as
$x$ increases and eventually turns to a crossover for some $x>0.3$. 
However, dimensional reduction is applicable only for temperatures
larger than the critical which means that physically relevant values
of $x$ are smaller than $0.2$.
The transition line was obtained by fitting different data points
with a polynomial and was found to be (see Figure \ref{phasexy})
\be
x y_c = 2/(9 \pi^2)(1 + 9/8 x + A x^{3/2} + B x^2 + C x^{5/2} + D x^3),
\label{fityc}
\ee
where 
$A=45.1(2.9),~B=-214.7(20.2),~
C=350.7(44.5),~D=-206.7(31.4)$.

The physics of the high temperature SU(2) gauge theory corresponds to some
line in $xy$-plane, the {\em line of 4d physics} $y_{4d}(x)$. This line
can be calculated using Eqs.(\ref{g3g})-(\ref{lag}) and for SU(2) it
has the following form
\be
y_{4d}(x)={(8-N_f)(4+N_f)\over144\pi^2x}+
{192-2N_f-7N_f^2-2N_f^3\over96(8-N_f)\pi^2}. \la{y_dr2}
\ee
The position of $y_{4d}(x)$ relative to $y_c(x)$ determines the physical phase of the 3d
adjoint Higgs model. It turns out the for physically interesting values
of $x$ $y_{4d}(x)<y_c(x)$ and the physical phase is the broken one. 
The corresponding
situation is shown in Figure \ref{phasexy}.
\begin{figure}[tb]

\vspace*{-0.5cm}

\epsfysize=18cm
\centerline{\epsffile{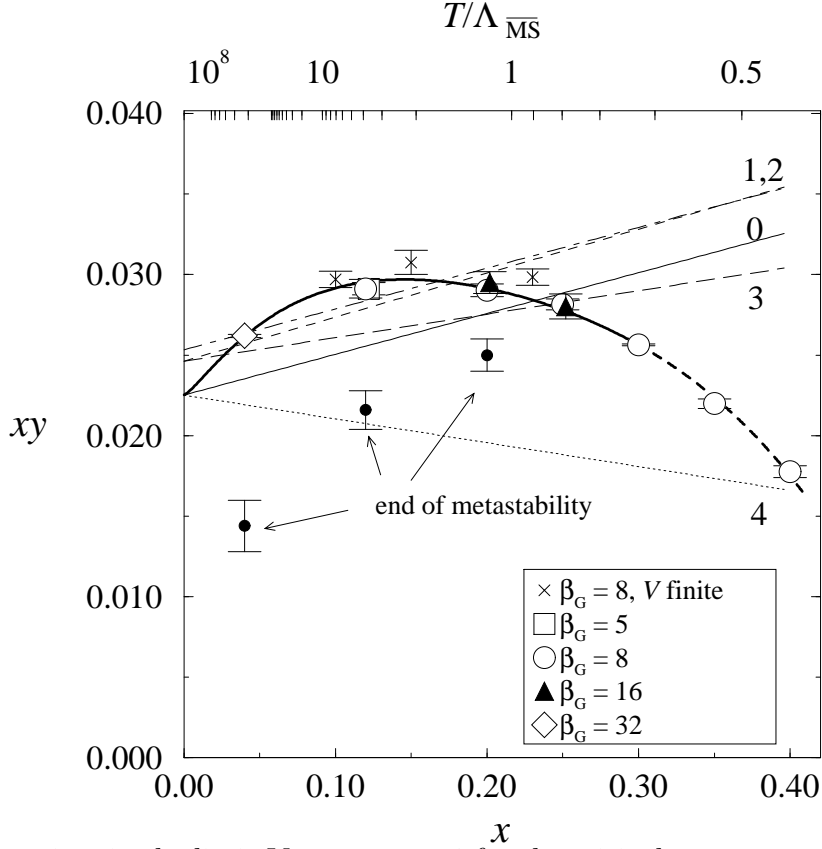}}

\vspace*{-7cm}

\caption[a]{Data points in the limit $V\to\infty,a\to0$ for
the critical  curve $y=y_c(x)$  (multiplied by $x$).  The thick line is
the fit to the $V=\infty$ extrapolated data.  
The dashed line
marks the region where the transition turns into a cross-over.
The straight lines are the 4d$\to$3d curves of eq.~(\ref{y_dr2})
marked by the value of $N_f$. The top scale shows the values of
$T/\lambdamsbar$ (Figure is taken from Ref. \cite{kajantie97b}).}
\la{phasexy}
\end{figure}

The fact that the broken phase corresponds to the high temperature
phase is self-contradictory because dimensional reduction is valid if
$A_0 \ll \sqrt{T}$, but in the broken phase $A_0 \sim 1/g$. However,
the line of 4d physics lies in the region where the symmetric phase is
still metastable for finite lattice volume as it is indicated in
Figure~\ref{phasexy}. Using this fact it was suggested in Ref.
\cite{kajantie97b} that one can perform simulations along the line
of 4d physics $y_{4d}(x)$ analyze samples in the symmetric branch of the
mixed phase. The obvious problem with this approach is that the metastable
phase disappears in the infinite volume limit. 

A different approach for fixing parameters appearing in Eq.
(\ref{act1}) was proposed in Ref. \cite{bielcoll}. 
This approach is based on matching  different quantities 
calculated with weak coupling expansion in the 4d lattice gauge theory 
and the corresponding
3d effective theory at some small physical distance, where the weak
coupling expansion is applicable.
In this approach the bare mass parameter is given by
\cite{bielcoll}
\be
h=g^2 \biggl(\tilde \Pi_{00}+2 L_0^{-1}\biggl[ 2 {1\over L_s^3} \sum_{{\bf
k}\ne 0} {1\over 4 \sum_{i=1}^3 \sin^2(k_i/2)}+{1\over 12
L_s^3}\biggr]\biggr),
\label{h4d}
\ee
where $L_0$ and $L_s$ are the temporal and spatial lattice sizes of
the original 4d lattice and $g$ is the 4d gauge coupling. $\tilde \Pi_{00}$
was calculated in Ref. \cite{irback91}. 
The other two couplings appearing in the lattice action are
given by 
\ba
&&
\beta={4 \over L_0 g^2 (1+\alpha g^2)}\\
&&
x={g^2\over(3 \pi^2 (1+\alpha g^2))},
\ea
where $\alpha$ is a slightly volume dependent factor.
The gauge coupling $g$ was found by matching the Polyakov loop
correlator measured in Monte-Carlo simulations to its perturbative
expression at distance $R=(4 T)^{-1}$ \cite{irback91}. The numerical
values of the coupling for lattices considered in Ref. \cite{bielcoll}
together with the physical temperature are summarized in the following Table
\footnote{The numerical values of $h$ given in \cite{bielcoll}
are not correct. The correct values of $h$ calculated from Eq.
(\ref{h4d}) are slightly different from those given in
Ref.\cite{bielcoll}.}
\begin{center}
\vskip0.5truecm
\begin{tabular}{|l|l|l|l|l|l|}
\hline
$~~~~~~~~~~~~~$  &$T/T_c$  &$~g^2$ &$\beta$  &$~~h~~$ &$~~x~~$\\
\hline
$4 \times 16^3$  &$3.5$    &$1.26$ &$13.70$  &$-0.28$ &$0.046$\\
$(\alpha=-0.06)$  &$6.0$    &$1.18$ &$14.60$  &$-0.27$ &$0.043$\\
$(\tilde \Pi_{00}=-0.0384)$ &$$    &$$ &$$ &$$ &$$\\
\hline
$4 \times 24^3$  &$2.0$    &$1.43$ &$12.25$  &$-0.31$ &$0.053$\\
$(\alpha=-0.059)$  &$3.5$    &$1.28$ &$13.54$  &$-0.28$ &$0.047$\\
$(\tilde \Pi_{00}=-0.0431)$  &$6.0$    &$1.19$ &$14.48$  &$-0.26$ &$0.043$\\
\hline
\end{tabular}
\vskip0.3truecm
Table 5: Parameters of the effective theories for different temperatures
and lattice volumes according Ref.\cite{bielcoll}.
\end{center}
Now we look for the physical phase in this approach.
This question was addressed in Ref. \cite{kark94} and it was concluded
that the physical space is the symmetric one. In Ref. \cite{kark94}
the values of $h$ corresponding to the phase transition were determined
for $4\times 12^3$, $4\times 16^3$ and $4 \times 24^3$ lattices and
found to be $h_{c}=-0.2623(1),~-0.2803(1)$
and $-0.30(1)$. Extrapolation to infinite volume
yields $-0.2943(2)$. 
Now using Eq. (\ref{hx}) we can evaluate $h_{c}$ from the result of the
polynomial fit for $y_c(x)$ given by Eq. (\ref{fityc}) 
at $x \sim 0.047$ and $\beta=13.54$, which yields $h_c=-0.2733(25)$.
Thus the values of $h$ corresponding to the transition are
underestimated.
We can estimate the infinite volume limit of $h$ defined by Eq. 
(\ref{h4d}), the calculation yields $h=-0.2699$. 
However, Eq. (\ref{h4d}) is only valid at 1-loop level, at 2-loop level
$h$ is given by Eq. (\ref{hx}) which yields $-0.2789$ which 
definitely corresponds to the broken phase.
Therefore the conclusion that the symmetric phase is the physical reached in
\cite{kark94} is sensitive to the approxiomation made in the analysis,
and thus is unreliable.

\subsection{Extracting the Screening Masses from the Propagators}

For the analysis of the large distance behaviour of the propagators
it is convenient to introduce the so-called local masses. They are
defined by the following relation
\be
{G_i(z)\over G_i(z+1)}={\cosh(m_i(z) (z-N_z/2))\over
\cosh(m_i(z)(z+1-N_z/2))},
\ee
where $i=D,T$ and the correspondig propagators are defined by Eqs.
(\ref{gdz}) and (\ref{gtz}) and $N_z$ is the extension of the lattice
in $z$ direction. If the propagators show exponential
decay starting from some value of $z$ the corresponding local
masses reach a plateau. Therefore the local masses are usefull in
defining the starting point of the fit interval. 
In order to determine properly the fit interval correlated fit
should be used together with the $\chi^2$ criteria. If one uses
simple uncorrelated fit the corresponding values of $\chi^2/d.o.f$
are very small which does not allow to choose the fit interval
properly.
In Figure
{\ref{mulocb8} the local magnetic masses measured on $16^2 \times
64$ lattice at $\beta=8,~x=0.09$ and $y=0.4007$ are shown.  
The filled squares denotes the data points which were used in the fit
and yields good $\chi^2/d.o.f$. The fit based on this data points
yields the value $0.422(6)g_3^2$ for the magnetic mass.
\begin{figure}
\epsfxsize=9cm
\epsfysize=7cm
\centerline{\epsffile{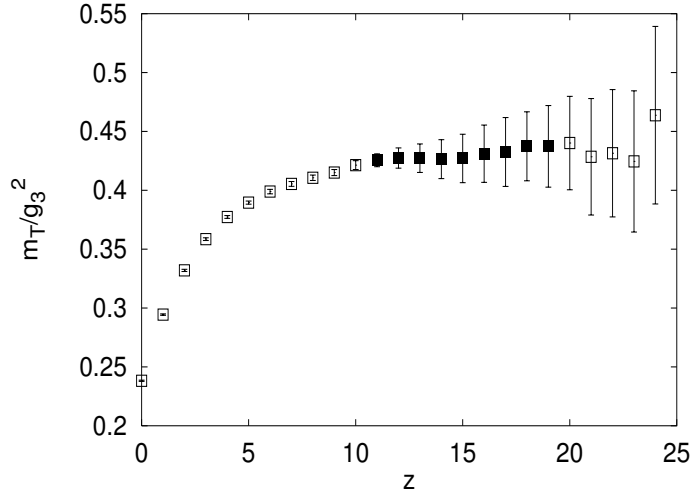}}
\caption{The local magnetic masses in units of $g_3^2$ as
function of $z$ measured at $\beta=8,~x=0.09$ and $y=0.4007$.
The filled squares denote the data points used for extracting the
magnetic screening mass}
\label{mulocb8}
\end{figure}
 
In Ref. \cite{karsch96a} the magnetic mass was measured in the symmetric
phase of the 3d SU(2) Higgs model. The magnetic mass for the 3d pure 
gauge theory was also determined there and its value was found to be
somewhat larger but close to the corresponding value of the 3d Higgs
model. In Ref. \cite{karsch96a} the magnetic masses were extarcted from
propagators using uncorrelated fit over the interval 
$[8,N_z-8]$ with the following fit anzatz
\be
G_T(z)=A \biggl(\exp(-m_T z)+\exp(-m_T (N_z-z))\biggr)+B
\ee
where the propagators were measured for $\beta=9$ 
on $16^2 \times N_z$ lattice, with
$N_z=32-128$. For $N_z=64$ the value of the magnetic mass was found to
be $0.39(2)g_3^2$ (see Table 1 in Ref. \cite{karsch96a}). 
Using the same procedure for our data we obtaine $m_T=0.408(8)g_3^2$
which is compatible with the result of Ref.\cite{karsch96a}.
However, most of simulations in Ref. \cite{karsch96a} were done
on $16^2 \times 32$ lattice. On this lattice the magnetic mass was
found to be $0.35(1)g_3^2$. 

In order to get in contact with the results of Ref.\cite{karsch96a}
we have performed simulations on $16^2 \times 32$ lattice at $\beta=9$
in 3d pure gauge theory and extarcted the magnetic mass using the 
procedure outlined above. The corresponding value of the magnetic mass
was found to be $0.39(3)g_3^2$. This value is smaller than
corresponding value obatined by us for $16^2 \times 64$ lattice
and this is probably due to the 
fact that on such lattices the magnetic mass cannot be reliably 
determined because the local masses do not show a plateau (see Figure 2
in Ref. \cite{karsch96a}). We think that the small difference between
this value and the corresponding value found in Ref.\cite{karsch96a} is
probably due to the presence of the Higgs fields. 

The above
analysis shows the importance of the finite size effects and the proper
procedure of extracting the screening masses. 
 
\subsection{The Dependence of the Screening Masses on the Choice of the 
Temperature
Scale and Lattice Spacing}

As it was already explained in section 4.1 the temperature scale in
the effective theory is set by parameter $x$ and its relation to the
renormalized 4d gauge coupling $g(T)$. In section 4.1 $g(T)$ was
determined by the 2-loop formula in $\overline{MS}$ scheme with $\mu=2
\pi T$. It seems to be interesting to examine to what extent the
numerical results on the screening masses presented in section 4.3 
depend on the definition of $g(T)$ and/or the choice of the
renormalization scale. In Figure \ref{mass1l} the temperature
dependence of the screening masses is shown for the case when the
temperature scale is set by the 1-loop level renormalized coupling 
\begin{figure}
\vspace{-2cm}
\epsfysize=7cm
\epsfxsize=9cm
\centerline{\epsffile{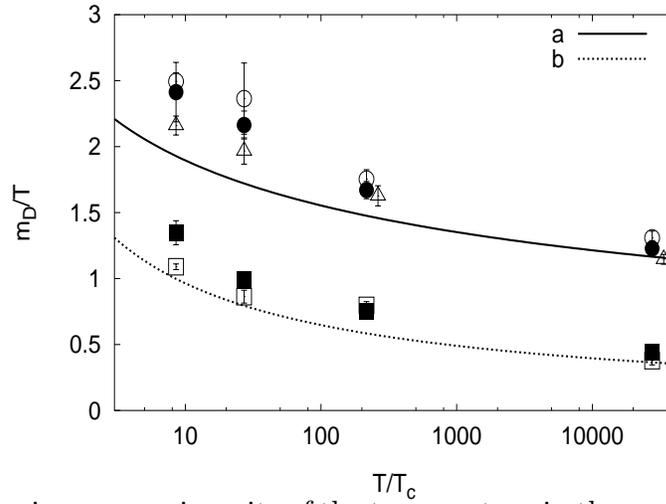}}
\vspace{-0.7cm}
\caption{The screening masses in units of the temperature  in
the case when the temperature scale is set by the 1-loop level
renormalized coupling (see text for details)
Shown are the Debye mass
$m_D$ for the first (filled circles) and the second (open circles)
set of $h$, and
the magnetic mass  $m_T$ for the first (filled squares)
and the second (open squares) set of $h$.
Also shown there are the values of the Debye mass measured in
metastable phase along the perturbative line of 4d physics
(open triangles).
The line (a) and line (b) represent the fit for the temperature
dependence of the Debye and the magnetic mass from 4d
simulations
from \cite{heller98}. Some data points at the temperature $T \sim 260
T_c$ and $ \sim 30000 T_c$ have been shifted in the temperature scale for
better visualization.}
\label{mass1l}
\end{figure}

\be
g^{-2}(T)={11\over 12 \pi^2} \ln{\mu\over \Lambda_{\overline{MS}}}
\label{gr1l}
\ee
with $\mu=2 \pi T$ and $\Lambda_{\overline{MS}}=T_c/1.06$.
As one can see from Figure \ref{mass1l} the screening masses are
systemetically overestimated compared to the 4d results in this case.

It was mentioned in section 4.4 that the square root of the 
spatial string tension in the
deconfined phase of the 4d SU(2) gauge theory is expected to scale
with $g_3^2\sim g(T)^2 T$. This fact can be used for non-perturbative 
definition of $g(T)$. In Ref. \cite{bali93} it was found that the
temperature dependence of the square root of the 
spatial string tension is well
described by the formula
\be
\sqrt{\sigma_s(T)}=(0.369 \pm 0.014)g^2(T) T
\ee
if the running coupling constant $g$ is chosen according to
\be
g^{-2}(T)={11\over 12 \pi^2} \ln{T\over \Lambda_T}+
{17\over 44 \pi^2}\ln\left[2\ln{T\over
\Lambda_T}\right].
\ee
with $\Lambda_T=0.076(13) T_c$ \cite{bali93}. 
This running coupling constant $g(T)$
is equivalent to the 2-loop running coupling in $\overline{MS}$ scheme if
the renurmalization scale is chosen to be $\mu \sim 13.3 T$.
However the data on the temperature dependence of the spatial string
tension can be also well fitted with the formula $\sqrt{\sigma_s(T)}=
(0.334 \pm 0.014)g^2(T) T$ if $g(T)$ is chosen according to
\be
g^{-2}(T)={11\over 12 \pi^2} \ln{T\over \Lambda_T}
\ee
with $\Lambda_T=0.050(10) T_c$ \cite{bali93}. 
This running coupling corresponds to
the running coupling constant in $\overline{MS}$ scheme with $\mu \sim 20
T$. The value of the string tension in 3d SU(2) gauge theory is 
$\sqrt{\sigma_3}=0.3340(25)g_3^2$ \cite{teper92}.
Now if the temperature scale is set by the running coupling $g(T)$
defined by Eq. (\ref{gr1l}) but with $\mu \sim 20 T$  very good
agreement between the 3d and 4d data is found.

From the above analysis it is clear that the definitions of $g(T)$ and
values of the normalization scale $\mu$ which yield good agreement
between 3d and 4d data are compatible with values of $g$ found
non-perturbatively from the analysis of the spatial string tension.
On the other hand the values of gauge coupling used in Ref.
\cite{bielcoll} are by factor of 2 smaller than these.

The matching analysis in section 4.3 was done at $\beta=16$.
It is important to check the sensitivity of these results 
to the choice of $\beta$ which defines the lattice spacing.
The dependence of the Debye masses on $\beta$ was analyzed 
for $x=0.05$ and for all three values of $y$ ($h$) used
in our matching analysis. The corresponding results are
shown in Table 6.
\begin{center}
\vskip0.3truecm
\begin{tabular}{|l|l|l|l|}
\hline
$\beta$ &$12$ &$16$ &$20$\\
\hline
$y=0.6721$ &$1.84(21)$ &$1.69(6)$ &$1.68(12)$\\
\hline
$y=0.5914$ &$1.57(6)$  &$1.68(4)$ &$1.65(6)$ \\
\hline
$y=0.4756$ &$-$       &$1.48(5)$ &$1.44(12)$\\
\hline
\end{tabular}
\vskip0.3truecm
Table 6: Debye masses for different values of $\beta$ for
$x=0.05$
\end{center}
As one can see from the results presented in  Table 6 no
lattice spacing dependence for the Debye mass can be observed.
A similar analysis should be done for the magnetic mass. But
for $\beta=20$ the magnetic mass can be measured reliably only for
lattices larger than $32 \times 64$. Using the fact that the
magnetic mass calculated in the 3d adjoint Higgs model is close
to the magnetic mass of 3d pure gauge theory one and based on results
presented in section 4.4 we can conclude that the magnetic mass
also does not depend on $\beta$.
\section{Self-energy Contributions in 3d fundamental + adjoint Higgs
model}
Below we list graphycally the additional diagrams contributing to 
the $A_0$ ({\bf a-i}), Higgs boson ({\bf j-l}) and the vector
boson ({\bf m}) self-energies and the vacuum expectation value ({\bf n}).
Here solid lines denotes the propagator of the Higgs field, dashed
lines corresponds to Goldstone fields, double line corresponds to the
adjoint Higgs field and finally the dotted line corresponds to 
ghost fields. Wawy lines as usually corresponds to vector fields.
\vskip0.4truecm
\unitlength1cm
\epsfysize=7.4cm
\centerline{
\epsfbox{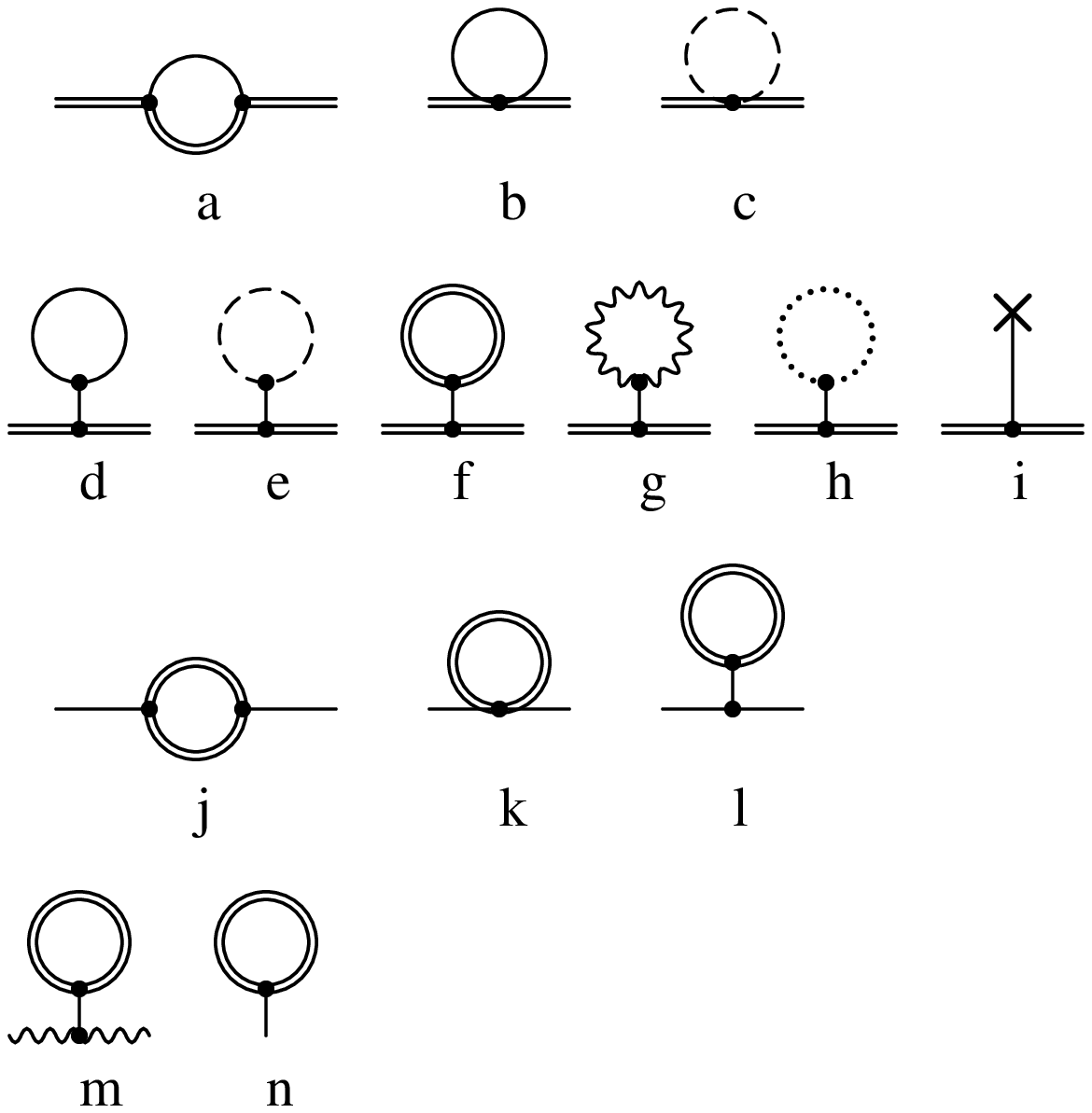}
}


\newpage
\section*{Acknowledgements}

A substantial part of the work presented in this thesis was carried
out at the University of Bielefeld under the supervision of Prof. Karsch.
Numerical Monte-Carlo simulations have been performed on CRAY T3E at
HLRZ J\"ulich and at High Performance Computing Center Stuttgart. 
I thank the Peregrination II Fund and Center of Interdisciplinary
Research (ZiF) for the financial support during my stay at the
University of Bielefeld. Travel grants from the doctoral program
of E\"otv\"os University are gratefully acknowledged.
I am grateful to Prof. Patk\'os for giving many usefull advice 
during the preparation of this thesis. Different parts of the work
presented in this thesis profited much from enjoyable discussions
with W. Buchm\"uller, Z. Fodor, A. Jakov\'ac, O. Philipsen and
K. Rummukainen. I am also grateful to O. Philipsen and K. Rummukainen
for sending some of their numerical data and e-mail correspondence.
I thank M. Oevers and Zs. Sz\'ep for collaboration.
Finally I thank my parents and my wife for their support during my
graduate studies.

\end{document}